\documentclass[prd,twocolumn,nofootinbib,superscriptaddress,amsmath,amssymb,groupedaddress]{revtex4-1}
\usepackage{mathtools}
\usepackage{graphics}
\usepackage{graphicx}
\usepackage{dcolumn}
\usepackage{bm}
\usepackage{dsfont} 
\usepackage{amsmath,amssymb}
\usepackage{verbatimbox}
\usepackage{hyperref}
\usepackage{tabularx}
\usepackage{epsf,epsfig}
\usepackage[normalem]{ulem}
\usepackage[usenames]{color}
\usepackage{multirow}
\usepackage{makecell}
\usepackage{diagbox}
\usepackage[abs]{overpic}
\allowdisplaybreaks
\hypersetup{
    colorlinks=true,
    linkcolor=blue,
    filecolor=magenta,      
    urlcolor=blue,
    citecolor=blue
}
\newcommand{\ISCO}{{\mbox{\tiny ISCO}}}
\newcommand{\EdGB}{{\mbox{\tiny EdGB}}}

\newcommand{\X}{{\mbox{\tiny X}}}

\newcommand{\EH}{{\mbox{\tiny EH}}}
\newcommand{\dCS}{{\mbox{\tiny dCS}}}
\newcommand{\ppN}{{\mbox{\tiny ppN}}}

\newcommand{\K}{{\mbox{\tiny K}}}

\newcommand{\PK}{{\mbox{\tiny PK}}}

\definecolor{red(ncs)}{rgb}{0.77, 0.01, 0.2}

\usepackage{array}
\newcolumntype{C}[1]{>{\centering\let\newline\\\arraybackslash\hspace{0pt}}m{#1}}

\newcolumntype{C}[1]{>{\centering\arraybackslash}m{#1}}

\begin{document}

\title{Asymptotically flat, parameterized black hole metric preserving Kerr symmetries}

\author{Zack Carson}
\author{Kent Yagi}

\affiliation{Department of Physics, University of Virginia, Charlottesville, Virginia 22904, USA}

\date{\today}


\begin{abstract}
Recently the Event Horizon Telescope Collaboration, with very-long baseline interferometric observations, resolved structure at the scale of $\sim5$ Schwarzschild radii about the center of M87$^*$, the supermassive black hole resident at the center of Messier 87.
This important observation has paved the way for testing what is known as the ``no-hair'' theorem, stating that isolated black holes are described by the Kerr metric, parameterized only by their mass and spin.
Generic, parameterized spacetimes beyond Kerr allow one to arbitrarily test the no-hair theorem for deviations from the Kerr result with no prior theoretical knowledge or motivation.
In this paper, we present such a new general, stationary, axisymmetric and asymptotically flat black hole solution with separable geodesic equations (thus preserving symmetries of a Kerr black hole), extending the previous work of Johannsen.
In this new metric, five free non-linear functions parameterically deviate from the Kerr result, allowing one to effectively transform to many alternative black hole solutions present in the literature.
We then derive analytic expressions for the Keplerian and epicyclic frequencies, the orbital energy and angular momentum, and the location of the innermost stable orbit of circular equatorial particle orbits. We also compute the image of the photon rings in the new spacetime, which correspond to the boundary of the black hole shadow image taken by the Event Horizon Telescope.
We finally compare each quantity for the Kerr result against various parameterizations of the metric, finding that, especially for highly rotating black holes, the two solutions disagree significantly.
Such a metric parameterization allows one to perform the no-hair tests in a model-independent way, and finally map constraints to specific alternative theories of gravity.
\end{abstract}

\maketitle


\section{Introduction}\label{sec:intro}
The no-hair theorem tells us that the spacetime surrounding isolated, uncharged black holes (BHs) is uniquely described by the Kerr metric.
This famous metric is asymptotically flat, stationary, axisymmetric, and is parameterized by only two BH parameters: the mass ($M$), and the spin ($a$).
Further, within this metric exists an event horizon masking the true coordinate singularity found within.
Supposing the no-hair theorem holds true, we can expect all BH observations, e.g. of photon orbits about them, to agree with those as described by the Kerr metric.
To date, there has been no sufficient evidence that points to otherwise~\cite{Menou:1997ys,Narayan:1997xv}, however only now have our capabilities advanced enough to accurately probe the near-field structure of these objects.
In the near future, we may have the opportunity to resolve effects that go beyond the standard Kerr model.

To date, several tests of the no-hair theorem have been enacted and proposed, as reviewed in~\cite{Yunes:2013dva,Gair:2012nm,Johannsen:2011mt,Bambi:2015ldr}.
Such tests include observations of pulsar-BH binaries~\cite{Liu_2012,Pfahl_2004,Wex_1999}, the orbits of SMBH stellar companions~\cite{Sadeghian:2011ub,Merritt:2009ex,Will_2008}, the electromagnetic accretion flows of SMBHs (continuume spectrum, iron lines, quasi-periodic oscillations, etc.)~\cite{Johannsen:2011dh,Bambi:2013sha,Bambi:2012at,Johannsen:2012ng,Bambi:2011jq,Bambi:2012ku,Johannsen:2010ru,Johannsen:2010xs,Psaltis_2011,Johannsen_2016,Krawczynski_2012,PhysRevD.83.103003,Johannsen_2010,Maselli:2014fca,Maselli:2017kic}, the quasinormal ringdown modes of a post-coalescence perturbed remnant BH~\cite{Berti:2007zu,Dreyer:2003bv,PhysRevD.73.064030}, and even the gravitational wave observations of extreme mass-ratio inspirals of super-massive and stellar-mass BHs~\cite{Vigeland:2011ji,Gair:2011ym,Apostolatos:2009vu,Gair:2007kr,Collins:2004ex,Glampedakis:2005cf,PhysRevD.84.064016,PhysRevD.81.024030,PhysRevD.78.102002,PhysRevD.69.082005,Barack:2006pq,Johannsen:2011mt,Mandel_2014,PhysRevD.77.064022,PhysRevD.56.1845,PhysRevD.52.5707,Isi:2019aib}.
No significant deviations from the Kerr result have been detected so far.

Recent developments in the very-long baseline interferometric (VLBI) array in the Event Horizon Telescope (EHT) have given us the unique opportunity to probe the spacetimes of supermassive BHs (SMBHs).
With the sole purpose of imaging SMBHs, the EHT and VLBI span the entire globe with an array of millimeter and sub-millimeter instruments, effectively creating an Earth-sized telescope~\cite{doeleman2009imaging}.
Currently operating with 8 telescopes, the EHT achieved the impressive feat of resolving the lensed photon orbits about Messier 87's central SMBH M87$^*$, with an angular resolution of $\sim50\mu$as~\cite{1435171,1435174,1435175,1435177,1435168}.
Even with these extraordinary results, strong deviations from the no-hair theorem have not been detected.
However, the EHT will continue to develop with the addition of new facilities, with a planned resolution increase of $\sim40\%$ over the next 3-5 years.
Along with the addition of new SMBH targets, this boost in resolution and image fidelity will further provide us with the ability to probe these extreme spacetimes.

While the Kerr BH metric, developed under the solutions to Einstein's theory of general relativity (GR), has had unprecedented success in describing our BH observations, we must continue to test the no-hair theorem.
In particular, to this day we have left unanswered several important observational questions regarding the nature of the universe, which could potentially be described by a new theory of gravity.
For example, the elusive ``dark energy'' and ``dark matter'' accelerating the expansion of our universe, and the rotation of our galaxies~\cite{Jain:2010ka,Salvatelli:2016mgy,Koyama:2015vza,Joyce:2014kja,Famaey:2011kh,Milgrom:DarkMatter,Milgrom:2008rv,Clifton:2011jh,Joyce:2014kja}, or the early universe's rapid inflationary period~\cite{Joyce:2014kja,Clifton:2011jh,Famaey:2011kh,Koyama:2015vza} and our current universe's extreme matter/anti-matter asymmetry~\cite{Clifton:2011jh,Famaey:2011kh}, or even the issue of unifying GR and quantum mechanics~\cite{Clifton:2011jh,Joyce:2014kja,Famaey:2011kh,Milgrom:2008rv,Jain:2010ka,Koyama:2015vza}, all remain unanswered.
Such questions could potentially be explained by a new theory of gravity, which would certainly exhibit itself within the strong-gravity BH spacetimes currently described by the Kerr metric.

In order to sufficiently test these spacetimes for the possibility of non-Kerr effects, we must first model a ``beyond-Kerr'' spacetime in a generic way~\cite{Collins:2004ex,Glampedakis:2005cf,Vigeland:2011ji,PhysRevD.81.024030,Johannsen:2011dh,Manko,Cardoso:2014rha,Rezzolla:2014mua,Konoplya:2016jvv,Konoplya:2020hyk}.
Ideally, each metric element should parametrically deviate from the Kerr metric separately, in such a way that the Kerr spacetime is obtained when all deviations vanish.
In this paper, we restrict ourselves to BHs preserving the symmetries of Kerr BHs, namely asymptotically flat, stationary, axisymmetric, and with separable geodesic equations.
The latter condition avoids chaotic particle orbits, and equates to the existence of a fourth constant of motion, in addition to the proper mass, the energy and angular momentum, and the so-called ``Carter constant''~\cite{PhysRevD.78.102002,Carter}.
The metric considered here is a more broad example of the general class of metrics presented in~\cite{Konoplya:2018arm} which admit separable Klein Gordon equations, and is reduced to the latter for certain assumptions on the beyond-Kerr functions presented here.
Several such metrics have been derived in the literature~\cite{Vigeland:2011ji,Johannsen:2015pca,Konoplya:2018arm}, each with one or more parametric deviations which reduce to Kerr when vanishing.
See also Ref.~\cite{Chen:2019jbs} where the authors obtained several separable spacetimes from the Newman-Janis algorithm, and present a Venn diagram showing the relationship among such metrics and others found in the literature.

Johannsen~\cite{Johannsen:2015pca} designed a Kerr-like BH solution to the Einstein field equations which is stationary, axisymmetric, asymptotically flat, and contains four constants of motion, and an event horizon.
His metric depends non-linearly on four free functions which parametrically deviate from the Kerr solution, and is general enough that it can be mapped to several other known BH solutions~\cite{Johannsen:2015pca}. Following this, Johannsen constrained several of the free-functions from weak-field solar system observations~\cite{Williams:2004qba}, and proceeded to derive expressions for several spacetime properties including the orbital energy and angular momentum, the Keplerian and epicyclic frequencies, and the location of the innermost stable circular orbit (ISCO).
Further, the same author derived expressions for the photon rings of the parameterized beyond-Kerr BH in~\cite{Johannsen:2015qca}.

In this paper, we follow the work of Johannsen~\cite{Johannsen:2015pca} and design a more generic, stationary and axisymmetric, asymptotically flat Kerr-like BH metric with separable structure.
The components of the inverse metric for a generic spacetime with separability structure were derived in~\cite{Benenti}. 
Such a metric has been used to construct a generic beyond-Kerr (inverse) metric with separable structure in~\cite{Yagi:2012ya} (Appendix B) and also recently in~\cite{Papadopoulos:2018nvd}, which contains five arbitrary functions of $r$ and five arbitrary functions of $\theta$.

We construct the new metric as follows. We first introduce the most generic deviation into the contravariant Kerr metric in such a way that the Hamilton-Jacobi separability condition is preserved.
The new metric is then simplified by imposing the constraint of asymptotic flatness at radial infinity. 
Further, we impose constraints consistent with the weak-field solar system tests as Johannsen did in~\cite{Johannsen:2015pca}, however we note that such constraints may not be explicitly valid in the strong-gravity regions surrounding BHs.

We follow this up by exploring several properties of the new spacetime.
We first locate the positions of the event horizon, Killing horizon, and ergosphere, finding that the former two reduce to the Kerr horizons, and the latter depends on just one of the 5 free functions found in the metric.
We additionally explore the spheroidicity conditions found in Ref.~\cite{Glampedakis:2018blj}, where we find the $\theta$-independent functions to admit Kerr-like spherical photon orbits.
Following this, we investigate the orbital properties of circular equatorial particle orbits, finding analytic expressions for the orbital energy and angular momentum, the Keplerian and epicyclic frequencies, and also the location of the ISCO.
We next derive analytic expressions for the photon rings as can be observed by e.g. the EHT, and present plots of the viewing plane as seen by a distant observer at radial infinity for several parameterizations of the metric.
We then demonstrate the effect each parameterization has on each of the above BH properties, and also investigate the presence of naked singularities emergent for certain parameterizations.
Finally, we produce the required mappings that relate the new metric to seven other BH solutions found in the literature~\cite{Papadopoulos:2018nvd,Randall:1999ee,Aliev:2005bi,Jai-akson:2017ldo,Ding:2019mal,Kerr-Sen,Kanti_EdGB,Maeda:2009uy,Sotiriou:2014pfa,Ayzenberg:2014aka,Jackiw:2003pm,Yagi_dCS,Yunes_dcs,Bardeen,Kumar:2019uwi,Pani:2011gy,Kumar:2020hgm}.

\subsection{Summary of the new metric}

The final metric for a BH with mass $M$ and spin $a = S/M$ with $S$ being the magnitude of the spin angular momentum is given by
\begin{align}
\allowdisplaybreaks
\nonumber g_{tt}&=- \frac{\tilde\Sigma \left(\Delta -a^2 A_2^2 \sin ^2\theta \right)}{\tilde \rho^4},\\
\nonumber g_{rr}&=\frac{\tilde\Sigma}{A_5\Delta},\\
\nonumber g_{\theta\theta}&=\tilde\Sigma,\\
\nonumber g_{\phi\phi}&=\frac{\tilde{\Sigma } \sin^2\theta \left[\left(a^2+r^2\right)^2A_1^2 -a^2 \Delta  \sin ^2\theta \right]}{\tilde \rho^4},\\
g_{t\phi}&=-\frac{a \tilde{\Sigma } \sin^2\theta  \left[ \left(a^2+r^2\right)A_0 - \Delta\right]}{\tilde \rho^4}, \label{eq:CYmetric}
\end{align}
%
with
\begin{widetext}
\begin{equation}
\tilde \rho^4 \equiv  \left[ \left(a^2+r^2\right)A_1 -a^2 A_2   \sin ^2\theta \right]^2 + a^2 \left(a^2+r^2\right)(A_0-A_1A_2) \sin ^2\theta \left\lbrack\frac{a^2+r^2}{\Delta} (A_0+A_1 A_2)-2  \right\rbrack,
\end{equation}
\end{widetext}
and
\begin{align}
\label{eq:Sigma_tilde}
\tilde\Sigma&\equiv\Sigma+f(r)+g(\theta),\\
\label{eq:Sigma}
\Sigma&\equiv r^2+a^2\cos^2\theta,\\
\label{eq:Delta}
\Delta&\equiv r^2+a^2-2Mr.
\end{align}
The arbitrary functions can be expanded about spatial infinity as
\begin{align}
A_i(r)&\equiv 1+ \sum\limits_{n=1}^\infty\alpha_{in}\left(\frac{M}{r}\right)^n, \quad (i=0,1,2,5),\\
f(r)&\equiv r^2\sum\limits_{n=1}^\infty \epsilon_n\left(\frac{M}{r}\right)^n,\\
g(\theta)&\equiv M^2\sum\limits^\infty_{n=0}\gamma_n P_n(\cos\theta),
\end{align}
for Legendre polynomials $P_n(\cos\theta)$.
Here, the parameters $\alpha_{in}$, $\epsilon_n$ and $\gamma_n$ control the amount of deviation from Kerr. We can set $\alpha_{01} = 0$ and $\alpha_{11} = \epsilon_{1}/2$ without loss of generality by rescaling $M$ and $a$. One can further impose $\epsilon_1 =\epsilon_2= \alpha_{51}=\alpha_{12}=g(\theta)=0$ to satisfy solar system bounds, though such weak-field constraints may not apply to spacetime outside of a BH. The difference from Johannsen's metric in~\cite{Johannsen:2015pca} is that we have introduced a new radial function $A_0$ that enters in $g_{tt}$, $g_{t\phi}$ and $g_{\phi\phi}$. The above metric reduces to the Johannsen one in the limit $A_0 \to A_1 A_2$. 
We believe this new metric is the most general stationary, axisymmetric, asymptotically flat, and separable spacetime.

\subsection{Organization}

Let us now present the outline of the following paper.
In Sec.~\ref{sec:CYst} we derive our new spacetime metric preserving Kerr symmetries, as well as compute the locations of the event horizon, Killing horizon, and ergosphere.
We follow this up in Sec.~\ref{sec:astrophysical} with a discussion on the theoretical underpinnings of the properties of the new spacetime, as well as their dependencies on the 5 deviation parameters of this metric.
This includes the Keplerian and epicyclic frequencies of orbiting particles, the orbital energy and angular momentum of orbiting particles, the location of the ISCO, and the photon rings.
Finally, the transformations that take one from this metric to other spacetimes found in the literature are described in Sec.~\ref{sec:mappings}.
Finally, we offer concluding remarks in Sec.~\ref{sec:conclusion}.
Throughout this paper, we have adopted geometric units such that $G=1=c$.
Additionally, we use derivative notation such that $\partial_\X\equiv \frac{\partial}{\partial X}$ and $\partial^2_\X\equiv \frac{\partial^2}{\partial X^2}$.


\section{A new metric preserving Kerr symmetries}\label{sec:CYst}
In this section, we present a new general spacetime metric with parameterized deviations beyond GR.
This spacetime preserves the Kerr symmetries, and is axisymmetric, stationary, and asymptotically flat.
We also show that the event and Killing horizons in such a spacetime coincides with those of a Kerr BH, and then we derive the locations of the ergosphere.
Finally, we compute the spheroidicity conditions of Ref.~\cite{Glampedakis:2018blj}, showing them to be independent of $\theta$, thus admitting Kerr-like spherical photon orbits.

\subsection{The metric}\label{sec:CYmetric}
Here we compute the new spacetime metric as used throughout this analysis.
We obtain this metric by following and modifying the analysis thoroughly done by Johannsen in Ref.~\cite{Johannsen:2015pca}.
There, a regular parameterized BH solution was created to be stationary, axisymmetric, asymptotically flat, and separable.
The latter property comes forth from the existence of a fourth constant of motion, the Carter-like constant~\cite{Carter}.
The metric presented here, while very similar to Johannsen's, is more general and admits an additional deviation function from the Kerr metric.

We begin with the Kerr metric for a rotating BH.
This well-known spacetime has a line element given by
\begin{align}
\nonumber &ds_\K^2=\\
\nonumber&-\left( 1-\frac{2Mr}{\Sigma} \right)dt^2+
\Sigma d\theta^2 -\frac{4Mar\sin^2\theta}{\Sigma}dtd\phi\\
 &+\frac{\Sigma}{\Delta} dr^2+\left( r^2+a^2+\frac{2Ma^2r\sin^2\theta}{\Sigma} \right)\sin^2\theta d\phi^2,
\end{align}
where
\begin{align}
\Sigma&\equiv r^2+a^2\cos^2\theta,\\
\Delta&\equiv r^2+a^2-2Mr,
\end{align}
with $(r,\theta,\phi)$ being the radial, polar, and azimuthal coordinates centered at the BH, and $M$, $a$ being the BH's total mass and spin.
Similar to Ref.~\cite{Johannsen:2015pca}, we introduce scalar deviation functions $f(r)$, $g(\theta)$, $A_{0}(r)$, $A_1(r)$, $A_2(r)$, $\bar A_{0}(\theta)$, $A_3(\theta)$, $A_4(\theta)$, and $A_6(\theta)$ into the contravariant Kerr metric
\begin{widetext}
\begin{align}
\nonumber g^{\alpha\beta}\frac{\partial}{\partial x^\alpha}\frac{\partial}{\partial x^\beta}=&-\frac{1}{\Delta\tilde\Sigma} \Bigg\lbrack (r^2+a^2)^{2}A_1(r)^{2}\left(\frac{\partial}{\partial t}\right)^{2}+a^{2} A_2(r)^{2}\left(\frac{\partial}{\partial \phi}\right)^{2}+ 2a(r^2+a^2)A_{0}(r) \frac{\partial}{\partial t}\frac{\partial}{\partial \phi}\Bigg\rbrack\\
&+\frac{1}{\tilde\Sigma\sin^2\theta} \Bigg\lbrack A_3(\theta)^{2}\left(\frac{\partial}{\partial\phi}\right)^{2}+a^{2} \sin^{4}\theta A_4(\theta)^{2}\left(\frac{\partial}{\partial t}\right)^{2}+ 2 a \sin^2\theta \bar A_{0}(\theta)\frac{\partial}{\partial t}\frac{\partial}{\partial \phi}  \Bigg\rbrack \nonumber \\
&+\frac{\Delta}{\tilde\Sigma}A_5(r)\left( \frac{\partial}{\partial r} \right)^2+\frac{1}{\tilde\Sigma}A_6(r)\left( \frac{\partial}{\partial\theta} \right)^2,\label{eq:CYcontravariant}
\end{align}
\end{widetext}
with 
\begin{equation}
\tilde\Sigma\equiv\Sigma+f(r)+g(\theta).
\end{equation}
Observe how this expression is similar to Eq.~(10) of Ref.~\cite{Johannsen:2015pca}, however the two additional scalar functions $A_0(r)$ and $\bar A_0(\theta)$ introduce more generality into the function. 
One recovers the metric in~\cite{Johannsen:2015pca} by setting $A_0 = A_1 A_2$ and $\bar A_0 = A_3 A_4$, while it reduces to the Kerr metric in the limit $A_i \to 1$ and $f \to 0$, $g \to 0$.
Such modifications guarantee the resulting Hamilton-Jacobi equations remain seperable, and a fourth constant of motion appears as thoroughly described in~\cite{Johannsen:2015pca}.
Additionally, following Ref.~\cite{Konoplya:2018arm} we find that our metric is a more broad example of the general class of metrics that admit separable Klein Gordon equations.
As was the case for the metric presented in~\cite{Johannsen:2015pca}, we find that with the additional assumption of $f(r)=(r^2+a^2)\left(\frac{A_1(r)}{A_2(r)}-1\right)$, our metric also reduces to one that allows for the separability of the Klein Gordon equations.

Next we define functional forms of our scalar deviation functions, and apply various constraints.
We expand the radial functions as a power series in $M/r$, $g(\theta)$ as a Legendre expansion, and ignore the remaining polar functions for now~\cite{Johannsen:2015pca}:
\begin{align}
A_i(r)&\equiv\sum\limits_{n=0}^\infty\alpha_{in}\left( \frac{M}{r} \right)^n, \hspace{5mm} (i=0,1,2,5),\\
f(r)&\equiv r^2\sum\limits_{n=0}^\infty\epsilon_n \left( \frac{M}{r} \right)^n\\
g(\theta)&\equiv M^2\sum\limits_{n=0}^\infty\gamma_{n} P_n(\cos\theta),
\end{align}
with Legendre polynomials $P_n(\cos\theta)$.
We note that the Legendre expansion of $g(\theta)$ differs from that presented in~\cite{Johannsen:2015pca}, where the author utilized a trigonometric expansion in powers of $\sin\theta$ and $\cos\theta$.
The Legendre expansion given here is a more natural choice given an axisymmetric spacetime metric, and gives unique choices on parameters $\gamma_n$, whereas degeneracies occur in the trigonometric expansion utilized previously.

We begin constraining the deviation parameters by imposing the condition of asymptotic flatness~\cite{heusler_1996,wald_1984}.
This corresponds to imposing that our metric line element must limit to 
\begin{equation}
ds_\infty^2=-\left( 1-\frac{2M}{r} \right)dt^2-\frac{4Ma}{r}\sin^2\theta dtd\phi+dr^2+r^2d\Omega^2,
\end{equation}
at spatial infinity $r\to \infty$ for $d\Omega\equiv d\theta^2+\sin^2\theta d\phi^2$.
Doing so reveals the need to constrain $\bar A_0(\theta)=A_3(\theta)=A_4(\theta)=A_6(\theta)=1$, as well as $\alpha_{00}=\alpha_{10}=\alpha_{20}=\alpha_{50}=1$ and $\epsilon_0=0$.
The asymptotic behavior of $g_{tt}$ and $g_{t\phi}$ become
\begin{align}
g_{tt} &= -1 + (2 + 2 \alpha_{11} -\epsilon_1)\frac{M}{r} + \mathcal{O}\left( \frac{M^2}{r^2} \right)\,, \\
g_{t\phi} &= -(2+\alpha_{01}) a  \frac{M}{r} \sin^2\theta + \mathcal{O}\left( \frac{M^2}{r^2} \right)\,.
\label{eq:g_tph_asympt}
\end{align}
Thus, we can rescale $M$ and $a$ to further set $\alpha_{01} = 0$ and $\alpha_{11} = \epsilon_{1}/2$ without loss of generality. The covariant form of the metric is summarized in Eq.~\eqref{eq:CYmetric}.

Next we consider imposing constraints obtained from the parameterized-post-Newtonian (ppN) framework~\cite{Will:2005va}\footnote{Apart from $\gamma_\ppN$ and $\beta_\ppN$ considered here, one could in principle consider other PPN parameters, including the one in~\cite{Alexander:2007zg} which is associated with the Lense-Thirring precession. We leave this possibility for future work.}.
This is done by further imposing that the metric for a non-spinning object must reduce to the line element given by
\begin{align}
\nonumber ds^2_\ppN=&-\left[1-\frac{2M}{r}+2(\beta_\ppN-\gamma_\ppN)\frac{M^2}{r^2}  + \mathcal{O}\left( \frac{M^3}{r^3} \right) \right]dt^2\\
&+\left[ 1+2\gamma_\ppN \frac{M}{r}  + \mathcal{O}\left( \frac{M^2}{r^2} \right)\right]dr^2 \nonumber \\
&+r^2 \left[ 1 + \mathcal{O}\left( \frac{M^2}{r^2} \right)\right]d\Omega,
\end{align}
for ppN parameters $\gamma_\ppN$ and $\beta_\ppN$, while the asymptotic behavior of the new metric is given by
\begin{align}
g_{tt} &= -1 + 2\frac{M}{r} + \left[\frac{M^2}{4}  (8\alpha_{12} + \epsilon_1^2 - 4 \epsilon_2)-g \right]\frac{M^2}{r^2} \nonumber \\
&+ \mathcal{O}\left( \frac{M^3}{r^3} \right)\,, \\
g_{rr} &= 1 + \left(2 - \alpha_{51} + \epsilon_1 \right)\frac{M}{r}+ \mathcal{O}\left( \frac{M^2}{r^2} \right)\,, \\
g_{\theta\theta} &= r^2 \left[1 + \epsilon_1\frac{M}{r}+ \mathcal{O}\left( \frac{M^2}{r^2} \right) \right]\,, \\
g_{\phi\phi} &= r^2 \sin^2\theta \left[1 + \epsilon_1\frac{M}{r}+ \mathcal{O}\left( \frac{M^2}{r^2} \right) \right]\,.
\end{align}
Given the strong observational constraints of $\beta_\ppN$~\cite{Williams:2004qba} and $\gamma_\ppN$~\cite{Bertotti:2003rm} from solar system experiments, one can further impose the conditions $\epsilon_1 = 0$ (which automatically sets $\alpha_{11}=0$), $\alpha_{51} = 0$ and $2\epsilon_2-2\alpha_{12}-g(\theta)/M^2=0$. The simplest choice of the last condition is $\epsilon_2=\alpha_{12}=g(\theta)=0$~\cite{Johannsen:2015pca}, which is what we consider in the main part of this paper.

Notice, however, that because Birkhoff's theorem is not guaranteed to hold in theories beyond GR, such ppN constraints obtained in the weak-field environment of the local solar system may not necessarily apply to the strong-gravity conditions present near the BHs considered here. 
This is indeed the case for BHs in e.g. Einstein-dilaton Gauss-Bonnet gravity, in which the BH exterior spacetime is different from that for stars due to the presence (absence) of the BH (stellar) scalar charge~\cite{Campbell:1991kz,Yunes:2011we,Yagi:2011xp,Sotiriou:2014pfa,Yagi:2015oca,Berti_ModifiedReviewSmall,Prabhu:2018aun}.
Thus, the presented constraints on $\epsilon_1$, $\alpha_{51}$ $\epsilon_2$, $\alpha_{12}$, and $g(\theta)$ may not necessarily hold, and App.~\ref{app:lowerOrder} provides a description of the effects of including such parameters in the metric.


\subsection{Location of the event horizon, Killing horizon, and ergosphere}\label{sec:horizons}

In this section we describe the locations of the event horizon, Killing horizon, and ergosphere in the new spacetime.
In particular, we note that the locations of each of these appear identically to those as presented in Ref.~\cite{Johannsen:2015pca}, thus we refer the reader there for a thorough description of each.

We begin by briefly describing the event horizon in both the new spacetime presented here, and Johannsen's spacetime.
The angular function $r_\EH\equiv H(\theta)$ is a solution to
\begin{equation}
g^{rr}+g^{\theta\theta}\left( \frac{dH}{d\theta} \right)^2=0,
\end{equation}
which results in the solution
\begin{equation}
\Delta A_5(H)+\left( \frac{dH}{d\theta} \right)^2=0.
\end{equation}
One solution is either the Kerr result ($\Delta=0$, $dH/d\theta = 0$)
\begin{equation}
r_\EH=M+\sqrt{M^2-a^2}
\end{equation}
or a solution to $A_5(r)=0$ with $dH/d\theta = 0$. 
The latter set is realized if $A_5$ diverges when $\Delta = 0$, which is the case for some example BH solutions that can be mapped to the metric presented in this paper. 
These solutions are the same as those for Johannsen's spacetime because $A_0(r)$ does not appear in either $g^{rr}$ or $g^{\theta\theta}$.
We see here that this expression only depends on the non-GR deviation function $A_5(r)$.

Next we find the location of the Killing horizon, which is located at the solution of $(g_{t\phi})^2-g_{tt} g_{\phi\phi}=0$.
Such an expression reduces down to $\Delta=0$, coinciding with the event horizon, in both the spacetimes considered here and by Johannsen.

Finally, the ergosphere exists at the roots of $g_{tt}=0$.
Because $A_0(r)$ only appears in the denominator of $g_{tt}$, the new solution is identical to that of Johannsen's, reducing to
\begin{equation}
\Delta=a^2A_2(r)^2\sin^2\theta
\end{equation}
in both cases, which is displayed in Fig.~1 of Ref.~\cite{Johannsen:2015pca}.
We see that this expression only depends on the non-GR deviation function $A_2(r)$. 


\subsection{The spheroidicity condition}\label{sec:spheroid}
In this section we compute the spheroidicity condition as detailed in the analysis of Ref.~\cite{Glampedakis:2018blj} for the new general metric presented in this chapter.
In the above investigation, the authors found the most general form of the ``spheroidicity condition''.
Such a condition defines non-equatorial circular orbits confined on a spheroidal shell described by $r_0(\theta)$.
In particular, we compute the spheroidicity condition found in Eq.~(14) of~\cite{Glampedakis:2018blj} as a function of $r_0(\theta)$ for the general metric presented here in Eq.~\eqref{eq:CYmetric}, with only the leading order terms of each beyond-Kerr function present.
We find the resulting condition to be
\begin{widetext}
\begin{align}
\nonumber 0 &= 3 \alpha _{13} a^6 M^3 \left(\alpha _{13} M^3+r_0^3\right)-a^5 b M^2 r_0 \left(5 \alpha _{13} \alpha _{22} M^3+3 \alpha _{13} M r_0^2+2 \alpha _{22} r_0^3\right)-a^4 r_0 \Big(M^4 \left(8 \alpha _{13} r_0^3-2 \alpha _{22}^2 b^2 r_0\right)\\
\nonumber &-2 \alpha _{22} b^2 M^2 r_0^3+7 \alpha _{13}^2 M^7-8 \alpha _{13}^2 M^6 r_0-7 \alpha _{13} M^3 r_0^4+M r_0^6+r_0^7\Big)+2 a^3 b M r_0^2 \Big(6 \alpha _{13} \alpha _{22} M^5-5 \alpha _{13} \alpha _{22} M^4 r_0\\
\nonumber&+4 \alpha _{13} M^3 r_0^2+3 \left(\alpha _{22}-\alpha _{13}\right) M^2 r_0^3-2 \alpha _{22} M r_0^4+r_0^5\Big)-a^2 r_0^3 \Big(b^2 \big(5 \alpha _{22}^2 M^5-3 \alpha _{22}^2 M^4 r_0+6 \alpha _{22} M^3 r_0^2-4 \alpha _{22} M^2 r_0^3\\
\nonumber& + M r_0^4-r_0^5\big)+10 \alpha _{13}^2 M^7-7 \alpha _{13}^2 M^6 r_0+8 \alpha _{13} M^4 r_0^3-5 \alpha _{13} M^3 r_0^4-2 M r_0^6+2 r_0^7\Big)-a b M r_0^4 \Big(-8 \alpha _{13} \alpha _{22} M^5\\
\nonumber&+5 \alpha _{13} \alpha _{22} M^4 r_0-4 \alpha _{13} M^3 r_0^2+\left(3 \alpha _{13}-2 \alpha _{22}\right) M^2 r_0^3+2 \alpha _{22} M r_0^4+2 r_0^5\Big)-r_0^5 \Big(3 \alpha _{13}^2 M^7-2 \alpha _{13}^2 M^6 r_0-\alpha _{13} M^3 r_0^4\\
&-3 M r_0^6+r_0^7\Big)
\end{align}
\end{widetext}
for orbital impact parameter $b$.
Similar to Johannsen's metric, we find the above condition to be independent of $\theta$, and we conclude that this metric admits $r_0=\text{const.}$ Kerr-like spherical photon orbits.


\section{Astrophysical implications}\label{sec:astrophysical}
In this section we present the various astrophysical implications emergent under a BH described by the new metric presented in this analysis.
Specifically, we derive expressions for the various astrophysical observables one might consider about such a BH.
Such properties include the Keplerian and epicyclic frequencies of orbiting particles, the orbital energy and angular momentum of particles' orbits, the location of the ISCO, and finally the photon orbits.


\subsection{Keplerian and epicyclic frequencies}\label{sec:frequencies}

Now let us describe the computation of the Keplerian and epicyclic frequencies $\nu_\phi$, $\nu_r$, and $\nu_\theta$.
The former frequency describes a particle's motion in the polar direction as observed at radial infinity, while the latter two describe the motion in the azimuthal and radial directions for perturbed orbits.

We begin by finding the Keplerian frequency $\nu_\phi=\Omega_\phi/2\pi$.
We start with the definition of $\Omega_\phi\equiv\dot{\phi}/\dot{t}$, which can be determined from the geodesic equations
\begin{equation}
\frac{d^2x^\alpha}{d\tau^2}=-\Gamma^\alpha_{\beta\gamma}\frac{dx^\beta}{d\tau}\frac{dx^\gamma}{d\tau},
\end{equation}
with Christoffel symbols $\Gamma^\alpha_{\beta\gamma}$ and proper time $\tau$.
Following Ref.~\cite{Johannsen:2015pca}, axi-symmetry and reflection symmetry of particles on circular equatorial orbits allow us to reduce this equation to
\begin{equation}
\partial_r  g_{rr} \dot{t}^2+2\partial_r g_{t\phi}\dot{t}\dot{\phi}+\partial_r g_{\phi\phi}\dot{\phi}^2=0.
\end{equation}
The above expression allows us to express the Keplerian frequency as
\begin{equation}
\Omega_\phi=\frac{-\partial_r g_{t\phi}\pm\sqrt{(\partial_r g_{t\phi})^2-\partial_r g_{tt}\partial_r g_{\phi\phi}}}{\partial_r g_{\phi\phi}}.\label{eq:KeplerianFrequency}
\end{equation}
We see that this expression only depends on the non-GR deviation functions $A_1(r)$, $A_2(r)$, $A_0(r)$, and $f(r)$.

Next we obtain expressions for the vertical and radial epicyclic frequencies describing the radial and polar motion of orbiting particles with mass $\mu$.
Following the derivation presented in Sec.~4 and~5 of~\cite{Abramowicz:2004tm}, the general epicyclic frequencies  observed with respect to the proper time of a comoving observer in the $X$-direction are given by
\begin{equation}
\omega_\X=\sqrt{\frac{\partial^2 \mathcal{U}_\text{eff}}{\partial X^2}},
\end{equation} 
for effective potential 
\begin{equation}
\mathcal{U}_\text{eff}=-\frac{1}{2}(g^{tt} E^2-2g^{t\phi} E L_z+g^{\phi\phi} L_z^2+\mu^2)\label{eq:Ueff}
\end{equation}
obtained from $p^\alpha p_\alpha=-\mu^2$ for the particle's four-momentum $p^\alpha$.
The resulting radial and vertical epicyclic frequencies observed at radial infinity are found to be
\begin{align}
\Omega_r&=\sqrt{\frac{(g_{tt}+\Omega_\phi g_{t\phi})^2}{2g_{rr}}\left( \partial^2_r g^{tt} - 2 L_z \partial^2_r g^{t\phi}+L_z^2 \partial^2_r g^{\phi\phi} \right)},\\
\Omega_\theta&=\sqrt{\frac{(g_{tt}+\Omega_\phi g_{t\phi})^2}{2g_{\theta\theta}}\left( \partial^2_\theta g^{tt} - 2 L_z \partial^2_\theta g^{t\phi}+L_z^2 \partial^2_\theta g^{\phi\phi} \right)},
\end{align}
where the angular momentum can be computed via Eq.~\eqref{eq:angMom} in the following section.
We see such epicyclic frequencies depend on the non-GR deviation functions $A_1(r)$, $A_2(r)$, $A_0(r)$, $A_5(r)$, and $f(r)$, while the vertical frequencies depend on all but $A_5(r)$.

\begin{figure*}[htb]
\begin{center}
\includegraphics[width=.3\textwidth]{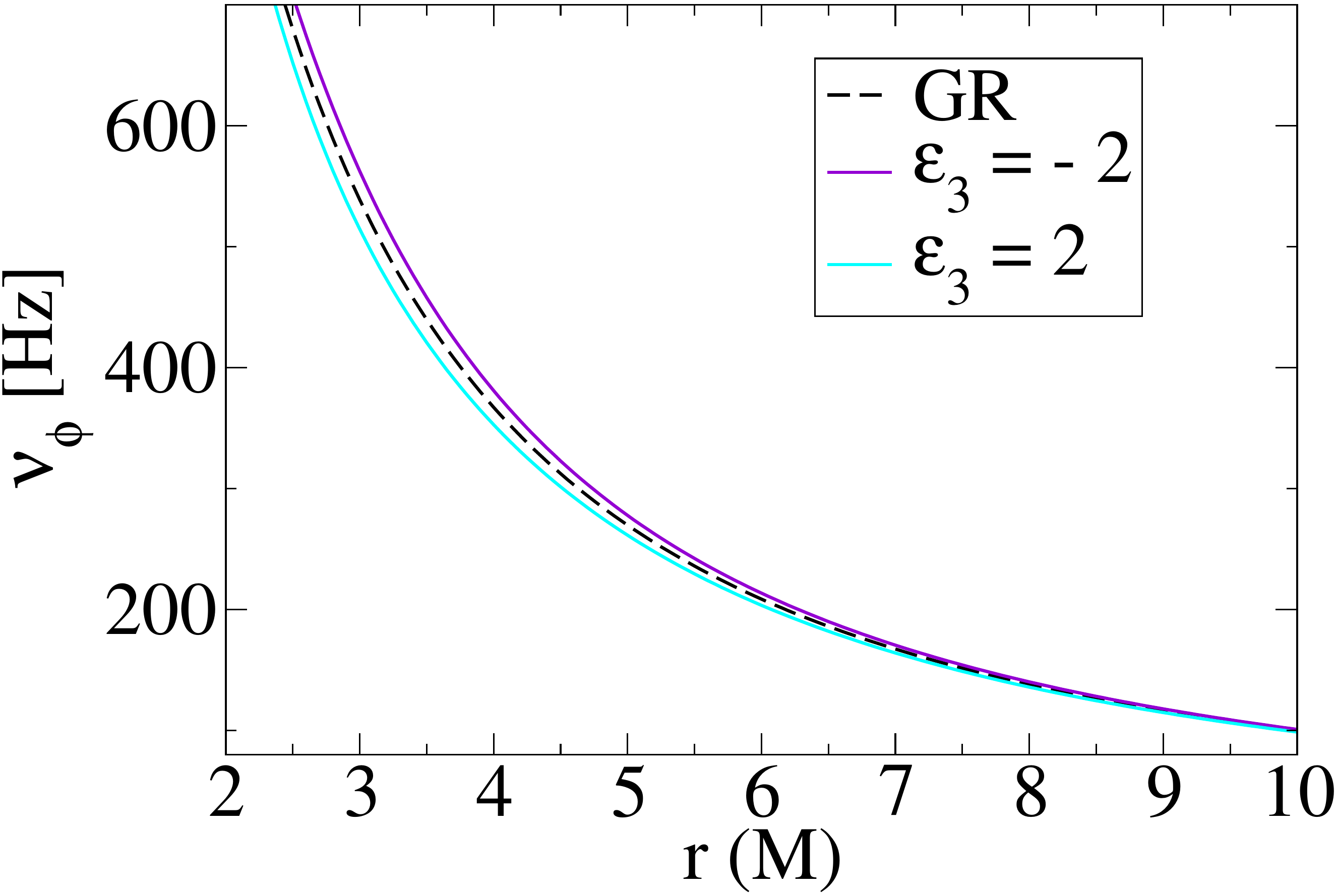}
\includegraphics[width=.3\textwidth]{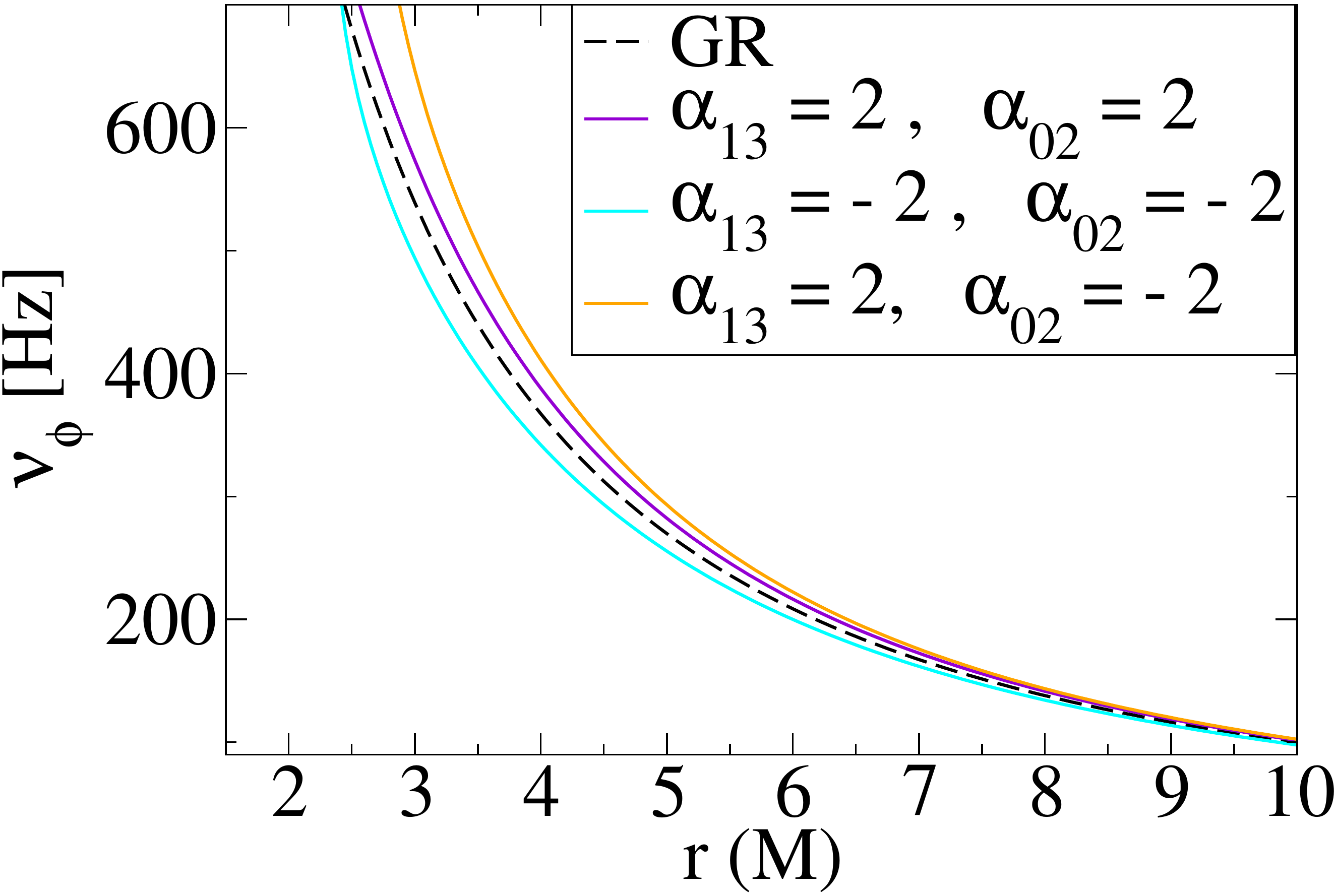}
\includegraphics[width=.3\textwidth]{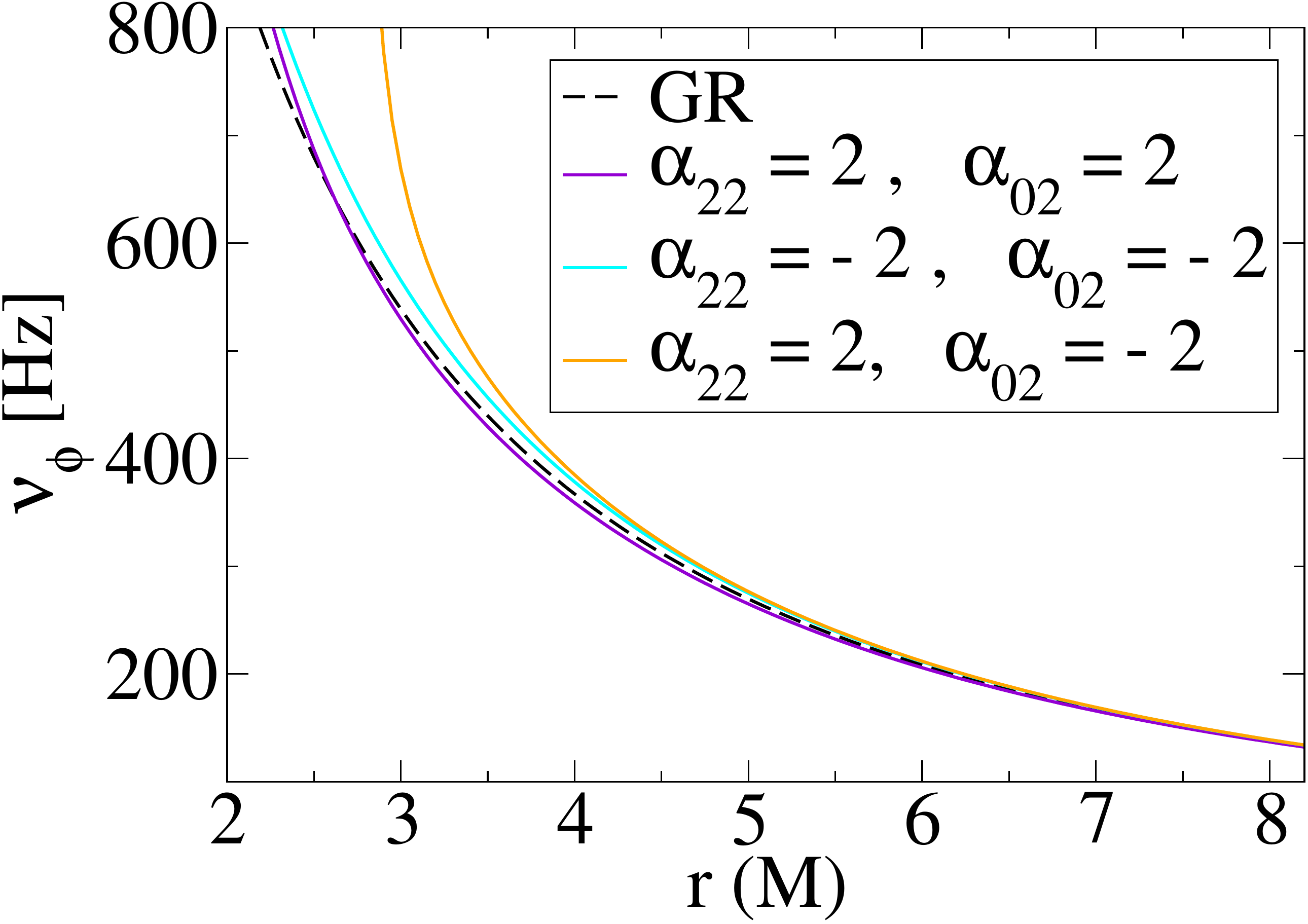}
\caption{Keplerian frequencies of a particle orbiting a $10\text{ M}_\odot$ BH with $a=0.8M$ on circular equatorial orbits for various values of the non-GR deviation parameters, while setting all remaining parameters to 0.
The frequencies are plotted for varying the $f(r)$ lowest-order parameter $\epsilon_3$ (left), varying the $A_1(r)$ and $A_0(r)$ lowest-order parameters $\alpha_{13}$ and $\alpha_{02}$ (center), and varying the $A_2(r)$ and $A_0(r)$ lowest-order parameters $\alpha_{22}$ and $\alpha_{02}$ (right).
Several cases with $\alpha_{22}=0$ while $\alpha_{02} \ne 0$ or $\alpha_{12}=0$ while $\alpha_{02} \ne 0$, or vice versa, produce BHs with naked singularities which are not shown here. The left-most plot agrees with that in Fig.~3 of~\cite{Johannsen:2015pca}.
}\label{fig:nuPhi}
\end{center}
\end{figure*}

\begin{figure*}[htb]
\begin{center}
\includegraphics[width=.3\textwidth]{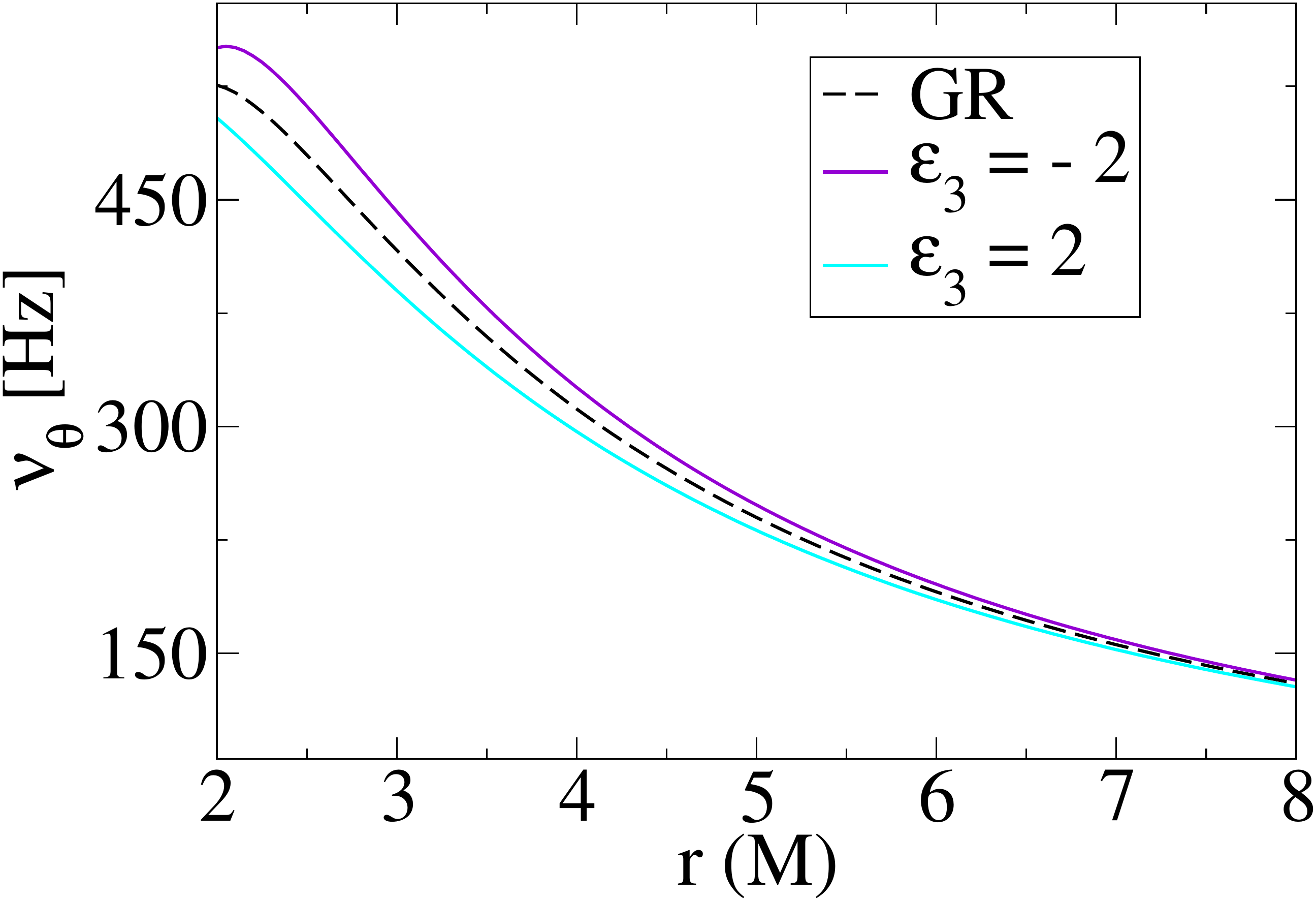}
\includegraphics[width=.3\textwidth]{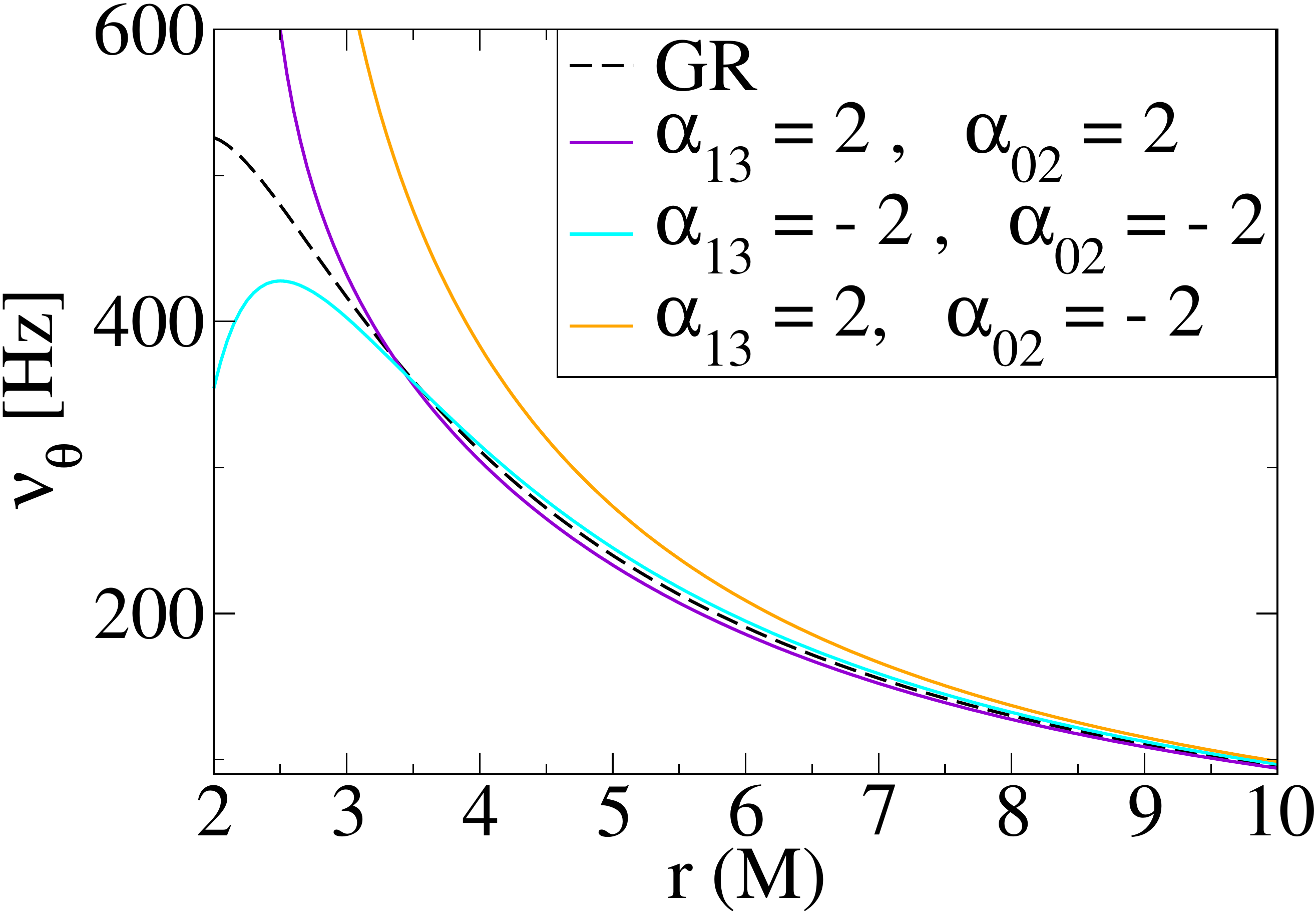}
\includegraphics[width=.3\textwidth]{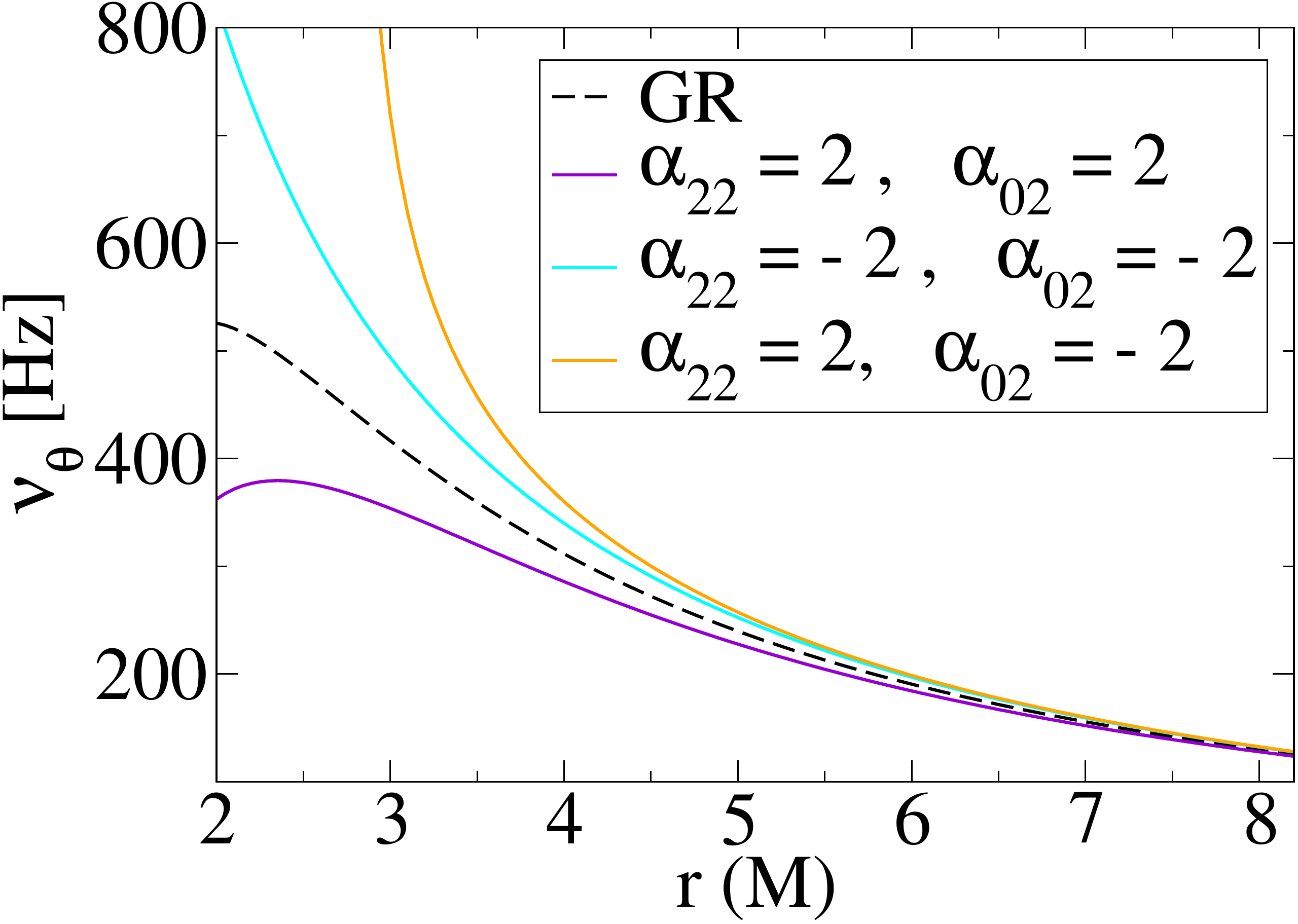}
\caption{Similar to Fig.~\ref{fig:nuPhi} but for the vertical epicyclic frequency $\nu_{\theta}$.
}\label{fig:nuTheta}
\end{center}
\end{figure*}

\begin{figure*}[htb]
\begin{center}
\includegraphics[width=.3\textwidth]{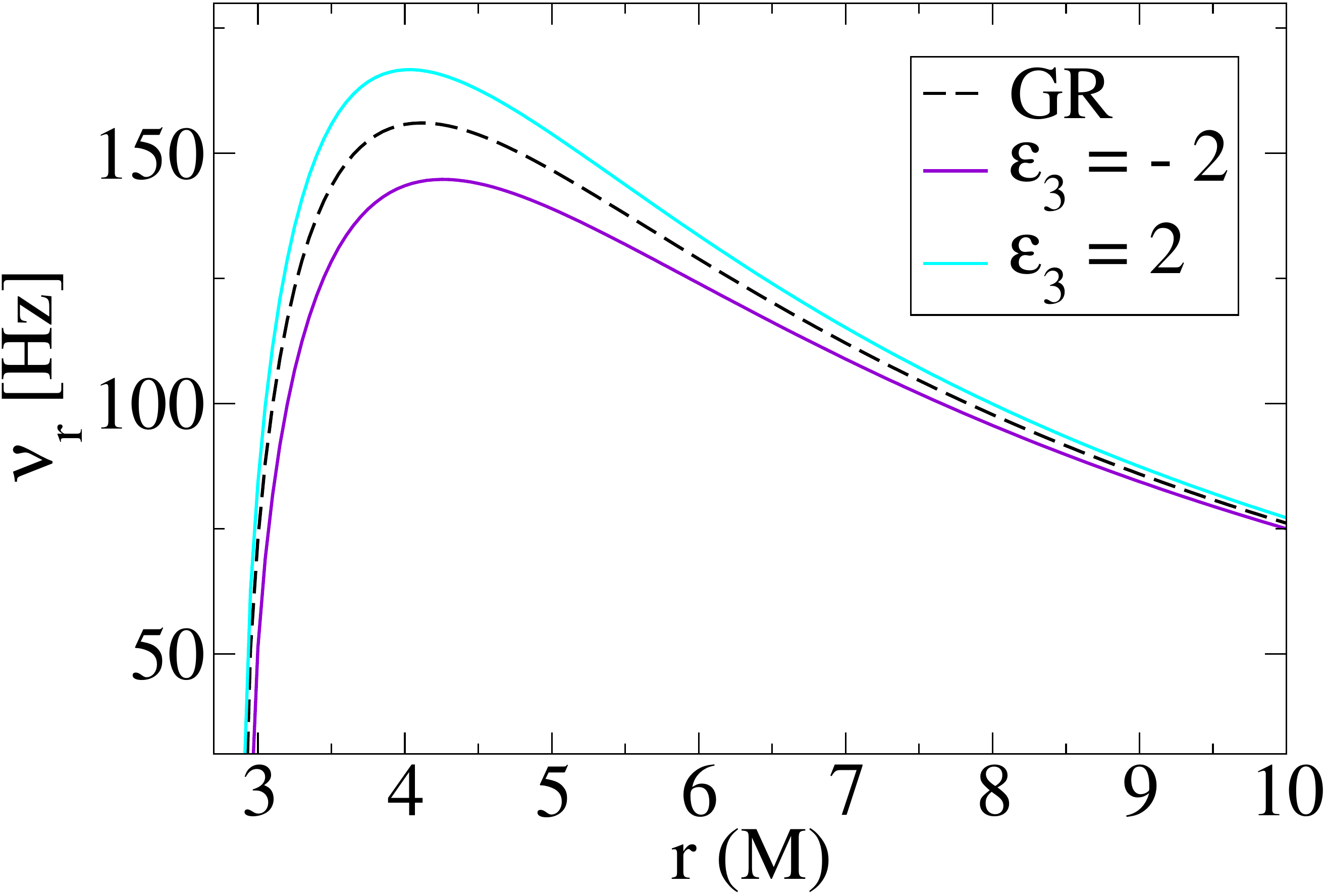}
\includegraphics[width=.3\textwidth]{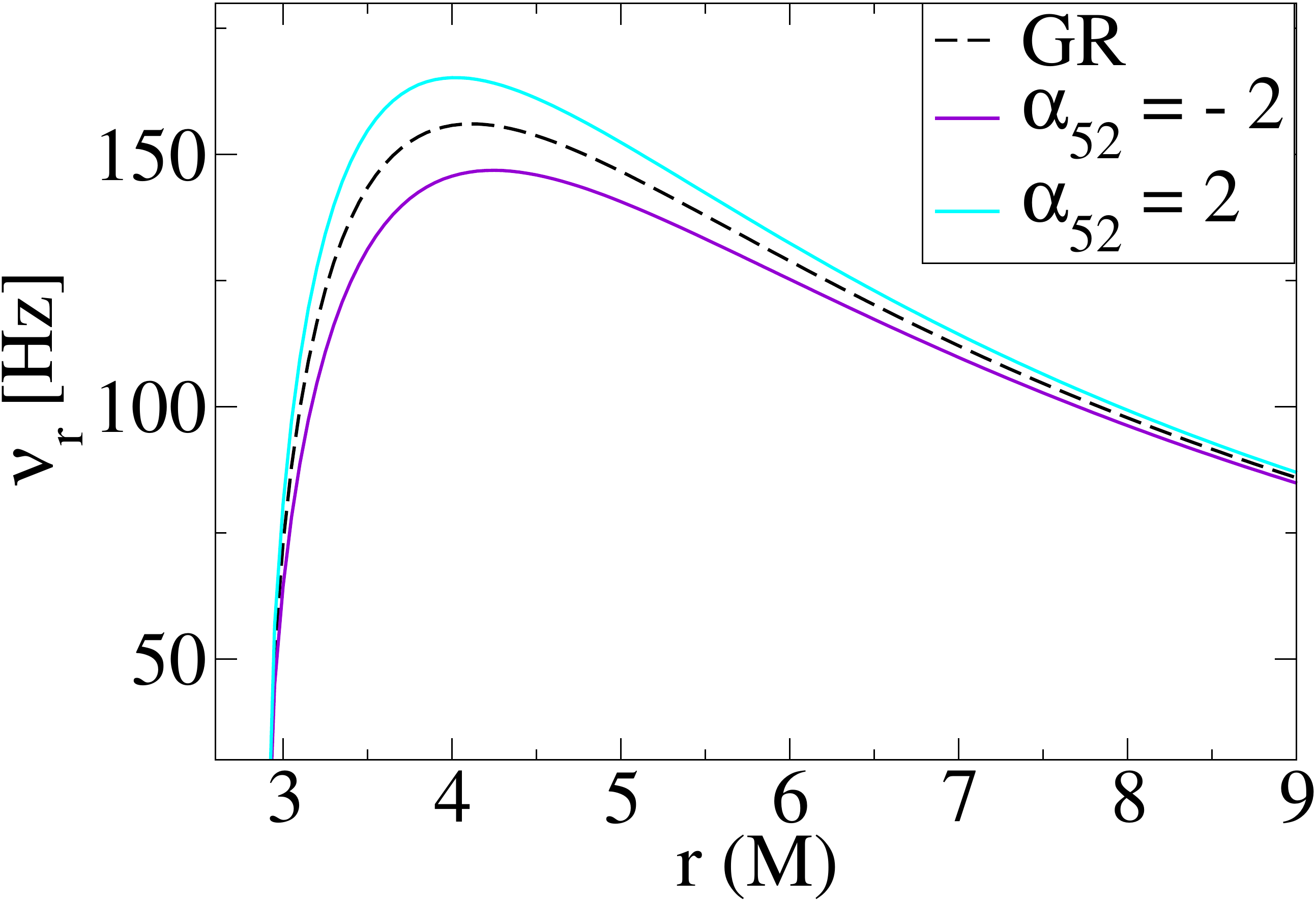}\\
\includegraphics[width=.3\textwidth]{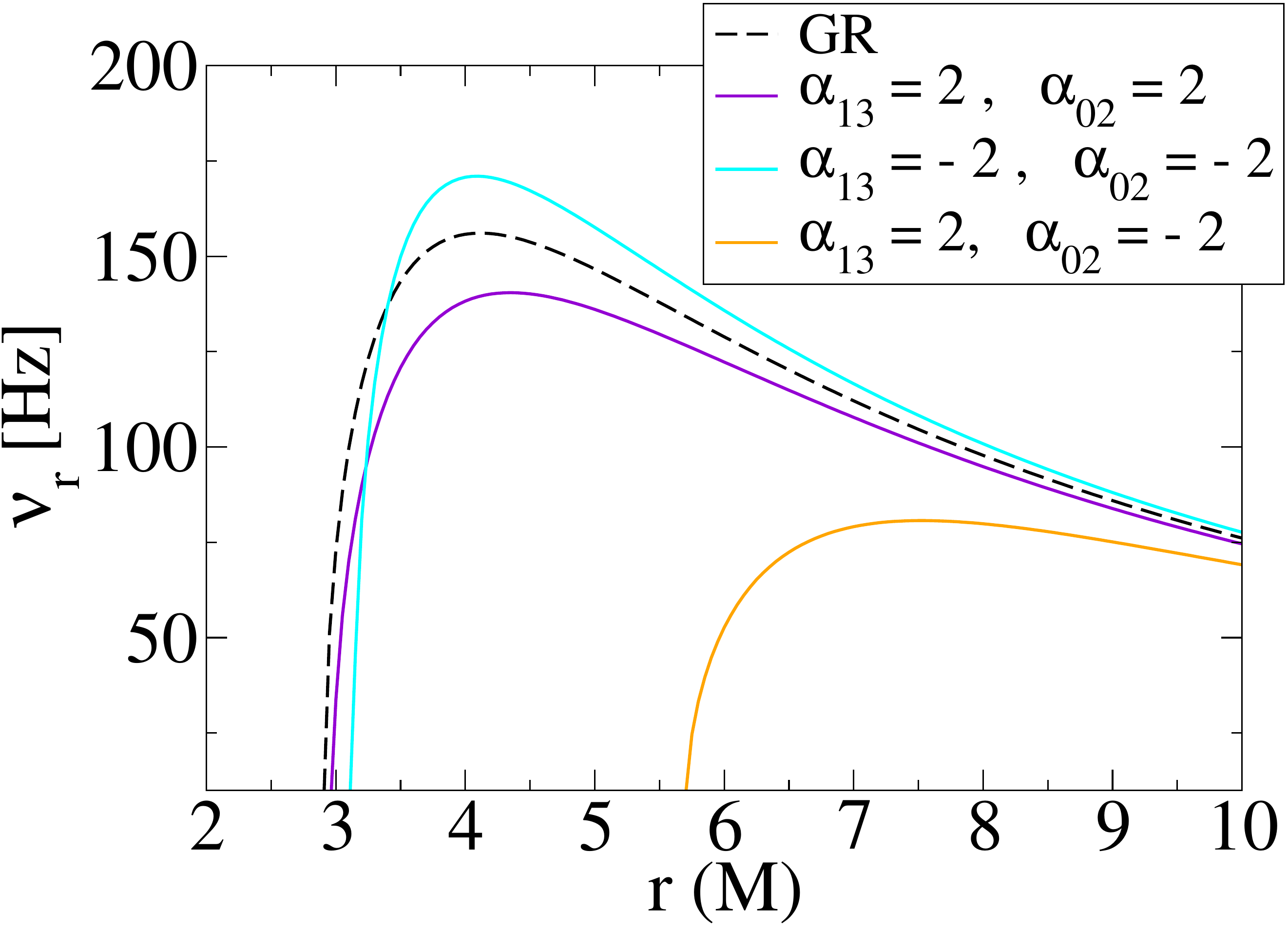}
\includegraphics[width=.3\textwidth]{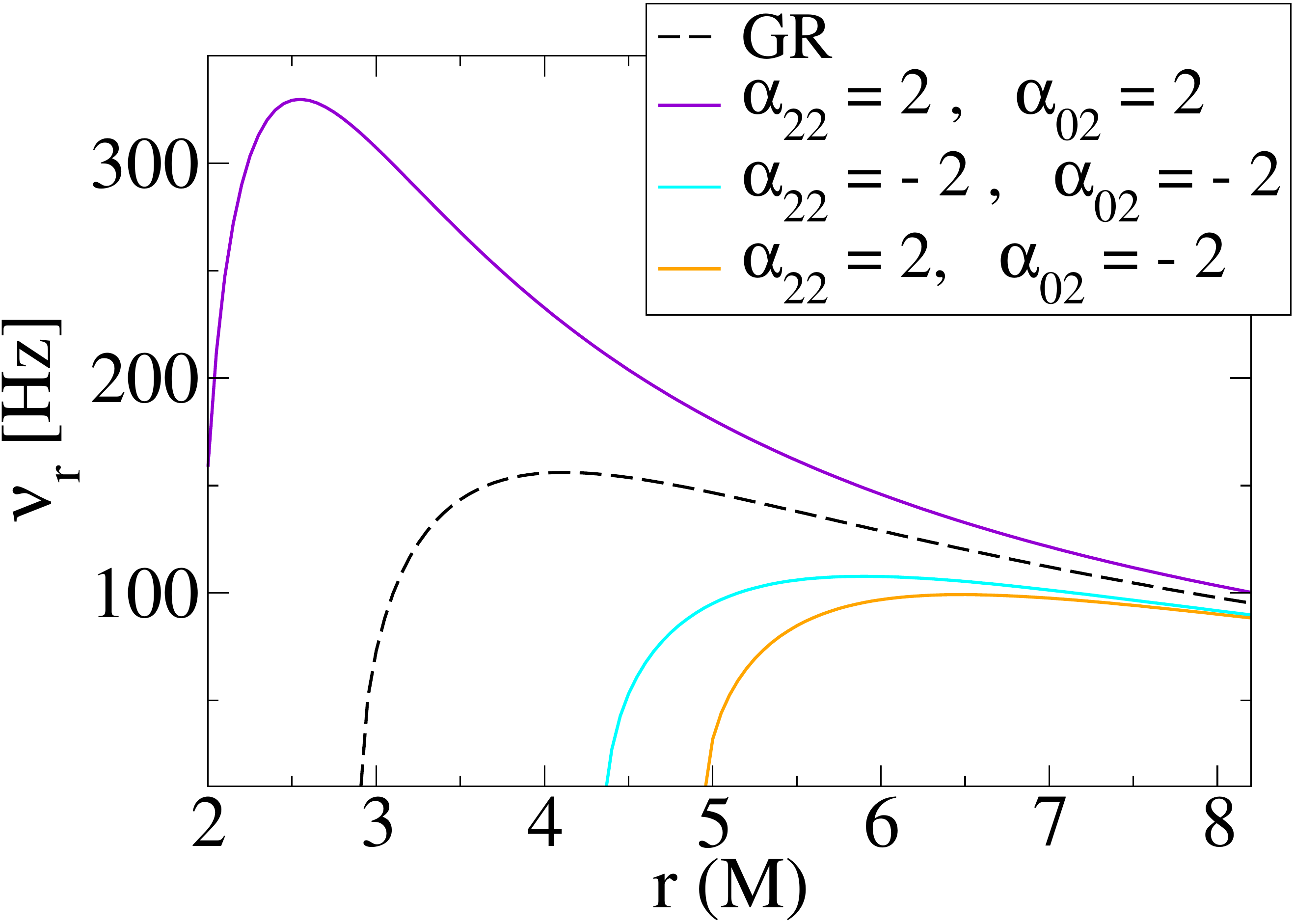}
\caption{
Similar to Fig.~\ref{fig:nuPhi} but for the radial epicyclic frequency $\nu_r$. 
Additionally plotted here (top right) is the dependence on the lowest-order parameter $\alpha_{52}$ of the deviation function $A_5(r)$ appearing only in the $g_{rr}$ metric element.
}\label{fig:nuR}
\end{center}
\end{figure*}

Finally, we plot the Keplerian and epicyclic frequencies $\nu_\X\equiv\Omega_\X/2\pi$ for various combinations of lower-order deviation paramters.
Here take note that certain combinations of deviation parameters produce naked singularities outside of the BH event horizon, as discussed further in App.~\ref{app:nakedSingularities}.
Such exotic singularities originate from disallowed combinations of parameters $A_1(r)$, $A_2(r)$, and $A_0(r)$.
Namely, we find that if $\alpha_{13}\ne0$ or $\alpha_{22}\ne0$, $\alpha_{02}$ must additionally be non-vanishing and of the same sign, else the photon orbit energies and angular momentum become discontinuous and negative, and photon orbits become open, letting photons escape to radial infinity as discussed in~\cite{Hioki:2009na,Papnoi:2014aaa}.
Here we vary only the lowest-order non-vanishing parameters present in the given expressions $\alpha_{13}$, $\alpha_{22}$, $\alpha_{02}$, and $\epsilon_3$ for the Keplerian and vertical epicyclic frequencies, and also $\alpha_{52}$ for the radial epicyclic frequency.
In each case, all other non-GR deviation parameters that are not specifically mentioned are set to be 0.
For a further analysis on the further-lower-order parameters assumed to vanish here, see App.~\ref{app:lowerOrder}.
In Fig.~\ref{fig:nuPhi} we plot the Keplerian frequencies $\nu_\phi$, while in Figs.~\ref{fig:nuTheta} and~\ref{fig:nuR} the vertical and radial epicyclic frequencies $\nu_\theta$ and $\nu_r$ are plotted for various non-vanishing parameters.
We observe that, in general, the parameters $\epsilon_3$ and $\alpha_{52}$ introduce very little change into the frequencies $\nu_\X$, while combinations of $\alpha_{02}$ and $\alpha_{13}$ or $\alpha_{22}$ have the power to significantly alter the ensuing trajectories. Observe how the frequencies (especially epicyclic ones) can deviate significantly from the Kerr case when varying the new parameter $\alpha_{02}$ introduced for the first time in this paper.


\subsection{Energy and angular momentum}\label{sec:EandLz}

%
%

In this section, we present the expressions for the orbital energy $E$ and angular momentum $L_z$ of a particle 
orbiting a BH described by the new metric.
We begin with the effective potential $\mathcal{U}_\text{eff}$ given in Eq.~\eqref{eq:Ueff}.
In the equatorial plane, circular orbits obey the expressions $\mathcal{U}_\text{eff}(r)=\frac{d \mathcal{U}_\text{eff}(r)}{dr}=0$.
When combined with the Keplerian frequency in Eq.~\eqref{eq:KeplerianFrequency}, and the Keplerian frequency written in terms of constants of motion
\begin{equation}
\Omega_\phi=\frac{p_\phi}{p_t}=-\frac{g_{t\phi} E+g_{tt} L_z}{g_{\phi\phi} E+g_{t\phi} L_z},
\end{equation}
we obtain expressions for the energy and angular momentum of a particle orbiting our central BH as
\begin{align}
E&=-\mu\frac{g_{tt}+g_{t\phi}\Omega_\phi}{\sqrt{-g_{tt}-2g_{t\phi}\Omega_\phi-g_{\phi\phi}\Omega_\phi^2}},\label{eq:energy}\\
L_z&=-\mu\frac{g_{t\phi}+g_{\phi\phi}\Omega_\phi}{\sqrt{-g_{tt}-2g_{t\phi}\Omega_\phi-g_{\phi\phi}\Omega_\phi^2}}.\label{eq:angMom}
\end{align}
We see that these expressions only depend on the non-GR deviation functions $A_1(r)$, $A_2(r)$, $A_0(r)$, and $f(r)$.

\begin{figure*}[htb]
\begin{center}
\includegraphics[width=.3\textwidth]{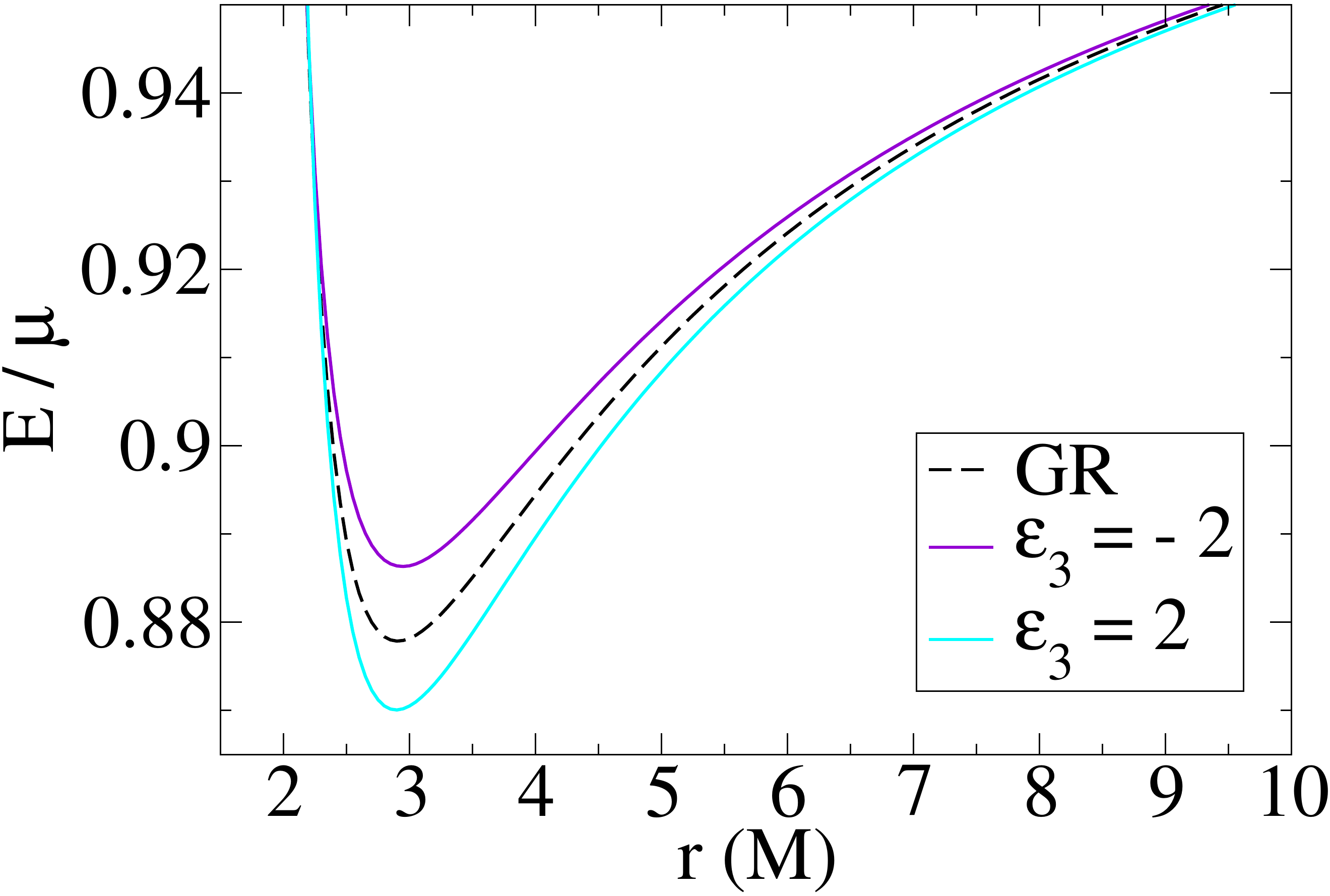}
\includegraphics[width=.3\textwidth]{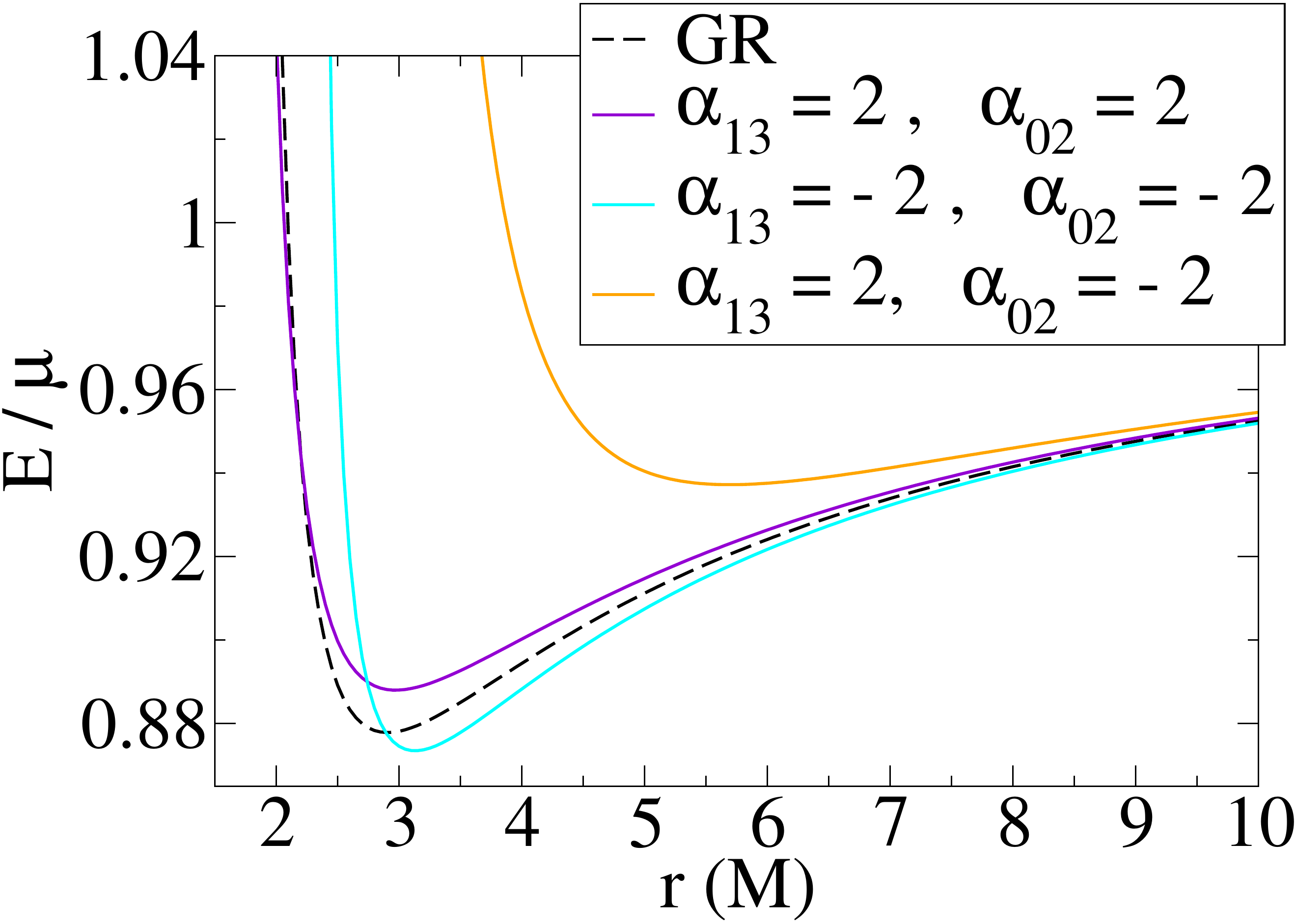}
\includegraphics[width=.3\textwidth]{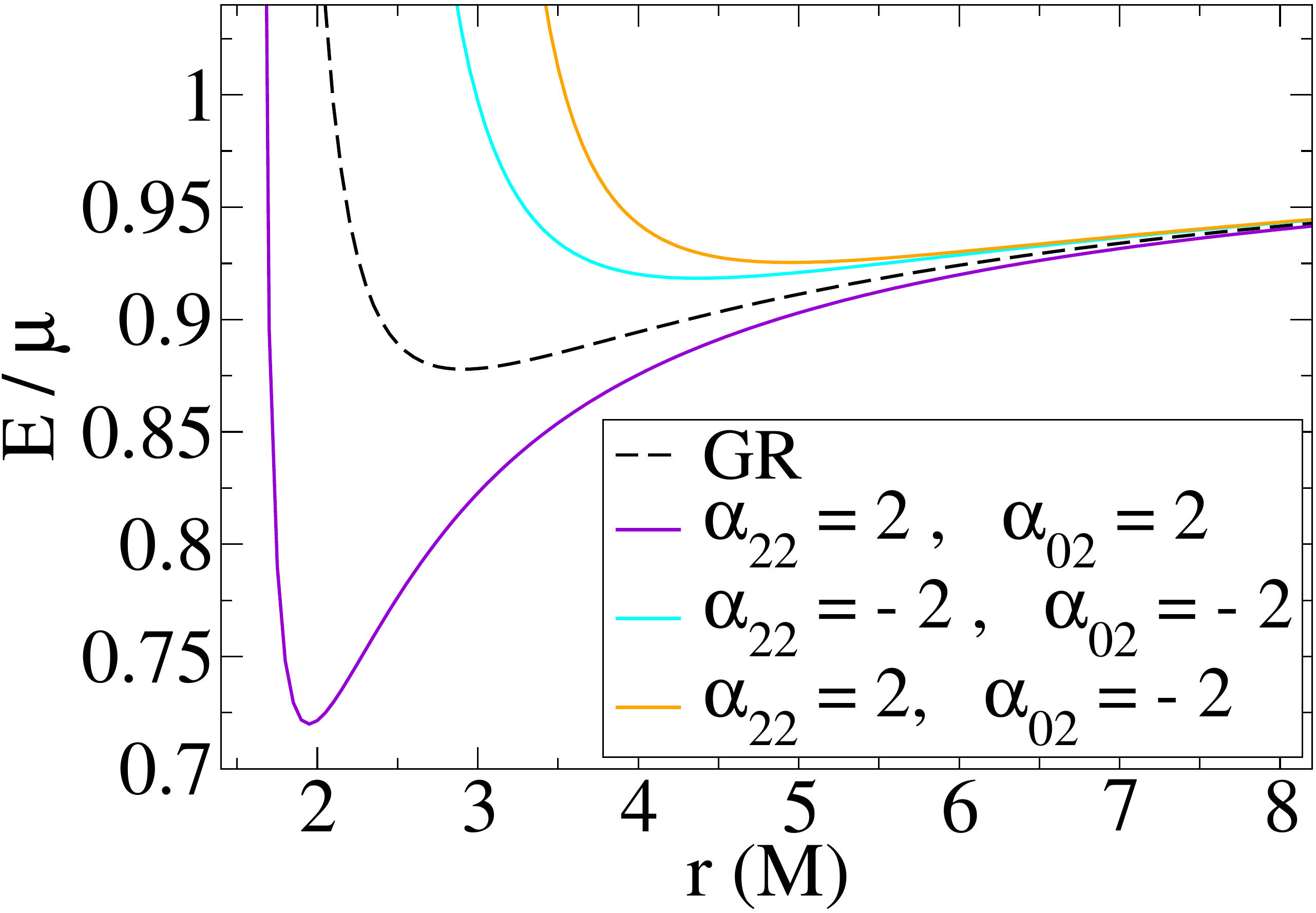}
\caption{Similar to Fig.~\ref{fig:nuPhi} but for the specific orbital energy $E/\mu$.
The left-most plot agrees with that in Fig.~4 of~\cite{Johannsen:2015pca}.
We note that the non-GR deviation parameters, especially the new one $A_0(r)$ introduced in this paper significantly impact the orbital energy of particles. 
}\label{fig:energy}
\end{center}
\end{figure*}

\begin{figure*}[htb]
\begin{center}
\includegraphics[width=.3\textwidth]{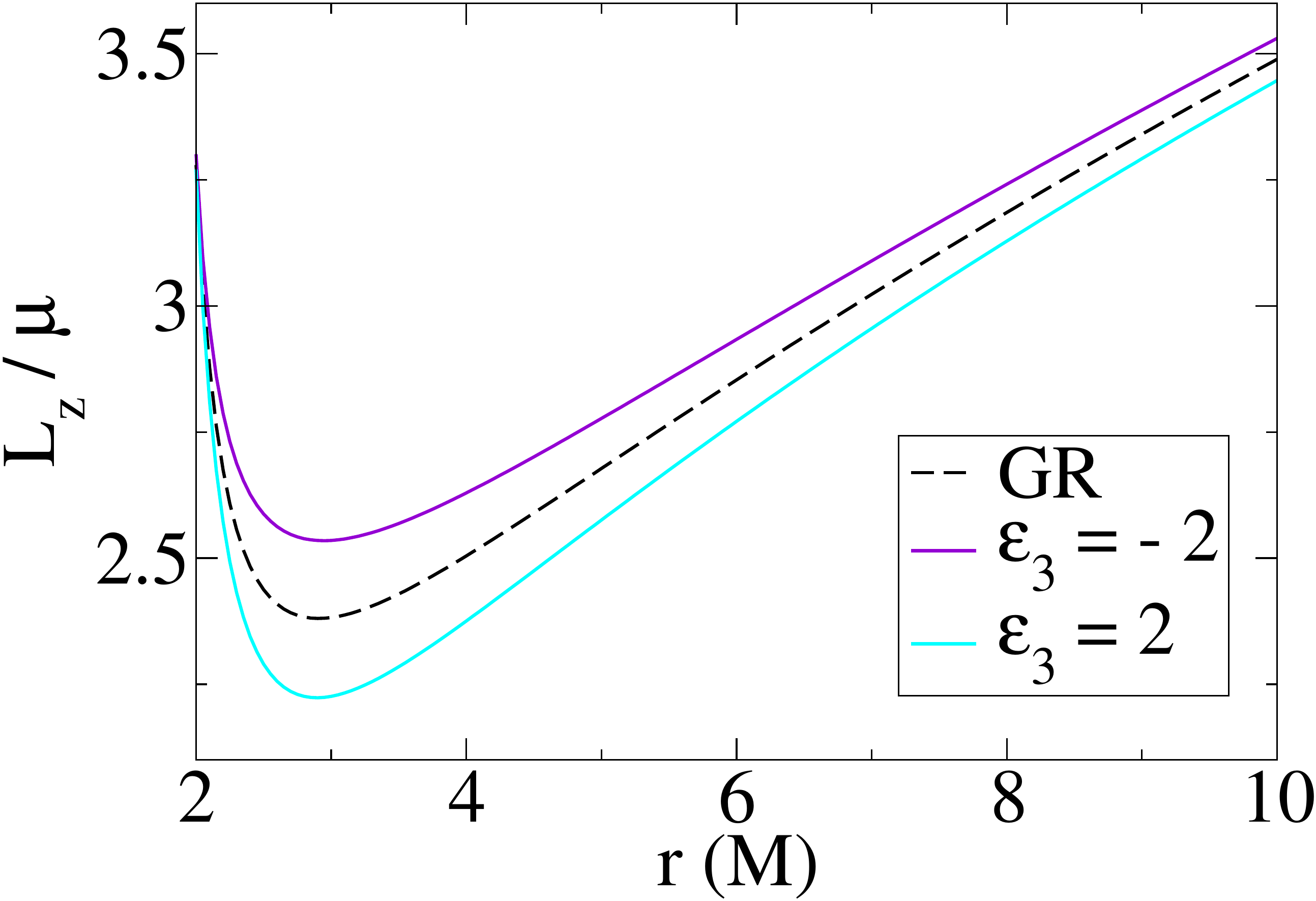}
\includegraphics[width=.3\textwidth]{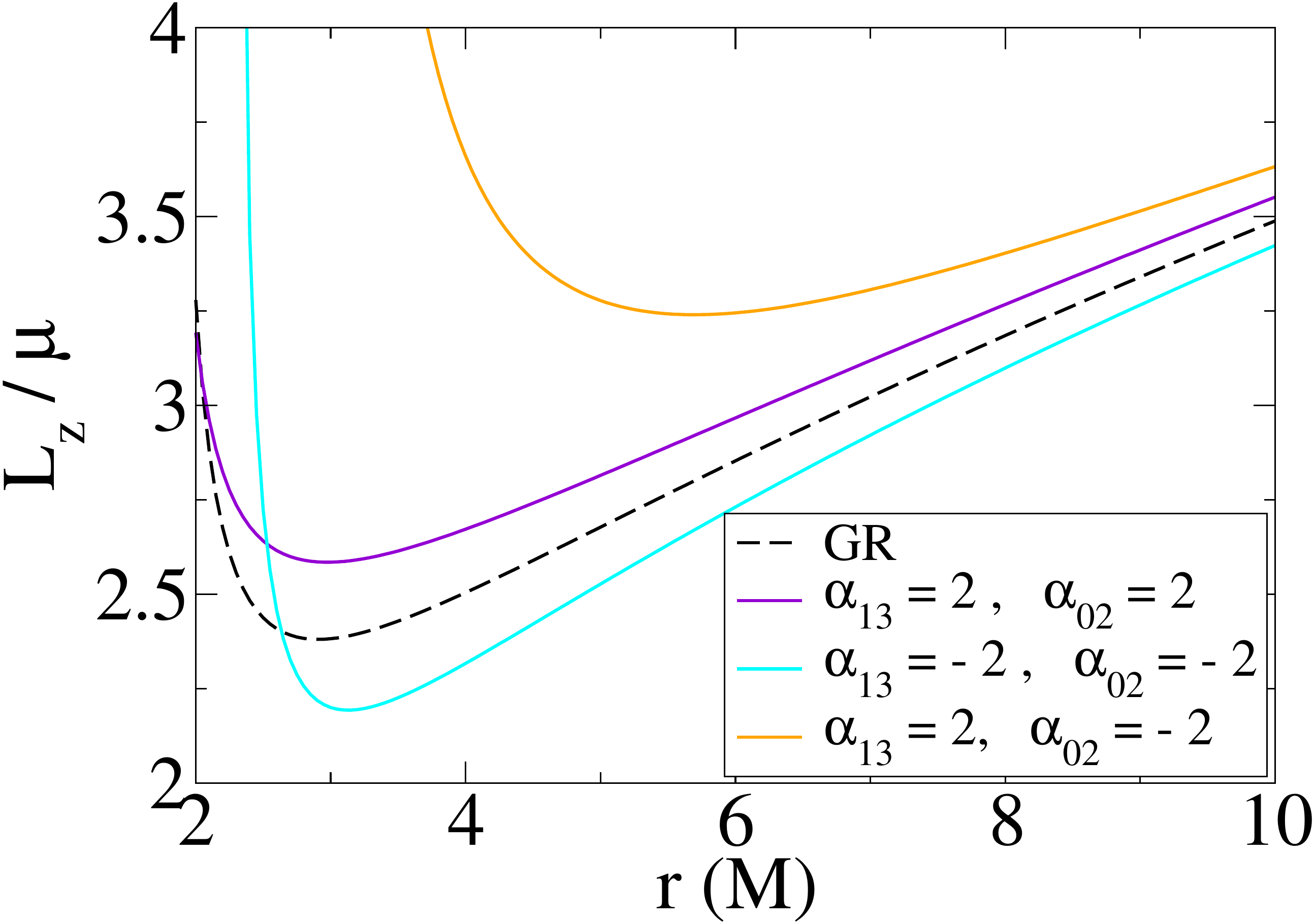}
\includegraphics[width=.3\textwidth]{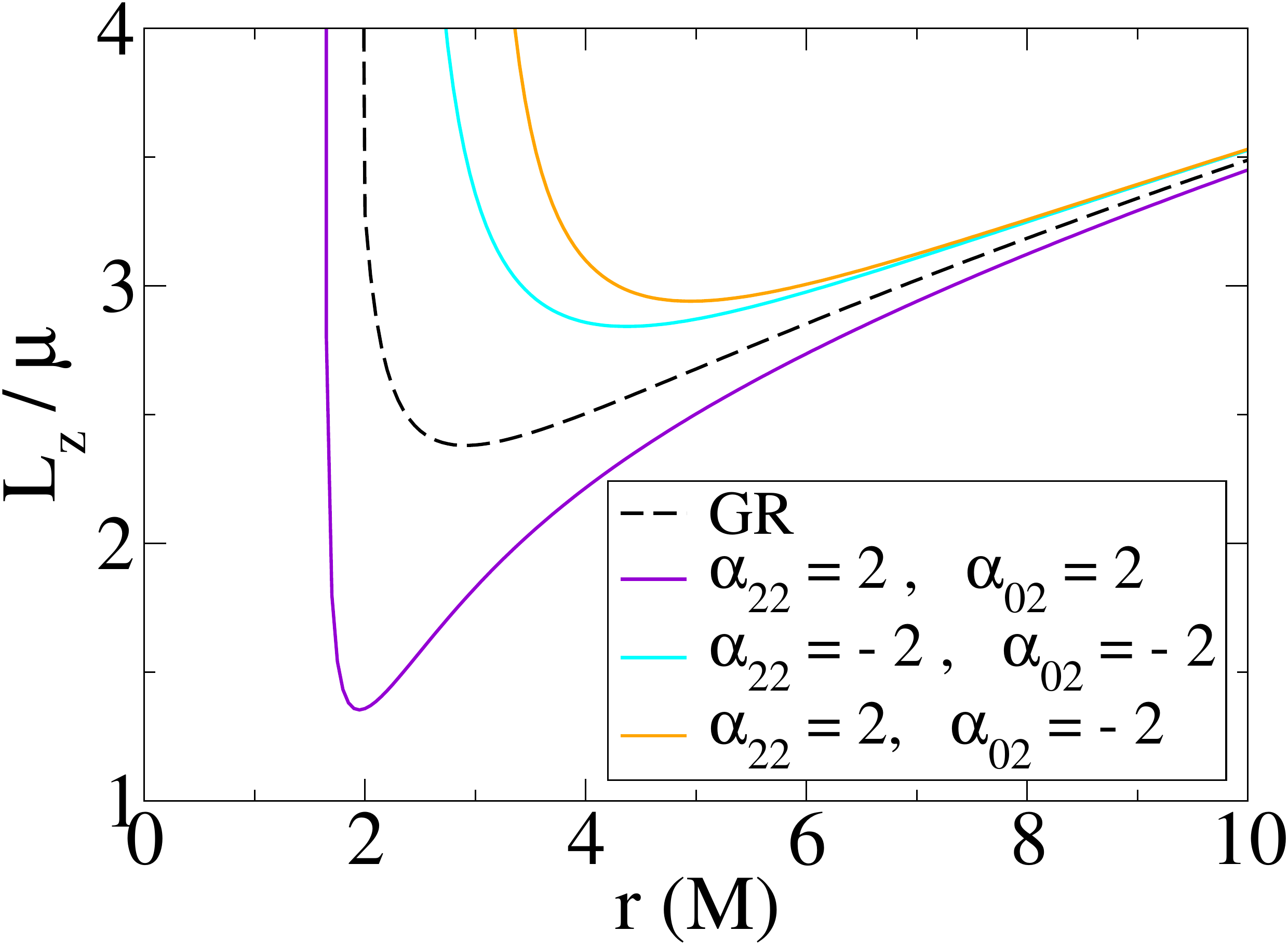}
\caption{Similar to Fig.~\ref{fig:nuPhi} but for the specific orbital angular momentum $L_z/\mu$. The left-most plot agrees with that in Fig.~5 of~\cite{Johannsen:2015pca}.
}\label{fig:momentum}
\end{center}
\end{figure*}

\begin{figure*}[!htbp]
\begin{center}
\includegraphics[width=.4\textwidth]{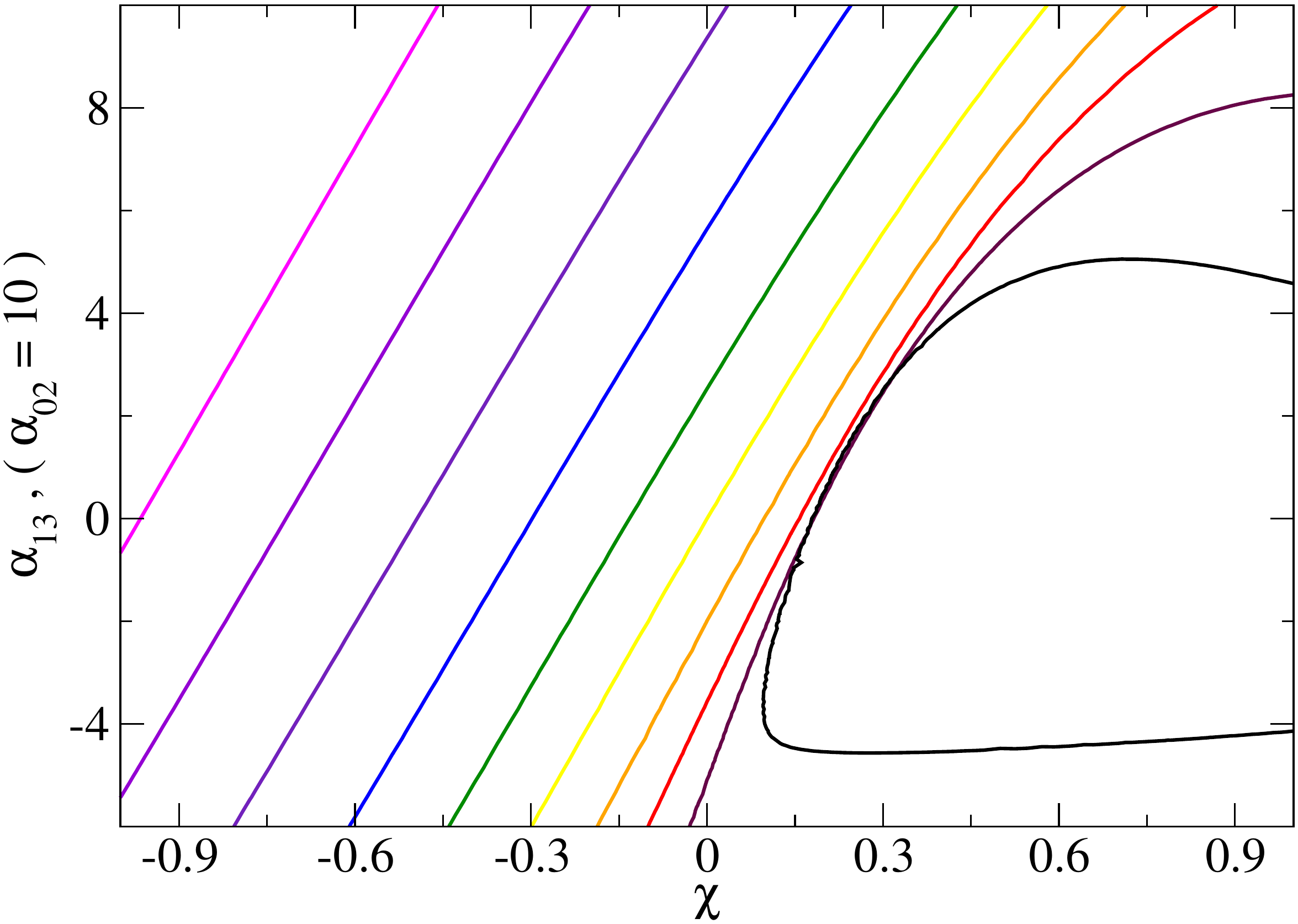}
\includegraphics[width=.4\textwidth]{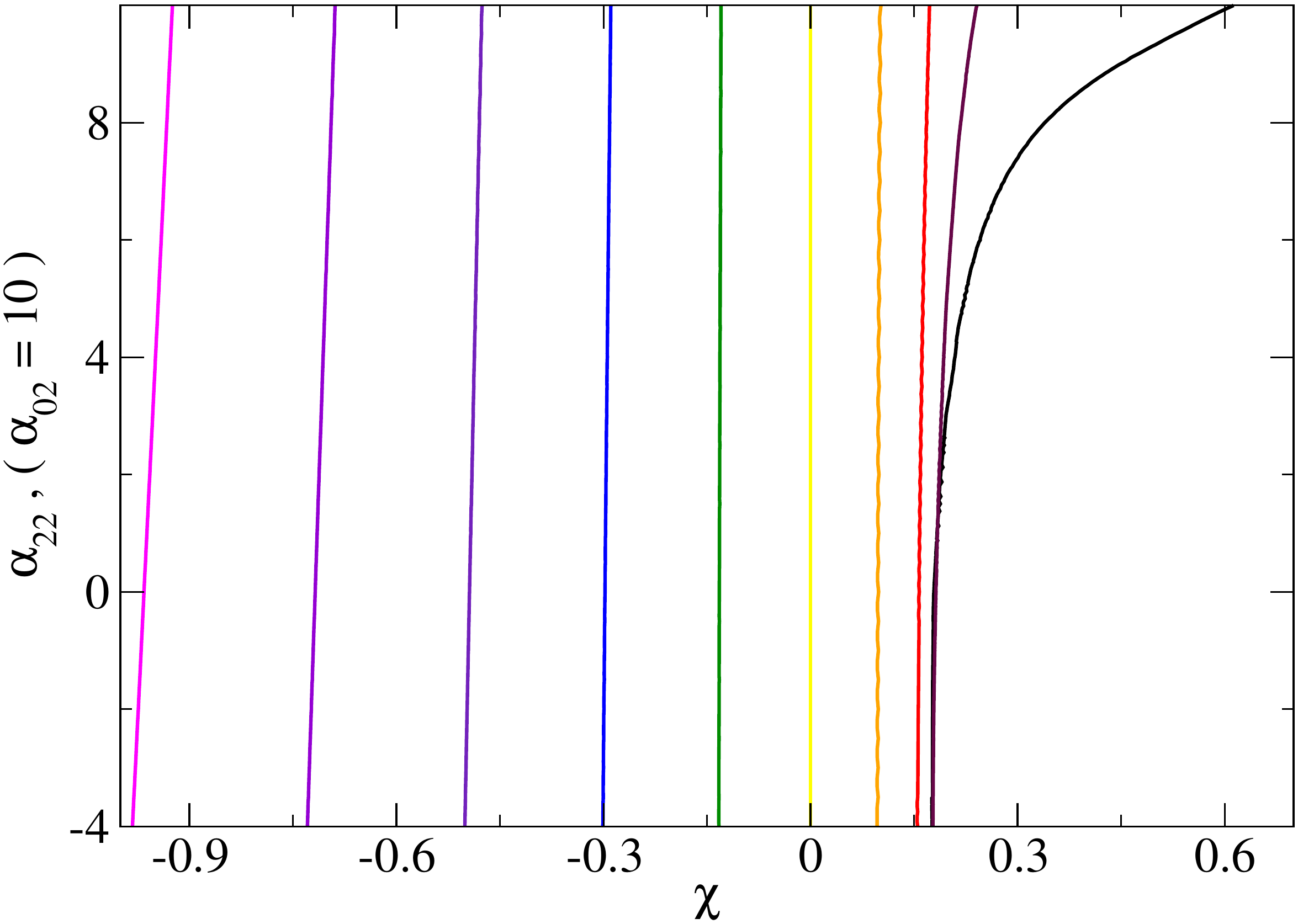}\\
\includegraphics[width=.4\textwidth]{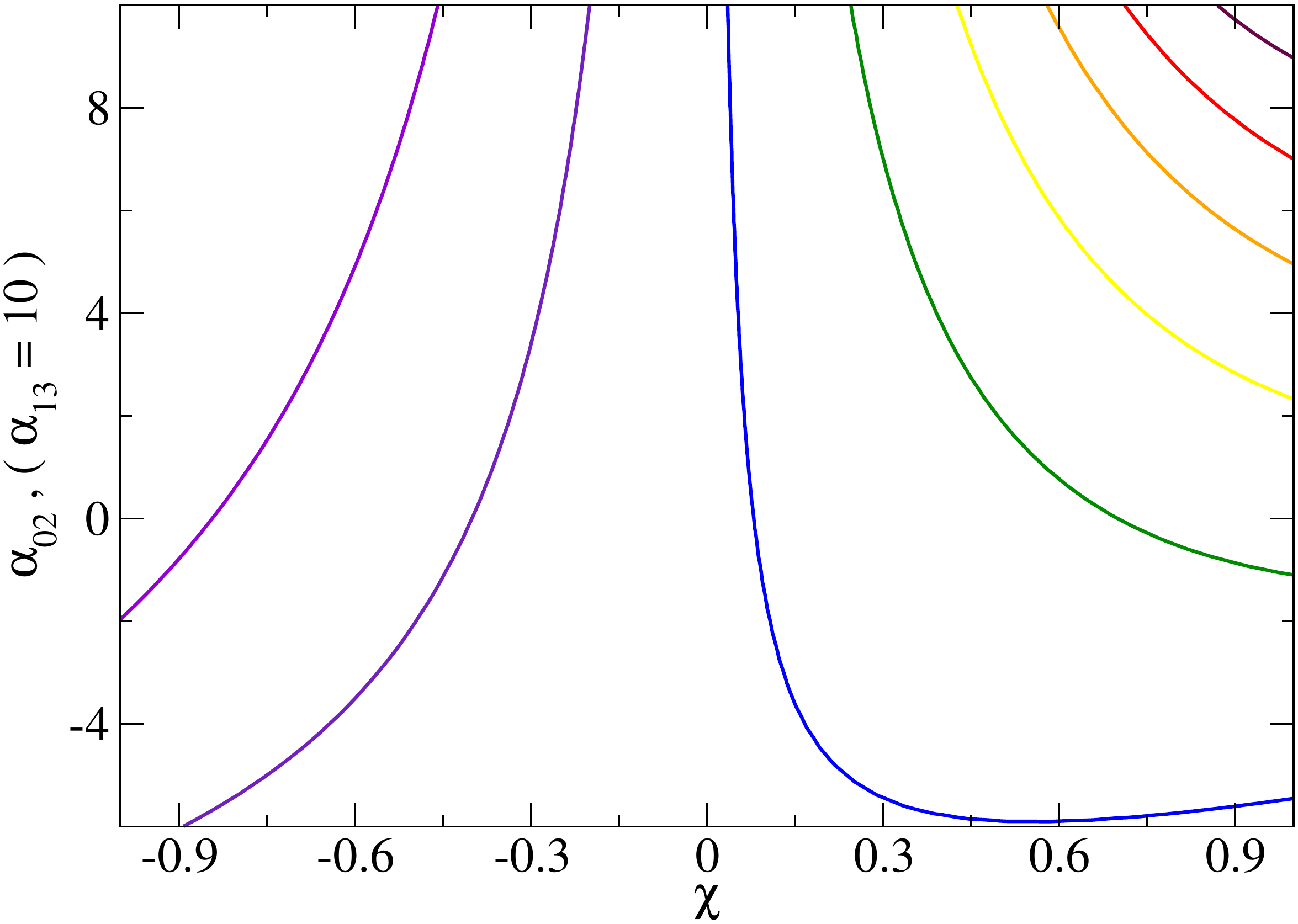}
\includegraphics[width=.4\textwidth]{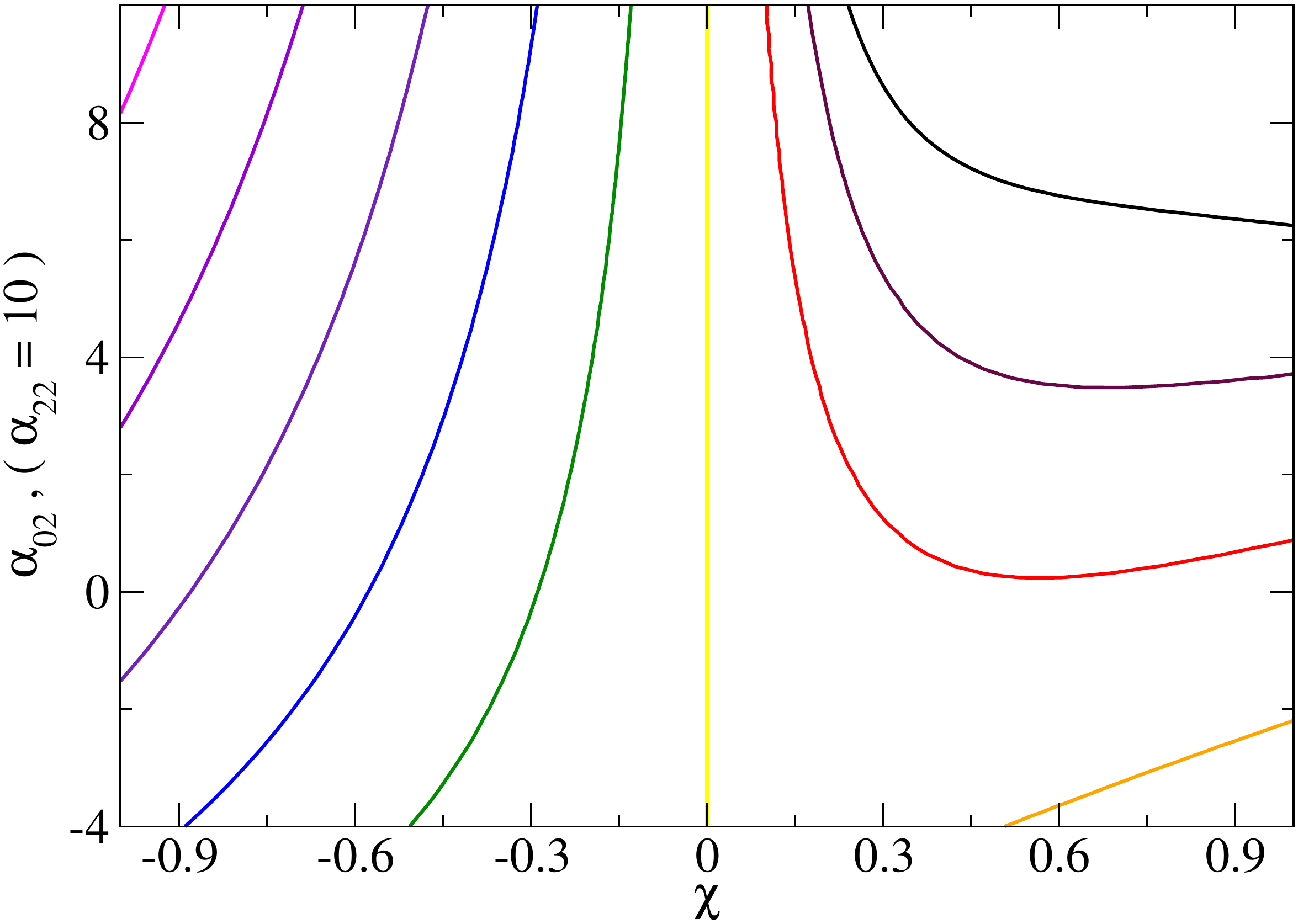}\\
\includegraphics[width=.4\textwidth]{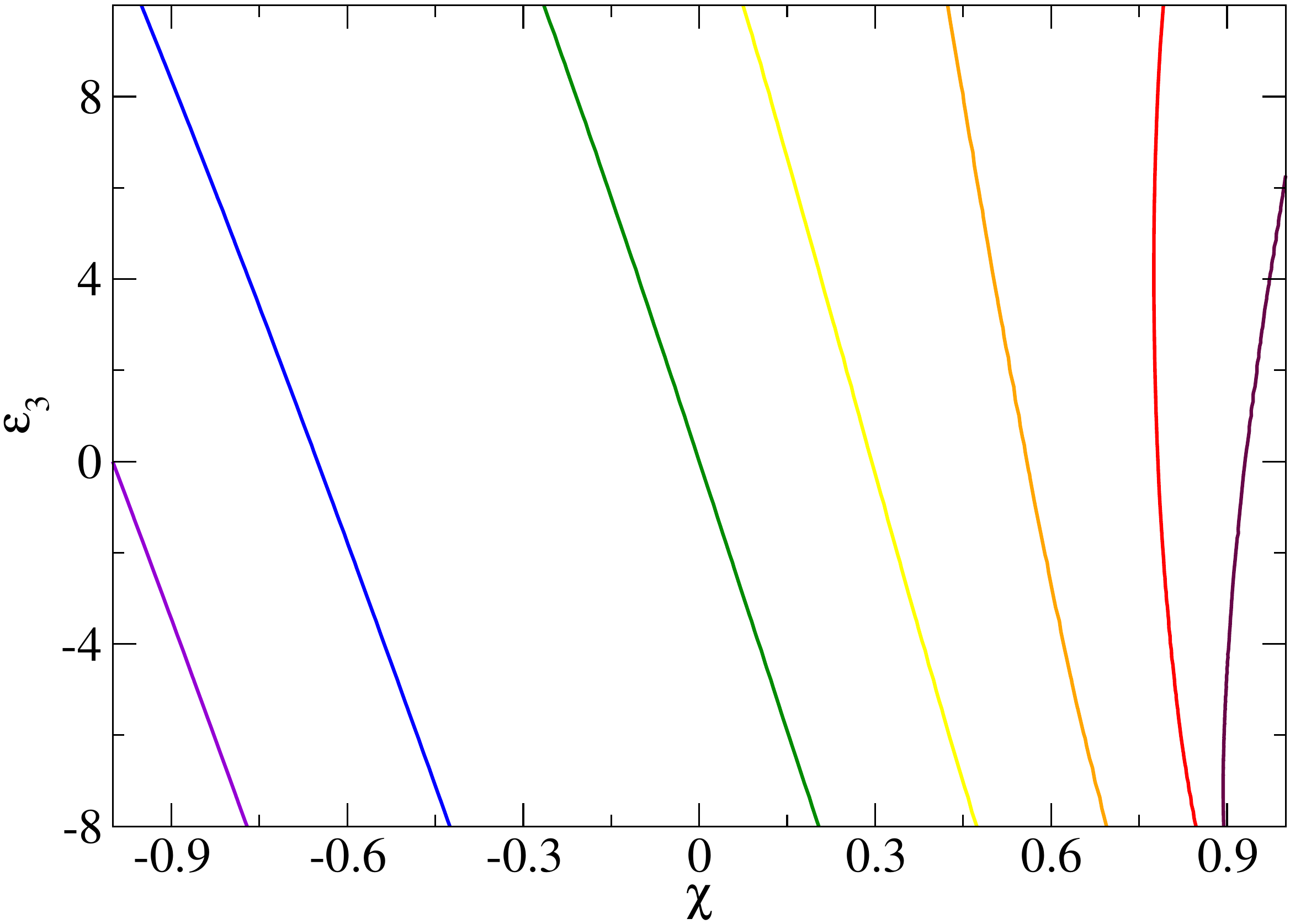}
\caption{(color online) Constant $r_\ISCO$ contours displaying their dependence on the unitless BH spin $\chi$ and non-GR deviation parameters $\epsilon_3$, $\alpha_{02}$, $\alpha_{13}$, and $\alpha_{22}$.
Such contours are presented (from right to left) for $r_\ISCO$ values of $2M$ (black), $3M$ (maroon), $4M$ (red), $5M$ (orange), $6M$ (yellow), $7M$ (green), $8M$ (blue), $9M$ (indigo), $10M$ (violet), and $11M$ (magenta). 
When varying the parameters $\alpha_{13}$, or $\alpha_{22}$, we fix $\alpha_{02}=10$, and vice versa, to avoid the presence of naked singularities.
}\label{fig:rISCO}
\end{center}
\end{figure*}

Now let us plot the resulting energy $E/\mu$ and angular momentum $L_z/\mu$ for various deviation parameters present in the expressions obtained above.
Here we vary only the lowest-order non-vanishing parameters present in the given expressions: $\alpha_{13}$, $\alpha_{22}$, $\alpha_{02}$, and $\epsilon_3$.
For a further analysis on lower-order parameters assumed to vanish here, see App.~\ref{app:lowerOrder}.
In Figs.~\ref{fig:energy} and~\ref{fig:momentum}, we plot the energy and angular momentum as a function of radius for a particle of mass $\mu$ on a circular orbit for several combinations of non-GR deviation parameters that produce BHs without naked singularities, where the energies and angular momenta become discontinuous and non-positive.
In each case, all non-GR parameters that are not specifically mentioned are set to be 0.
We see that in general, non-GR parameters (including the new parameter $\alpha_{02}$) have a significant impact on the energy and angular momentum of orbiting particles. and also the ISCO radius (minimum point of the energy curves) that we will describe in more detail in the next section.


\subsection{Innermost stable circular orbits}\label{sec:ISCO}

In this section, we compute the location of the ISCO.
In particular, the ISCO occurs at the minimum stable point of the orbital energy E of a particle with a circular orbit, namely 
\begin{equation}
\frac{dE}{dr}\Big|_{r=r_\ISCO}=0.
\end{equation} 
Because the dependence of this solution is very complicated in terms of the lower-order deviation parameters, we here plot contours of constant $r_\ISCO$ for varying unitless BH spins $\chi\equiv a/M$, and magnitude of deviation parameters.
Figure~\ref{fig:rISCO} does just this for 5 different classes of non-vanishing deviation parameters, taking note that the parameters $\alpha_{13}$ or $\alpha_{22}$ can not be the sole non-vanishing parameter unless $\alpha_{02}$ is also non-vanishing, else naked singularities appear as discussed previously and in App.~\ref{app:nakedSingularities}.
Thus, to vary $\alpha_{13}$ or $\alpha_{22}$, we fix $\alpha_{02}=10$ and vice versa, in order to check the $r_\ISCO$ dependence on individual non-GR parameters.

Now we discuss the ISCO dependence on the lower-order non-GR parameters $\epsilon_3$, $\alpha_{02}$, $\alpha_{13}$, and $\alpha_{22}$ as seen in Fig.~\ref{fig:rISCO}.
When varying the parameter $\epsilon_3$, we see that for $\chi<0.8$ the ISCO is mildly dependent on non-GR perturbations.
When varying $\alpha_{22}$ we observe that $r_\ISCO$ stays almost constant for any given value of $\alpha_{22}$ except for very large spins.
As for $\alpha_{13}$, we see that the location of the ISCO depends very strongly on the non-GR parameter.
Finally we observe that for BHs with non-vanishing spin, the dependence of $r_\ISCO$ on $\alpha_{02}$ becomes increasingly stronger for increasingly larger BH spins $\chi$.


\subsection{Photon rings}\label{sec:shadows}
In this section, we obtain solutions describing the orbit of a photon about a BH described by the new metric with various non-vanishing deviation parameters.
Following Refs.~\cite{Carter,Johannsen:2015pca,Johannsen:2015qca}, we begin with the Hamilton-Jacobi function
\begin{equation}
S\equiv - \frac{1}{2}\mu\tau-E t+L_z\phi+S_r(r)+S_\theta(\theta)
\end{equation} 
for particle mass $\mu$, proper time $\tau$, orbital energy $E$, angular momentum $L_z$, and generalized radial and polar functions $S_r(r)$ and $S_\theta(\theta)$.
We compute the Hamilton-Jacobi equations
\begin{equation}
-\frac{\partial S}{\partial\tau}=\frac{1}{2}g^{\alpha\beta}\frac{\partial S}{\partial x^\alpha}\frac{\partial S}{\partial x^\beta},
\end{equation}
to obtain
\begin{align}
\nonumber &-\mu ^2 \left(a^2 \cos ^2(\theta )+f(r)+g(\theta)+r^2\right)=\\
&\nonumber\frac{1}{\Delta}\Bigg\lbrack-a^4 A_1^2 E^2+2 a^3 A_0 E L_z-2 a^2 A_1^2 E^2 r^2-a^2 A_2^2 L_z^2\\
\nonumber&+a^2 \Delta  E^2 \sin ^2\theta +2 a A_0 E L_z r^2-2 a \Delta  E L_z-A_1^2 E^2 r^4\\
\nonumber &+A_5 \Delta ^2 \left(\frac{\partial S_r}{\partial r}\right)^2+\Delta  L_z^2 \csc ^2\theta +\Delta  \left(\frac{\partial S_\theta}{\partial \theta}\right)^2\Bigg\rbrack.
\end{align}

Next we separate the Hamilton-Jacobi equations, using the separation constant
\begin{align}
\nonumber C=&-\mu ^2 -\left(f(r)+r^2\right)-\frac{1}{\Delta}\Big\lbrack-a^4 A_1^2 E^2+2 a^3 A_0 E L_z\\
\nonumber&-2 a^2 A_1^2 E^2 r^2-a^2 A_2^2 L_z^2+2 a A_0 E L_z r^2-A_1^2 E^2 r^4\\
&+A_5 \Delta ^2 \left( \frac{\partial S_r}{\partial r} \right)^2\Big\rbrack,\\
\nonumber C=& a^2 E^2 \sin^2\theta+\mu ^2 (g(\theta)+a^2\cos ^2\theta)-2 a E L_z\\
&+L_z^2 \csc^2\theta+\left(\frac{\partial S_r}{\partial r} \right)^2.
\end{align}
We then define the Carter-like constant of motion $Q\equiv C-(L_z-a E)^2$ which gives us a solution for $S_r(r)$ (and $S_\theta(\theta)$, not displayed here) as
\begin{align}
S_r(r)&=\pm\int dr\frac{1}{\Delta}\sqrt{\frac{R(r)}{A_5(r)}},\\
\nonumber R(r)&\equiv a^4 A_1^2 E^2-2 a^3 A_0E L_z+2 a^2 A_1^2 E^2 r^2+a^2 A_2^2 L_z^2\\
\nonumber &-a^2 \Delta  E^2-2 a A_0 E L_z r^2+2 a \Delta  E L_z+A_1^2 E^2 r^4\\
&-\Delta  f(r) \mu ^2-\Delta  L_z^2-\Delta  Q-\Delta  \mu ^2 r^2,
\end{align}
where the different signs represent particles with prograde and retrograde motion, respectively. 
This proves that the new metric presented here indeed has a separable structure, and thus generalizing Johannsen's~\cite{Johannsen:2015pca}.

Finally, we compute the generalized momenta $p_\alpha$ given by 
\begin{equation}
p_\alpha=\frac{\partial S}{\partial x^\alpha}.
\end{equation}
In particular, we focus on the radial momenta, given in covariant and contravariant form as 
\begin{align}
p_r=\pm \frac{1}{\Delta}\sqrt{\frac{R(r)}{A_5(r)}},\\
p^r=\pm \frac{A_5(r)R(r)}{\tilde\Sigma}.\label{eq:momentum}
\end{align}
Following Ref.~\cite{Johannsen:2015qca}, the impact parameters $x'$ and $y'$~\cite{Bardeen_1973} describing the image plane from an observer's point of view at infinity with an inclination angle $i$ can be found to be
\begin{equation}
x'=-\frac{\xi}{\sin i}, \hspace{4mm} y'=\pm\sqrt{\eta+a^2\cos^2 i-\xi^2\cot^2i}.\label{eq:photonRings}
\end{equation}
In the above expression, the new invariant parameters $\xi\equiv L_z/E$ and $\eta\equiv Q/E^2$ have been constructed entirely out of constants of motion.

Now we describe the solutions $(\xi,\eta)$ which describe the photon rings of a black hole.
Such new constants of motion are conserved along the null geodesics, and thus can be simply solved for at the special case of a circular-orbit for simplicity.
Here, the radial photon momentum $p^r$ found in Eq.~\eqref{eq:momentum} as well as its radial derivative must vanish.
Because $\tilde\Sigma$ and $A_5(r)$ are both non-negative, this results in the system of equations
\begin{align}
R(r)&=0, \quad
\frac{dR(r)}{dr}=0,\label{eq:SolveEqs}
\end{align}
with the full re-parameterized expression for $R(r)$ for an orbiting photon ($\mu=0$) given by
\begin{align}
\nonumber R(r)=&a^4 A_1(r)^2-2 a^3 A_0(r) \xi +2 a^2 A_1(r)^2 r^2\\
\nonumber &+a^2 A_2(r)^2 \xi ^2-a^2 \Delta -2 a A_0(r) \xi  r^2\\
 &+2 a \Delta  \xi+ A_1(r)^2 r^4-\Delta  \eta -\Delta  \xi ^2.
\end{align}
With this, one can simultaneously solve Eq.~\eqref{eq:SolveEqs} for $\xi$ and $\eta$ to give parameterized expressions for the image of the photon rings in Eq.~\eqref{eq:photonRings}.
We do not present such results here since they are quite lengthy. In the Kerr limit, however, they are found to correctly reduce to the GR expressions found in~\cite{Bardeen}
\begin{align}
\xi^\K&=-\frac{r^2(r-3M)+a^2(r+M)}{a(r-M)},\\
\eta^\K&=\frac{r^3\lbrack 4a^2M-r(r-3M)^2 \rbrack}{a^2(r-M)^2}.
\end{align} 
We also note that the expressions for $\xi$ and $\eta$ only depend on the non-GR deviation functions $A_1(r)$, $A_2(r)$, and $A_0(r)$.

Now we compute the image of the photon rings about a BH described by the new metric.
In particular, we focus our attention on only the lowest order parameters $\alpha_{13}$, $\alpha_{22}$, and $\alpha_{02}$, as was done in~\cite{Johannsen:2015pca,Johannsen:2015qca}.
However, we refer the reader to App.~\ref{app:lowerOrder} for an analysis of the inclusion of lower-order parameter $\alpha_{12}$, which was assumed to vanish in the ppN framework.
There we also consider photon rings in the EdGB theory of gravity, where we investigate the validity of the $1/r$ expansion in the EdGB metric.
Here we vary each parameter $\alpha_{13}$, $\alpha_{22}$, and $\alpha_{02}$, while avoiding the cases with emergent naked singularities (See App.~\ref{app:nakedSingularities}), else closed photon rings do not appear and the photons escape to radial infinity as discussed in~\cite{Hioki:2009na,Papnoi:2014aaa}.
This means that if $\alpha_{13} \ne 0$ or $\alpha_{22} \ne 0$, then $\alpha_{02}$ must also be non-vanishing and of the same sign.

\begin{figure}[htb]
\begin{center}
\includegraphics[width=.4\textwidth]{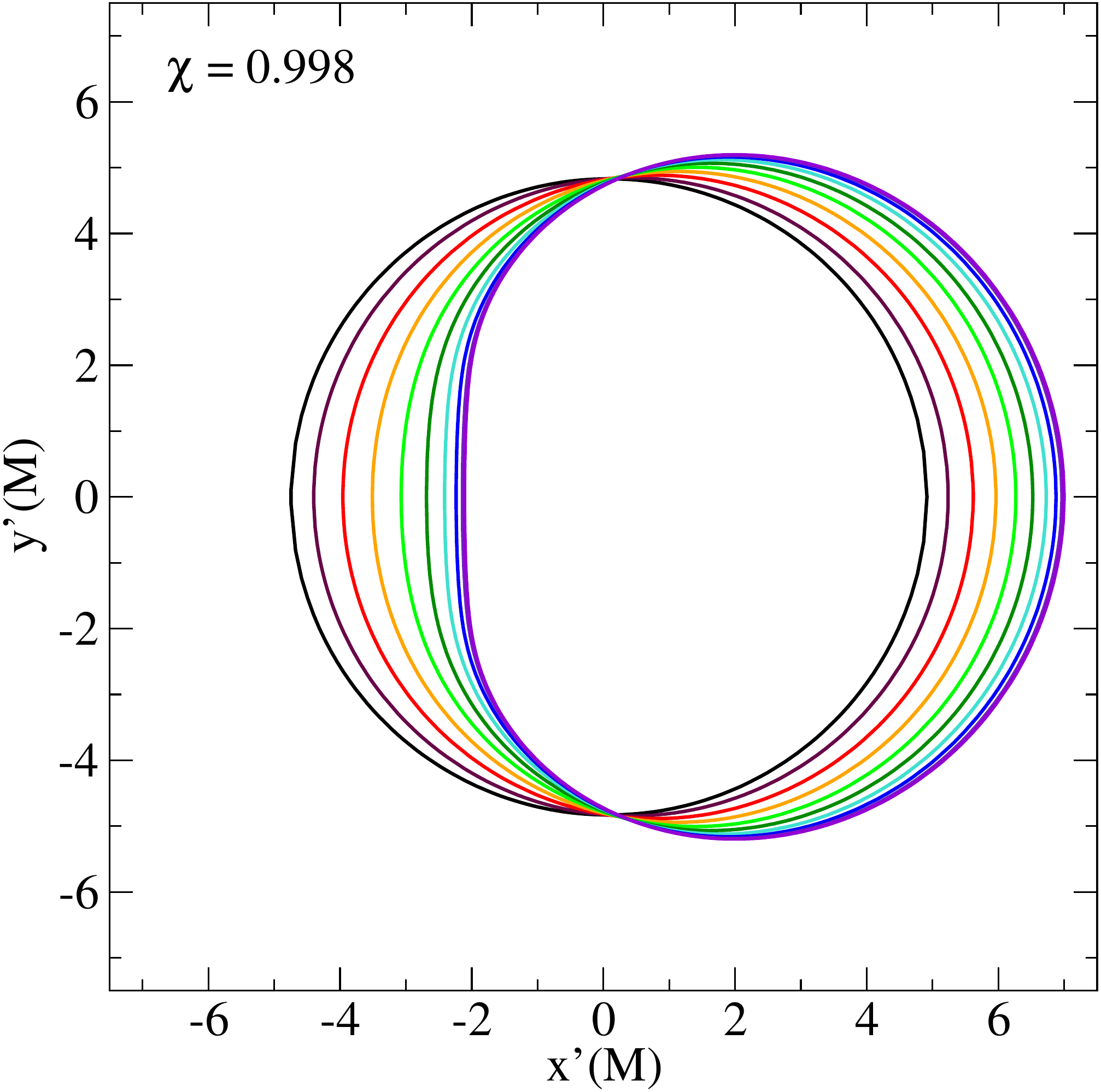}
\includegraphics[width=.4\textwidth]{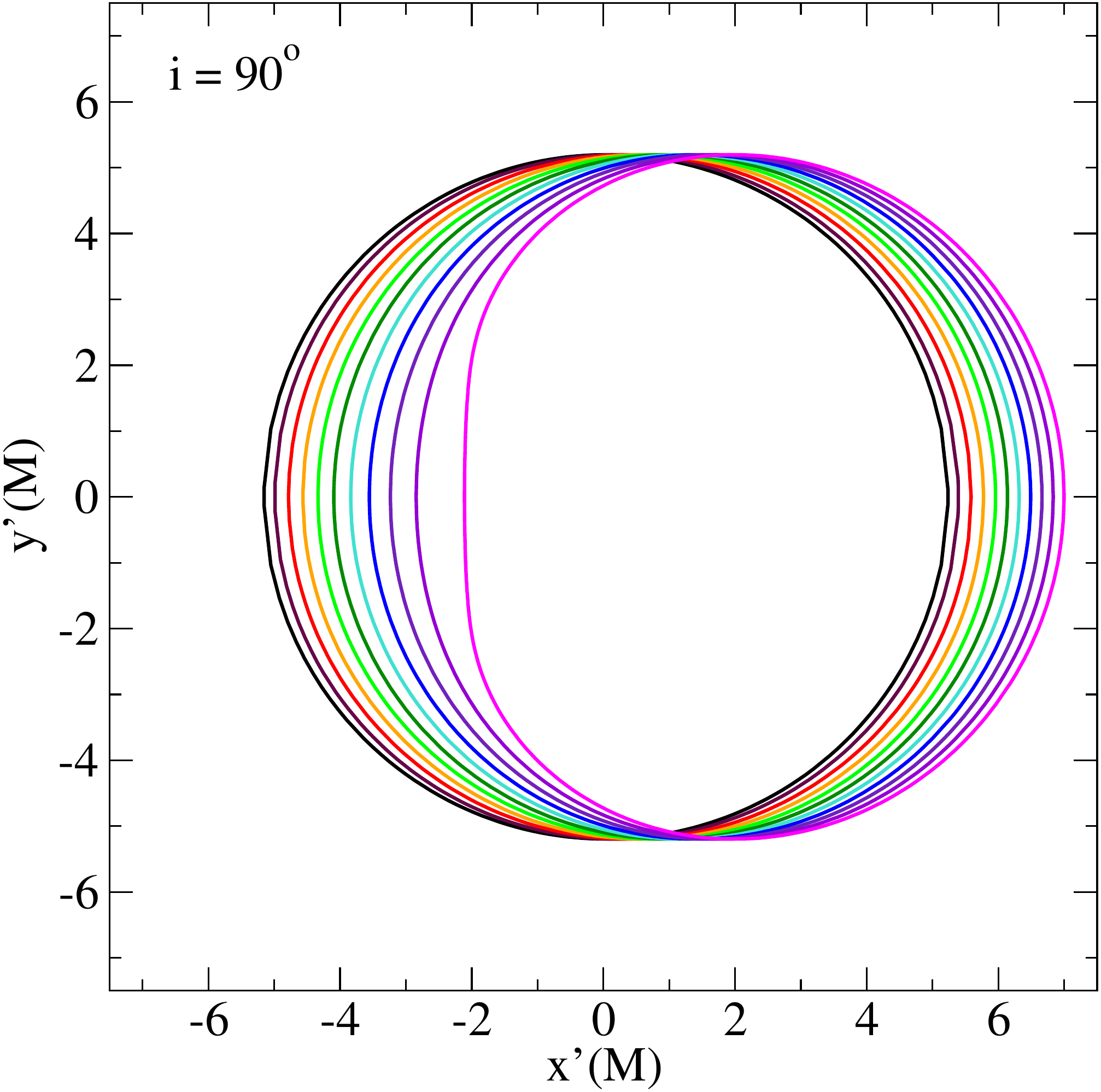}
\caption{(color online) Images of photon rings of a Kerr BH seen by an observer at infinity (all deviation parameters vanishing in the new metric) for increasing degrees of inclination at a fixed BH spin (top), and increasing BH spins at a fixed inclination (bottom).
For the former, we fix the BH spin at the extreme case of $\chi=0.998$ for demonstration purposes, and increase the inclination angle going left to right from $i=0^\circ$ to $i=90^\circ$ in intervals of $10^\circ$.
For the latter, we fix the inclination angle at the extreme case of $i=90^\circ$, and increase the BH spin going left to right from $\chi=0$ to $\chi=0.998$ in intervals of $0.1$.
}\label{fig:BHshadowsInclinationSpin}
\end{center}
\end{figure}

Figure~\ref{fig:BHshadowsInclinationSpin} shows the image of the photon ring as it depends on the BH's spin ($\chi$), and the observers inclination angle ($i$) for the case of all deviation parameters vanishing (Kerr) for simplicity.
We observe that for a fixed inclination angle, increasing the BHs spin serves to increasingly displace and deform the photon's orbit.
Similarly, for a fixed BH spin, the inclination dilutes the displacement and deformation gained from the rotating BH for all but the highest angles. 
Such displacement and deformation as a function of inclination only reaches the maximum value allowable by the spin, with none present for a static non-rotating BH.

\begin{figure*}[!htbp]
\begin{center}
\includegraphics[width=.4\textwidth]{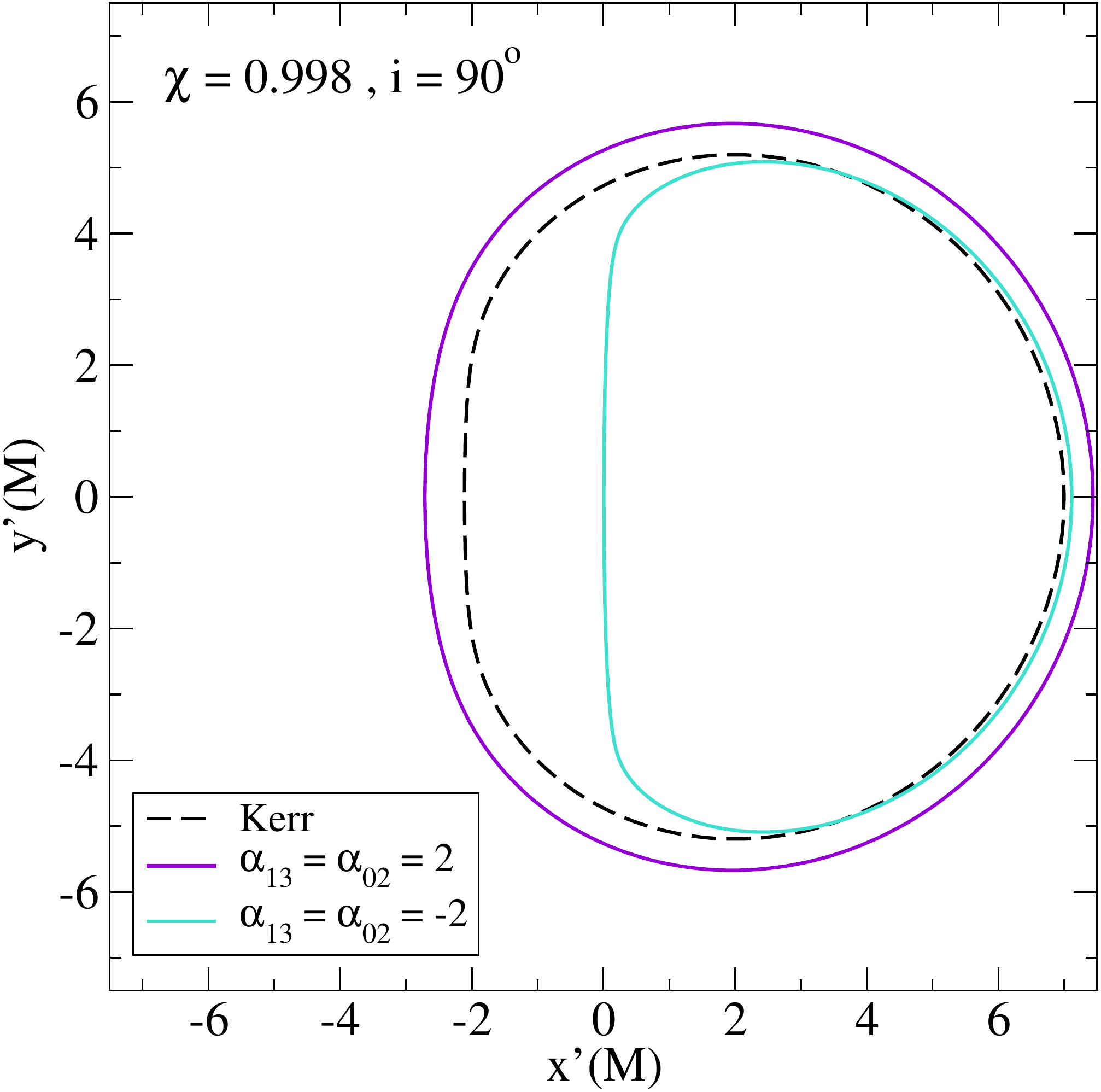}
\includegraphics[width=.4\textwidth]{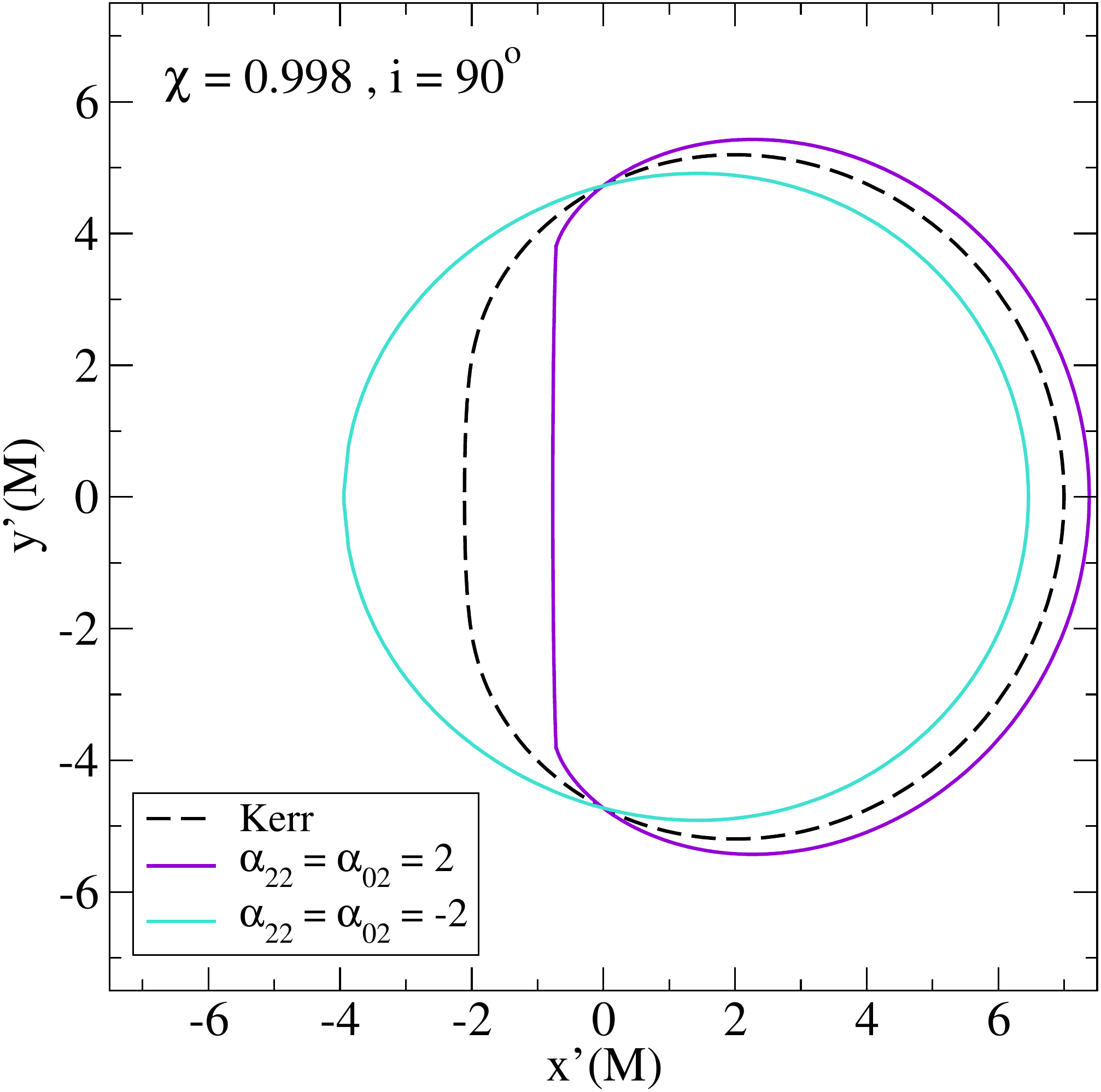}\\
\includegraphics[width=.4\textwidth]{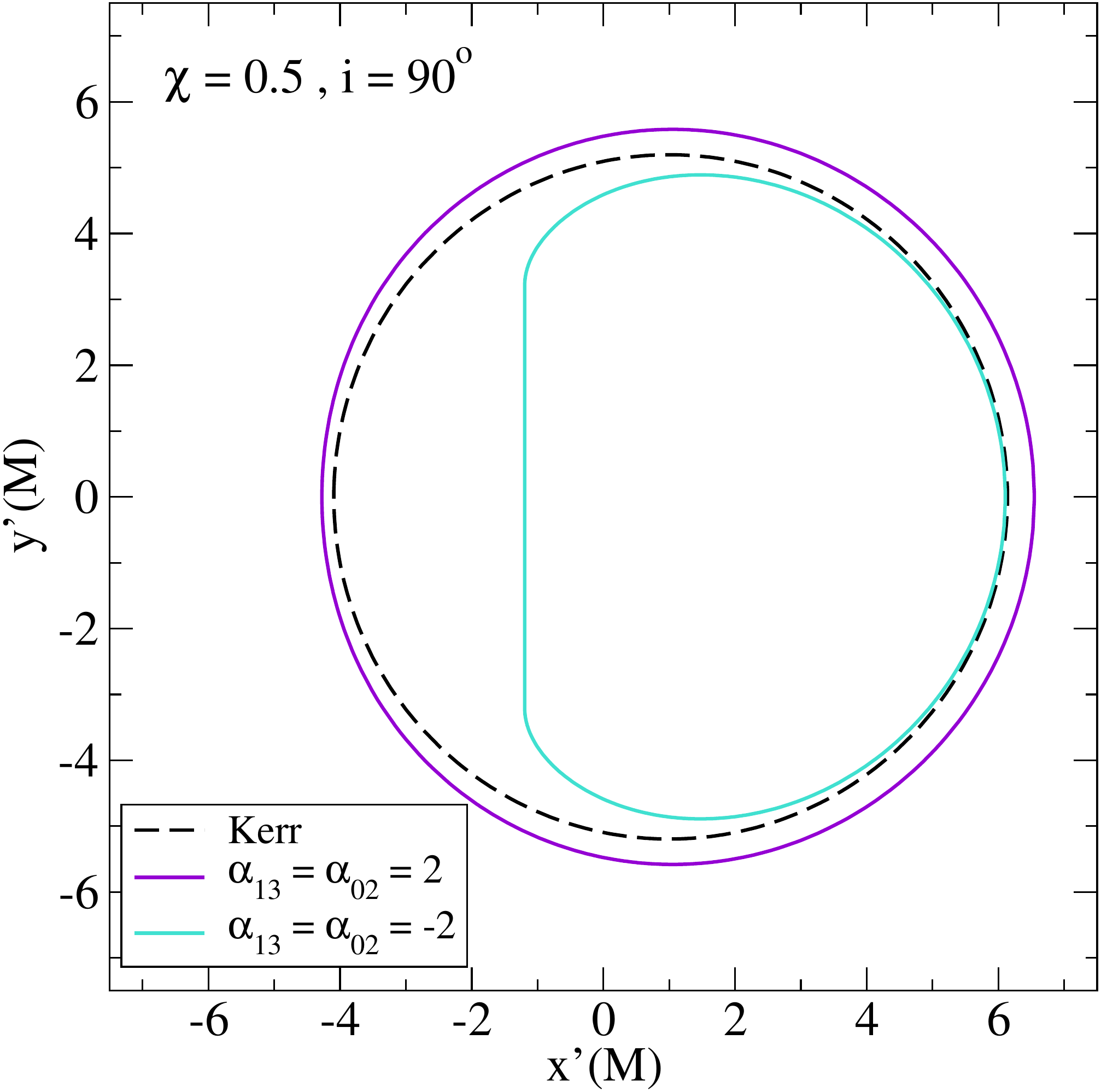}
\includegraphics[width=.4\textwidth]{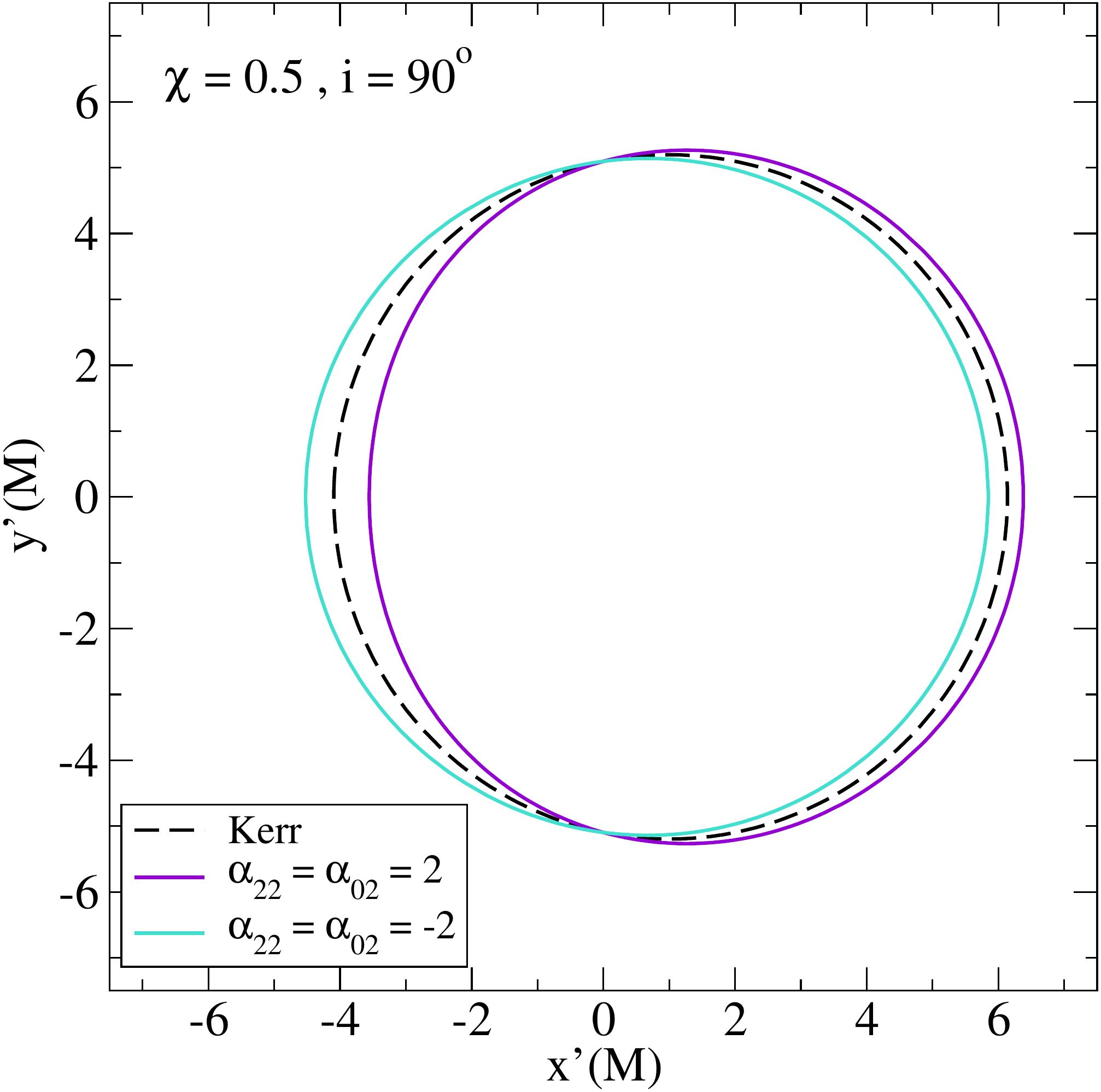}\\
\includegraphics[width=.4\textwidth]{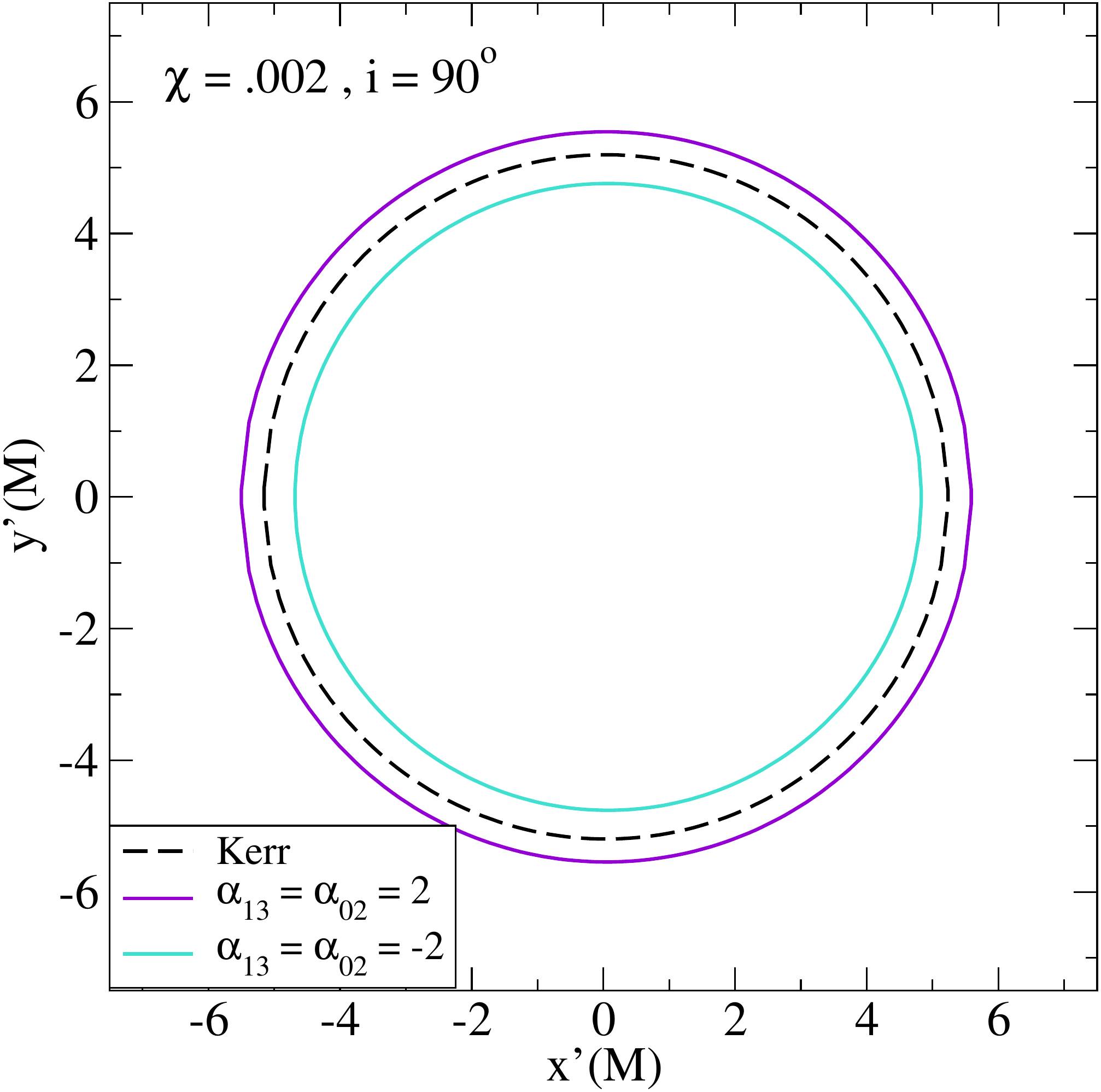}
\includegraphics[width=.4\textwidth]{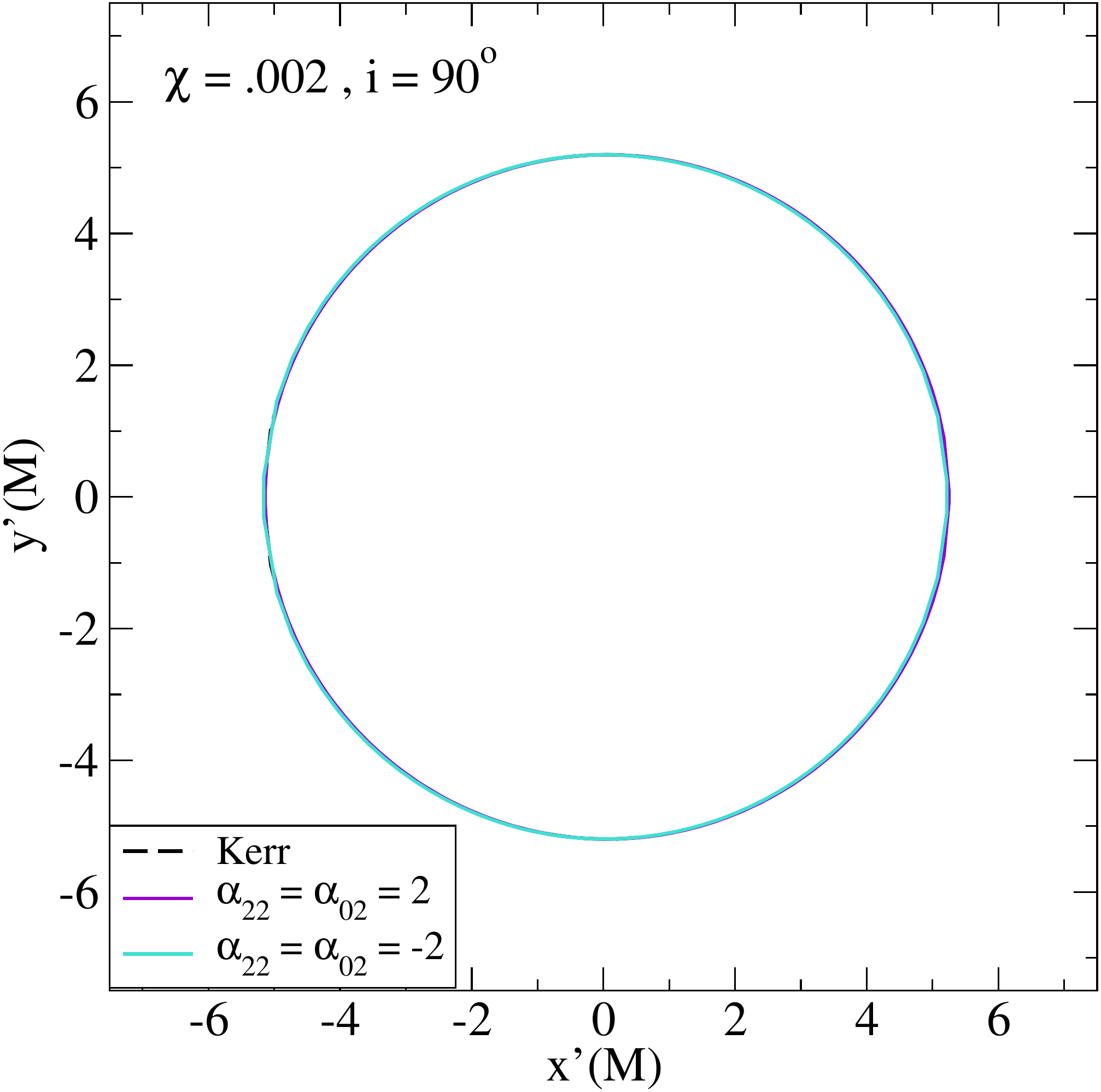}\\
\caption{(color online) Images of photon rings about a high-spin $\chi=0.998$ (top), medium-spin $\chi=0.5$ (middle), and low-spin $\chi=0$ (bottom) BH, for various non-vanishing GR deformation parameters $\alpha_{13}$, $\alpha_{22}$, and $\alpha_{02}$.
We avoid combinations of such parameters that produce naked singularities and let $\alpha_{13}=\alpha_{02}=\pm2$ (left) and $\alpha_{22}=\alpha_{02}=\pm 2$ (right).
The inclination is fixed at the extreme case of $i=90^\circ$ in every scenario, for demonstration purposes.
In the bottom-right panel, we observe that the parameters $\alpha_{22}$ and $\alpha_{02}$ do not significantly alter the photon orbit for slowly-rotating BHs. 
}\label{fig:BHshadowsNonGR}
\end{center}
\end{figure*}

Finally, we compute the images of the closed photon rings about a BH for several non-vanishing deviation parameters in Fig.~\ref{fig:BHshadowsNonGR}.
Specifically, for highly-rotating BHs ($\chi=0.998$), moderately-rotating BHs ($\chi=0.5$), and slowly-rotating BHs ($\chi=0.002$), we generate the photon rings for different non-vanishing values of $\alpha_{13}$, $\alpha_{22}$, and $\alpha_{02}$.
We observe that the effect of increasing $\alpha_{13}$ and $\alpha_{02}$ acts to increase the image size, and negative values of each parameter works to deform the image.
The latter becomes less apparent as the spin decreases, while the former still holds true for even low BH spins.
Next we see that non-vanishing values of $\alpha_{22}$ and $\alpha_{02}$ only marginally affect the image size, but highly deform the orbits for fast-rotating BHs.
In this case (and not in the case of non-vanishing $\alpha_{13}$ and $\alpha_{02}$), we see that positive values of the parameters work to deform the image inwards, while negative values distort outwards.
We conclude with the remark that, especially for highly-rotating and/or highly-inclined observations, that BHs with deviations from Kerr are highly distinguishable from the exact Kerr result.
This is because the deviation parameters $\alpha_{22}$ and $\alpha_{02}$ (corresponding to free functions $A_2(r)$ and $A_0(r)$) are associated with modifications to the angular portions ($\phi$-components) of the contravariant metric in Eq.~\eqref{eq:CYcontravariant}.


\section{Transformation of the new metric to other spacetimes}\label{sec:mappings}

In this section we present the maps that take one from the new metric presented here to several other deformed spacetimes present in the literature.
In particular, we focus on the following theories and spacetimes:

\begin{enumerate}

\item the separable deformed spacetime: 

\begin{itemize}

\item spacetime in Papadopoulos and Kokkotas~\cite{Papadopoulos:2018nvd};

\item parameters: generic deviation parameters $\mathcal{A}_i$ and $\mathcal{B}_i$ for $i=1\dots5$;

\end{itemize}

\item the string-inspired RS2 Braneworld~\cite{Randall:1999ee}:

\begin{itemize}

\item spacetime in~\cite{Aliev:2005bi};

\item parameters: the tidal charge $\beta$;

\end{itemize}

\item the heterotic string theory:

\begin{itemize}

\item spacetime in Kerr \& Sen~\cite{Kerr-Sen};

\item parameters: deviation parameter $b$ related to the magnetic dipole moment;

\end{itemize}

\item Einstein-dilaton-Gauss-Bonnet (EdGB) gravity~\cite{Kanti_EdGB,Maeda:2009uy,Sotiriou:2014pfa}: 

\begin{itemize}

\item spacetime in~\cite{Pani:2011gy,Ayzenberg:2014aka};

\item parameters: the dimensionless coupling constant $\zeta_\EdGB$;

\end{itemize}

\item dynamical Chern-Simons (dCS) gravity~\cite{Jackiw:2003pm,Alexander_cs} 

\begin{itemize}

\item spacetime in~\cite{Yunes_dcs,Yagi_dCS,Pani:2011gy};

\item parameters: the dimensionless coupling constant $\zeta_\dCS$;

\end{itemize}

\item quantum-corrected regular BHs:

\begin{itemize}

\item spacetime in Bardeen~\cite{Bardeen};

\item parameters: deviation $g$ controlling the amount of BH ``regularity'';

\end{itemize}

\item  the Kalb-Ramond BH with Kalb-Ramond parameter $s=1$ and $s=2$:

\begin{itemize}

\item spacetime in~\cite{Kumar:2020hgm};

\item parameters: the Kalb-Ramond Lorentz-violating parameter $\Gamma$.

\end{itemize}

\end{enumerate}

Now let us provide a brief overview of the procedure used to find such mappings between the new spacetime presented here and the ones (X) listed above.
Using a computer-algebra software, this is done by first equating each metric element $g_{\alpha\beta}=g_{\alpha\beta}^\X$ to solve for the the functions $A_0(r)$, $A_1(r)$, $A_2(r)$, $A_5(r)$, and $f(r)$ found in the new metric as a function of the GR deformation parameters present in metric X.
Such functions are then expanded in powers of $M/r$ about $r=\infty$ to obtain the mappings between non-GR parameters $\alpha_{0n}$, $\alpha_{1n}$, $\alpha_{2n}$,  $\alpha_{5n}$, and $\epsilon_n$ and non-GR parameters in metric $X$.
We present all such mappings in Table~\ref{tab:maps}, for the three lowest-order non-vanishing parameters from each free function.
We note that here, we do not take into account the ppN bounds mentioned in Ref.~\cite{Johannsen:2015pca}, and instead allow all lower-order parameters to enter.

\renewcommand{\arraystretch}{1.2}
\begin{table*}[!htbp]
\centering
\resizebox{.6\textwidth}{!}{%
\begin{tabular}{|c| l|}
\hline
&\\[-1em]
BH spacetime & Deviation parameters\\
&\\[-1em]
\hline
\hline
&\\[-1em]
\multirow{5}{*}{separable deformed spacetime~\cite{Papadopoulos:2018nvd}} & $f(r)=A_1^\PK(r)-r^2$, \\
& $A_1(r)=\frac{\sqrt{-A_5^\PK(r)\Delta}}{a^2+r^2}$, \\
& $A_2(r)=\frac{\sqrt{-A_3^\PK(r)\Delta}}{a}$, \\
& $A_5(r)=\frac{A_2^\PK(r)}{\Delta}$, \\
& $A_0(r)=-\frac{A_4^\PK(r)\Delta}{a(a^2+r^2)}$ \\
&\\[-1em]
\hline
&\\[-1em]
\multirow{4}{*}{RS2 Braneworld~\cite{Randall:1999ee}} & $\alpha_{10}=1$, $\alpha_{12}=-\frac{\beta}{2M^2}$, $\alpha_{13}=-\frac{\beta}{M^2}$, \dots, \\
& $\alpha_{20}=1$, $\alpha_{22}=-\frac{\beta}{2M^2}$, $\alpha_{23}=-\frac{\beta}{M^2}$, \dots, \\
& $\alpha_{50}=1$, $\alpha_{52}=\frac{\beta}{M^2}$, $\alpha_{53}=\frac{2\beta}{M^2}$, \dots, \\
& $\alpha_{00}=1$, $\alpha_{02}=-\frac{\beta}{M^2}$, $\alpha_{03}=-\frac{2\beta}{M^2}$, \dots \\
&\\[-1em]
\hline
&\\[-1em]
\multirow{5}{*}{heterotic string (Kerr-Sen)~\cite{Kerr-Sen}} & $\epsilon_1=\frac{2b}{M}$, \dots, \\
& $\alpha_{10}=1$, $\alpha_{11}=\frac{b}{M}$, $\alpha_{12}=-\frac{b^2+4bM}{2M^2}$, \dots, \\
& $\alpha_{20}=1$, $\alpha_{21}=-\frac{b}{M}$, $\alpha_{22}=\frac{3b^2-4bM}{2M^2}$, \dots, \\
& $\alpha_{50}=1$, $\alpha_{51}=\frac{2b}{M}$, $\alpha_{52}=\frac{4b}{M}$, \dots, \\
& $\alpha_{00}=1$, $\alpha_{02}=-\frac{4b}{M}$, $\alpha_{03}=\frac{8b(b-M)}{M^2}$, \dots \\
&\\[-1em]
\hline
&\\[-1em]
\multirow{4}{*}{EdGB gravity~\cite{Ayzenberg:2014aka}} & $\alpha_{10}=1$, $\alpha_{13}=-\frac{\zeta}{6}$, $\alpha_{14}=-\frac{14\zeta}{3}$, \dots, \\
& $\alpha_{20}=1$, $\alpha_{23}=-\frac{13\zeta}{30}$, $\alpha_{24}=-\frac{16\zeta}{3}$, \dots, \\
& $\alpha_{50}=1$, $\alpha_{52}=\zeta$, $\alpha_{53}=3\zeta$, \dots, \\
& $\alpha_{00}=1$, $\alpha_{03}=-\frac{3\zeta}{5}$, $\alpha_{04}=-10\zeta$, \dots \\
&\\[-1em]
\hline
&\\[-1em]
\multirow{4}{*}{dCS gravity~\cite{Yagi_dCS,Yunes_dcs}} & $\alpha_{10}=1$,\\
& $\alpha_{20}=1$, $\alpha_{24}=-\frac{5\zeta}{8}$, $\alpha_{25}=-\frac{15\zeta}{14}$, \dots, \\
& $\alpha_{50}=1$, \\
& $\alpha_{00}=1$, $\alpha_{04}=-\frac{5\zeta}{8}$, $\alpha_{05}=-\frac{15\zeta}{14}$, \dots \\
&\\[-1em]
\hline
&\\[-1em]
\multirow{4}{*}{quantum-corrected (Bardeen)~\cite{Bardeen}} & $\alpha_{10}=1$, $\alpha_{13}=-\frac{3g^2}{2M^2}$, $\alpha_{14}=-\frac{3g^2}{M^2}$, \dots,\\
& $\alpha_{20}=1$, $\alpha_{23}=-\frac{3g^2}{2M^2}$, $\alpha_{24}=-\frac{3g^2}{M^2}$, \dots, \\
& $\alpha_{50}=1$, $\alpha_{53}=\frac{3g^2}{M^2}$, $\alpha_{54}=\frac{6g^2}{M^2}$, \dots, \\
& $\alpha_{00}=1$, $\alpha_{03}=-\frac{3g^2}{M^2}$, $\alpha_{04}=-\frac{6g^2}{M^2}$, \dots \\
&\\[-1em]
\hline
&\\[-1em]
\multirow{8}{*}{Kalb-Ramond~\cite{Kumar:2020hgm}} \multirow{4}{*}{($s=1$)} & $\alpha_{10}=1$, $\alpha_{12}=-\frac{\Gamma}{2M^2}$, $\alpha_{13}=-\frac{\Gamma}{M^2}$, \dots, \\
& $\alpha_{20}=1$, $\alpha_{22}=-\frac{\Gamma}{2M^2}$, $\alpha_{23}=-\frac{\Gamma}{M^2}$, \dots, \\
& $\alpha_{50}=1$, $\alpha_{52}=\frac{\Gamma}{M^2}$, $\alpha_{53}=\frac{2\Gamma}{M^2}$, \dots, \\
& $\alpha_{00}=1$, $\alpha_{02}=-\frac{\Gamma}{M^2}$, $\alpha_{03}=-\frac{2\Gamma}{M^2}$, \dots \\[0.5em]
\cline{2-2}
&\\[-1em]
\multirow{4}{*}{\color{white}Kalb-Ramond~\lbrack00\rbrack \color{black}($s=2$)} & $\alpha_{10}=1$, $\alpha_{11}=-\frac{\Gamma}{2M}$, $\alpha_{12}=\frac{\Gamma(3\Gamma-8M)}{8M^2}$, \dots,\\
& $\alpha_{20}=1$, $\alpha_{21}=-\frac{\Gamma}{2M}$, $\alpha_{22}=\frac{\Gamma(3\Gamma-8M)}{8M^2}$, \dots, \\
& $\alpha_{50}=1$, $\alpha_{51}=\frac{\Gamma}{M}$, $\alpha_{52}=\frac{2\Gamma}{M}$, \dots, \\
& $\alpha_{00}=1$, $\alpha_{01}=-\frac{\Gamma}{M}$, $\alpha_{02}=\frac{\Gamma(\Gamma-2M)}{M^2}$, \dots \\[0.5em]
\hline
\end{tabular}
}
\caption{Mappings from the new metric presented in this paper to several other BH solutions in related works.
For the separable deformed spacetime, we show the transformation between arbitrary functions rather than expanded coefficients, where $A_i^\PK(r)$ correspond to the beyond-GR functions in~\cite{Papadopoulos:2018nvd}.
The mappings to EdGB and dCS gravity are only valid up to linear order in BH spin and first order in deformation parameters $\zeta$, and the mapping to Bardeen is only valid to quartic order in deviation parameter $g$.
Parameters which are missing correspond to those that are vanishing in the series expansion.
}\label{tab:maps}
\end{table*}

We now discuss the results presented in Table~\ref{tab:maps}.
We first note that in the separable deformed spacetime presented by Papadopoulos and Kokkotas, we only find the transformation from the arbitrary functions presented here $A_i(r)$, $f(r)$ to the ones ($A_i^\PK(r)$) found in~\cite{Papadopoulos:2018nvd}.
Additionally, Ref.~\cite{Johannsen:2015pca} states that the RS2 Braneworld metric could not be related to Johannsen's metric, and in Ref.~\cite{Johannsen:2015qca} the author found a mapping by introducing a new non-GR parameter $\beta$ such that $\Delta \to \Delta+\beta$.
However, we have found that both Johannsen's metric and the metric presented here could be mapped to RS2 Braneworld as shown in Table~\ref{tab:maps}, with some difficulty.
In both the EdGB and dCS theories of gravity, the mapping is only valid up to first order in spin $\chi$ and coupling parameter $\zeta$.
We note that all the mappings presented in Table~\ref{tab:maps} for the new metric can also be mapped to Johannsen's.

We finish this section by noting the high versatility of this new metric, with the ability to map to many BH solutions found in the literature. Having said this, we point out there are many BH solutions that cannot be mapped to the new metric found here. Such metrics include BHs in 
Einstein-scalar gravity~\cite{Bogush:2020lkp} and Einstein-Maxwell dilaton theory~\cite{Jai-akson:2017ldo}, the Bumblebee metric~\cite{Ding:2019mal}, and slowly-rotating BHs in EdGB and dCS gravity~\cite{Yagi_dCS,Ayzenberg:2014aka}.
These metrics do not contain separable geodesic equations, thus no Carter-like constant is present and the geodesic equations may become chaotic~\cite{Cardenas-Avendano:2018ocb}.
Such fundamental differences manifest themselves as an inability to transform to Johannsen's metric or the metric presented here, and can be seen by the appearance of angular functions within the mapping functions of $f(r)$ and $A_i(r)$.


\section{Conclusion and Discussion}\label{sec:conclusion}
The no-hair theorem tells us that isolated BHs create a spacetime described by the famous Kerr metric.
In this metric, we can predict the shape and size of photon rings seen by a far-away observer that depend only on the central BH's mass and spin.
While several tests to date have confirmed this hypothesis~\cite{Yunes:2013dva,Gair:2012nm,Liu_2012,Pfahl_2004,Wex_1999,Sadeghian:2011ub,Merritt:2009ex,Will_2008,Johannsen:2011dh,Bambi:2013sha,Bambi:2012at,Johannsen:2012ng,Bambi:2011jq,Bambi:2012ku,Johannsen:2010ru,Johannsen:2010xs,Psaltis_2011,Johannsen_2016,Krawczynski_2012,PhysRevD.83.103003,Johannsen_2010,Berti:2007zu,Dreyer:2003bv,PhysRevD.73.064030,Vigeland:2011ji,Apostolatos:2009vu,Gair:2007kr,Collins:2004ex,Glampedakis:2005cf,PhysRevD.84.064016,PhysRevD.81.024030,PhysRevD.78.102002,PhysRevD.69.082005,Barack:2006pq,Johannsen:2011mt,Mandel_2014,PhysRevD.77.064022,PhysRevD.56.1845,PhysRevD.52.5707,Isi:2019aib}, what if small deviations from the Kerr metric yet exist in nature?
That is one question the EHT with the VLBI aim to answer, by using an effective earth-sized telescope to accurately map photon rings about SMBHs located at the center of galaxies.

In this manuscript, we have extended the important analysis done by Johannsen in~\cite{Johannsen:2015pca} to design a more general Kerr-like BH solution.
This new metric can be interpreted as the most general stationary, axisymmetric, and asymptotically flat spacetime we can create with intact separable geodesic equations.
Such a new metric is parameterized non-linearly by 5 free functions $A_i(r)$ for $i=0,1,2,5$ and $f(r)$ which deviate from the Kerr metric, and reproduce the exact Kerr metric when vanishing.
This general metric can be mapped to a large range of BH solutions found in the literature, as demonstrated for seven different cases~\cite{Papadopoulos:2018nvd,Randall:1999ee,Aliev:2005bi,Jai-akson:2017ldo,Ding:2019mal,Kerr-Sen,Kanti_EdGB,Maeda:2009uy,Sotiriou:2014pfa,Ayzenberg:2014aka,Jackiw:2003pm,Yagi_dCS,Yunes_dcs,Bardeen,Kumar:2019uwi,Pani:2011gy,Kumar:2020hgm}. 
The metric has been shown to produce an event horizon and Killing horizon coexistent with the Kerr one. 
Finally, we looked at the spheroidicity conditions found in Ref.~\cite{Glampedakis:2018blj}, finding a $\theta$-independent function which admits a Kerr-like spherical photon orbits.

Now that we have a new, general metric in hand, we proceeded to calculate several properties of the ensuing spacetime.
In particular, we focused on circular equatorial particle orbits and found analytic expressions for the orbital energy and angular momentum of such a particle, along with its Keplerian and epicyclic frequencies for perturbed radial and vertical orbits.
We plot these quantities for several different parameterizations of the new metric for comparison against the Kerr result to show the effect of the parameterized deviations.
We then compute the location of the ISCO, once again comparing the results for several parameterizations against the Kerr result.

We finally shift our attention to the orbits of photons about BHs described by the new metric presented here.
By following the analysis of Johannsen in~\cite{Johannsen:2015qca}, we derive analytic expressions for thin photon orbit solutions, called ``photon rings''.
The images of such orbits are observable by, e.g. the EHT, and are extremely timely due to the recent image of the lensed photon orbits about the SMBH M87$^*$~\cite{1435171,1435174,1435175,1435177,1435168}, with future resolution and fidelity improvements imminent.
We then compare the photon rings about BHs with several different parameterizations against the Kerr result.
We find that, especially for highly-rotating BHs and/or highly inclined observation angles, the non-Kerr photon rings indeed distinguish themselves prominently against the standard Kerr result.

Future work includes constraining the new metric found in this paper with current and future observations. For example, one can use observations of X-ray continuum spectrum and iron line emissions from accretion disks around BHs to constrain some of the parameters, as already done in~\cite{Choudhury:2018zmf,Tripathi:2018lhx,Tripathi:2019bya,John:2019rhj,Tripathi:2019fms}.
Another way to constrain the metric is to use future gravitational-wave observations. 
For example, extreme-mass ratio inspirals can probe accurately spacetime around BHs~\cite{Vigeland:2011ji,Gair:2011ym,Apostolatos:2009vu,Gair:2007kr,Collins:2004ex,Glampedakis:2005cf,PhysRevD.84.064016,PhysRevD.81.024030,PhysRevD.78.102002,PhysRevD.69.082005,Barack:2006pq,Johannsen:2011mt,Mandel_2014,PhysRevD.77.064022,PhysRevD.56.1845,PhysRevD.52.5707,Isi:2019aib}. Other possibilities include pulsars orbiting around BHs~\cite{Liu_2012,Liu:2014uka}, stars orbiting around the center of Sgr A$^*$~\cite{Sadeghian:2011ub,Merritt:2009ex,Will_2008} and low-mass X-ray binaries with BHs~\cite{Yagi_EdGB}.
One could additionally repeat the analysis done in~\cite{Cardenas-Avendano:2019zxd} with the new metric presented here, where the authors compared current and future gravitational wave and x-ray constraints on deformed spacetime metric parameters.

In addition, future work on the presented topic includes a detailed investigation into a finite stress-energy tensor that sources the beyond-Kerr nature presented here, if such a metric corresponds to a non-vacuum spacetime.
In particular, one could compute Einstein's Equations with the new metric considered in this paper assuming GR, and attribute the purely beyond-Kerr components to an additional stress-energy tensor.
While the key purpose of this investigation was to determine an arbitrary, theory agnostic spacetime model, a source term for such effects is interesting to study, and e.g. determine if the energy conditions are satisfied.
We found that the stress-energy tensor and energy conditions required for the beyond-Kerr corrections presented here are lengthy and complicated given the number of arbitrary beyond-Kerr functions, and do not provide any immediately meaningful observations. 
We leave a further analysis on this point for future work.

Additionally, one could introduce a stress-energy tensor corresponding to the accretion disk found outside of e.g. M87$^*$ in the EHT observations, find a black hole solution with such an accretion disk perturbatively and see if such a solution can be mapped to the beyond-Kerr metric presented in this paper (or with any other beyond-Kerr spacetimes).
Such future work could be used to probe properties of the disk with the arbitrary free parameters utilized in this paper.


\section*{Acknowledgments}\label{acknowledgments}
We thank George Pappas for insightful comments on the spheroidicity conditions on the spacetime metric presented here.
Z.C. and K.Y. acknowledge support from NSF Award PHY-1806776. K.Y. would like to also acknowledge support by a Sloan Foundation Research Fellowship, the Ed Owens Fund, the COST Action GWverse CA16104 and JSPS KAKENHI Grants No. JP17H06358.

\appendix

\section{The effects of lower-order parameters and validity of $1/r$ expansion}\label{app:lowerOrder}
In this section we investigate the effects of lower-order parameters on the direct observables in the new spacetime: the Keplerian and epicyclic frequencies, and the photon rings.
Such lower-order parameters were originally assumed to be vanishing by Johannsen in Ref.~\cite{Johannsen:2015pca} due to strong constraints on the ppN parameters~\cite{Williams:2004qba}.
However, such constraints were obtained via observations of the local, weak-field solar system and may not hold true in the strong-gravity  regimes present near BHs, where the spacetime is not guaranteed to even be similar to that surrounding a star.
Thus, in this section we revive these neglected parameters:
\begin{itemize}
\item the first order parameter $\epsilon_1$ from the function $f(r)$\footnote{We note that to avoid a rescaling of the observable BH mass $M$, when $\epsilon_1\neq0$, we must set $\alpha_{11}=\epsilon_1/2$.}; 
\item the lowest-order parameter $\alpha_{12}$ from the $A_1(r)$ function; 
\item the lowest-order parameter $\alpha_{51}$ from the function $A_5(r)$; 
\item the first- and second-order parameters $\gamma_1$ and $\gamma_2$ from the function $g(\theta)$\footnote{We additionally consider the second-order parameter $\gamma_{2}$ corresponding to terms of $\cos^2 \theta$. This is because in all example metrics considered, typically $\cos^2\theta$ enters at first order in $g(\theta)$, as is the case of the Kerr-Sen metric.}.
\end{itemize} 

\begin{figure*}[htb]
\begin{center}
\includegraphics[width=.3\textwidth]{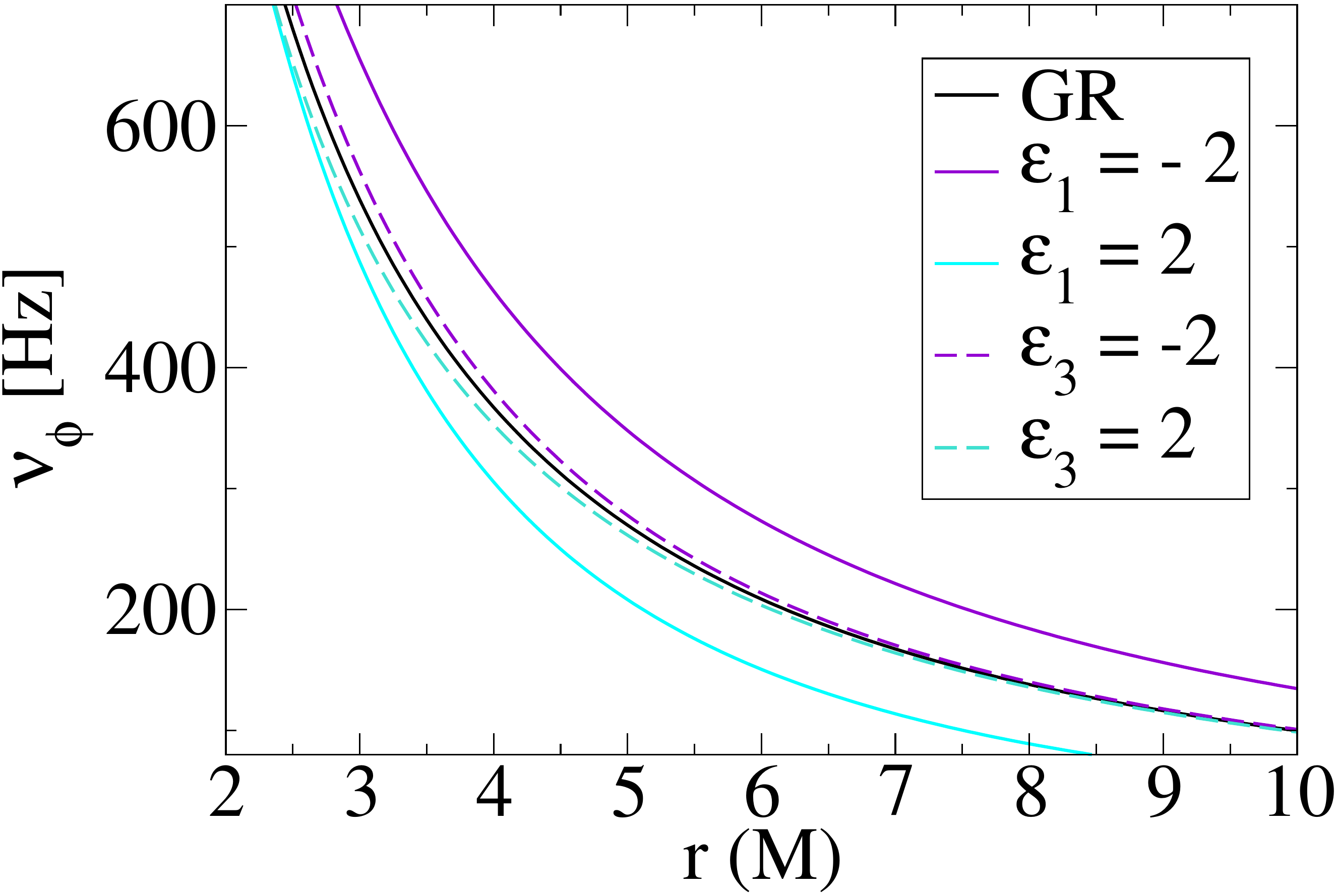}
\includegraphics[width=.3\textwidth]{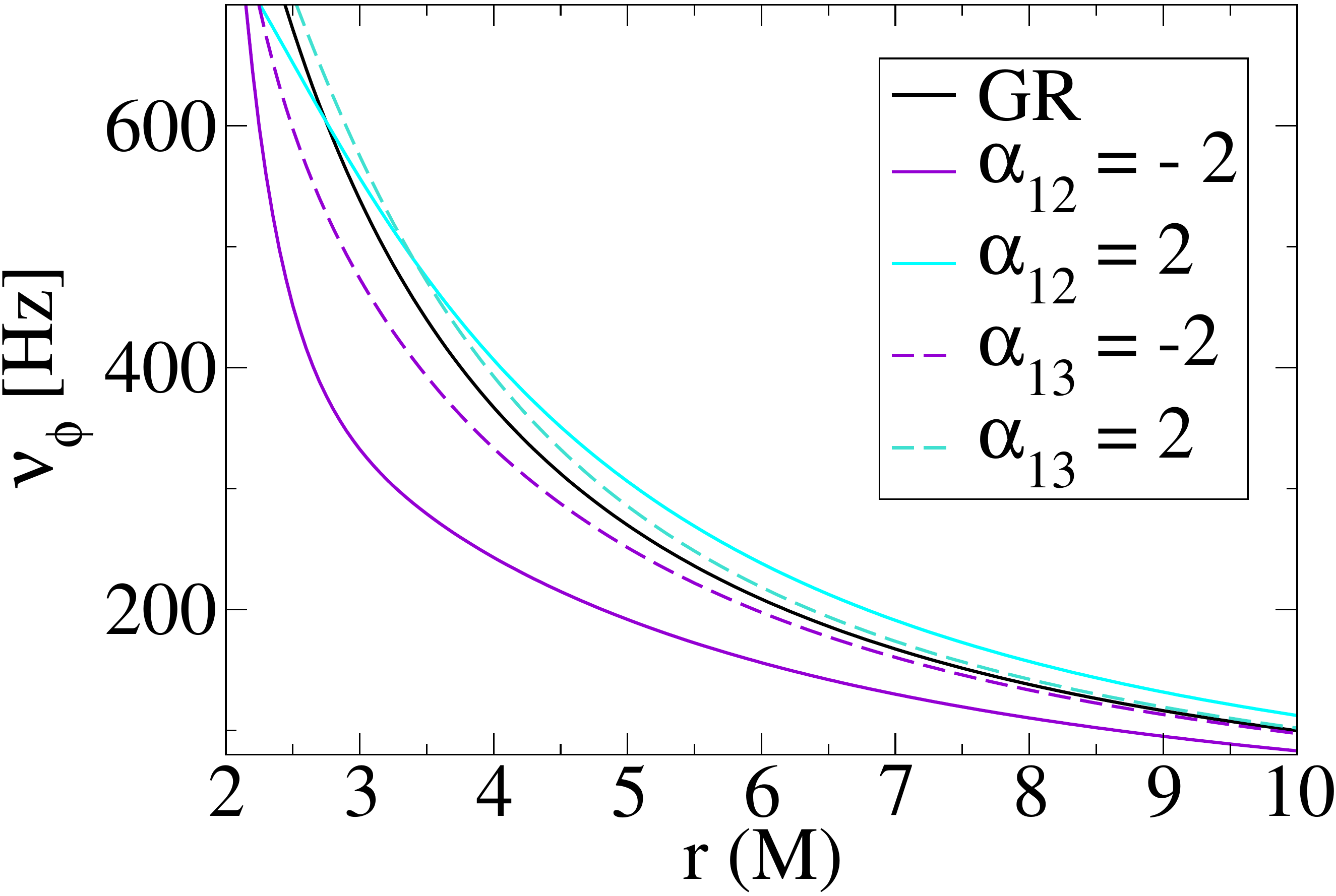}\\
\includegraphics[width=.3\textwidth]{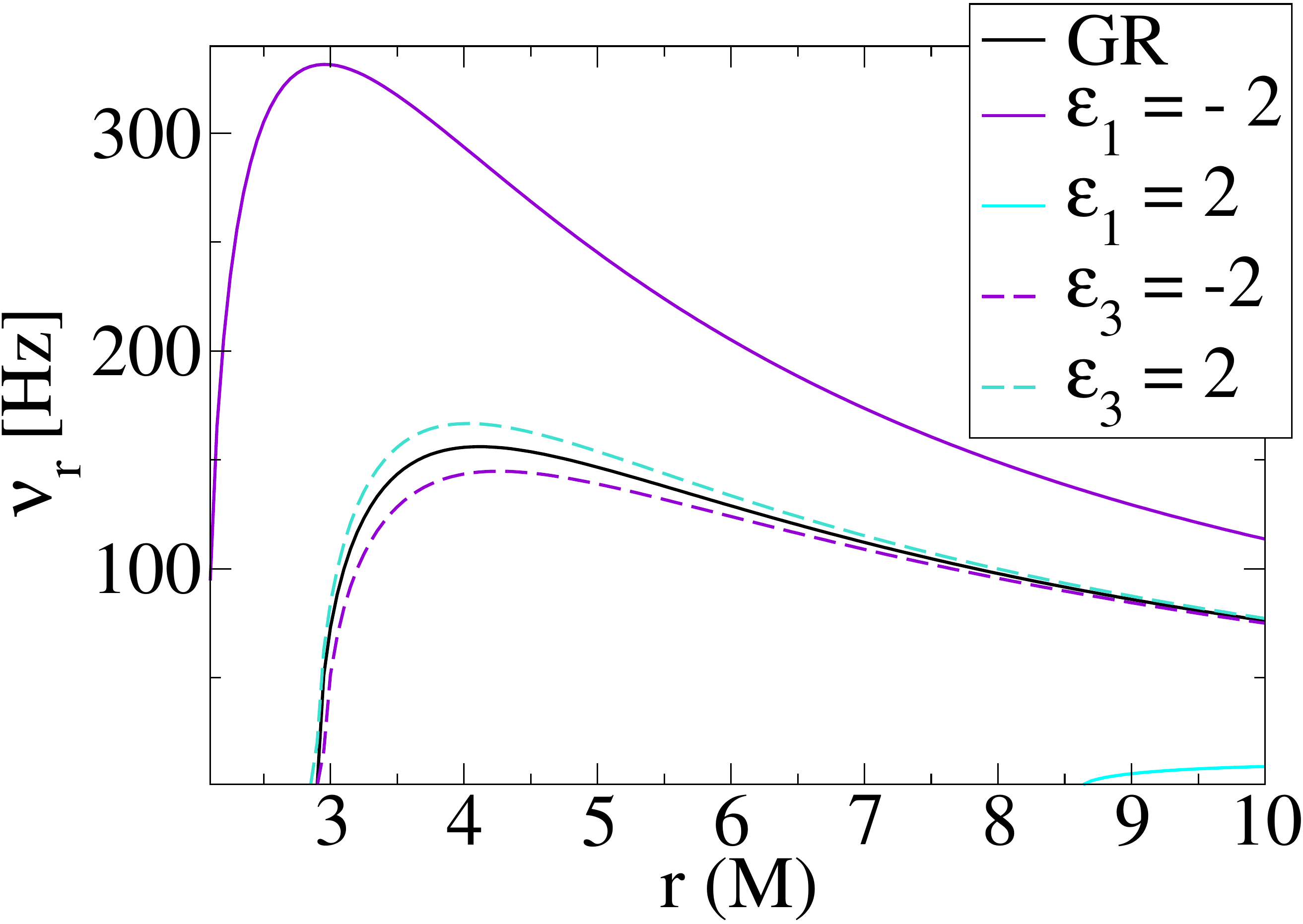}
\includegraphics[width=.3\textwidth]{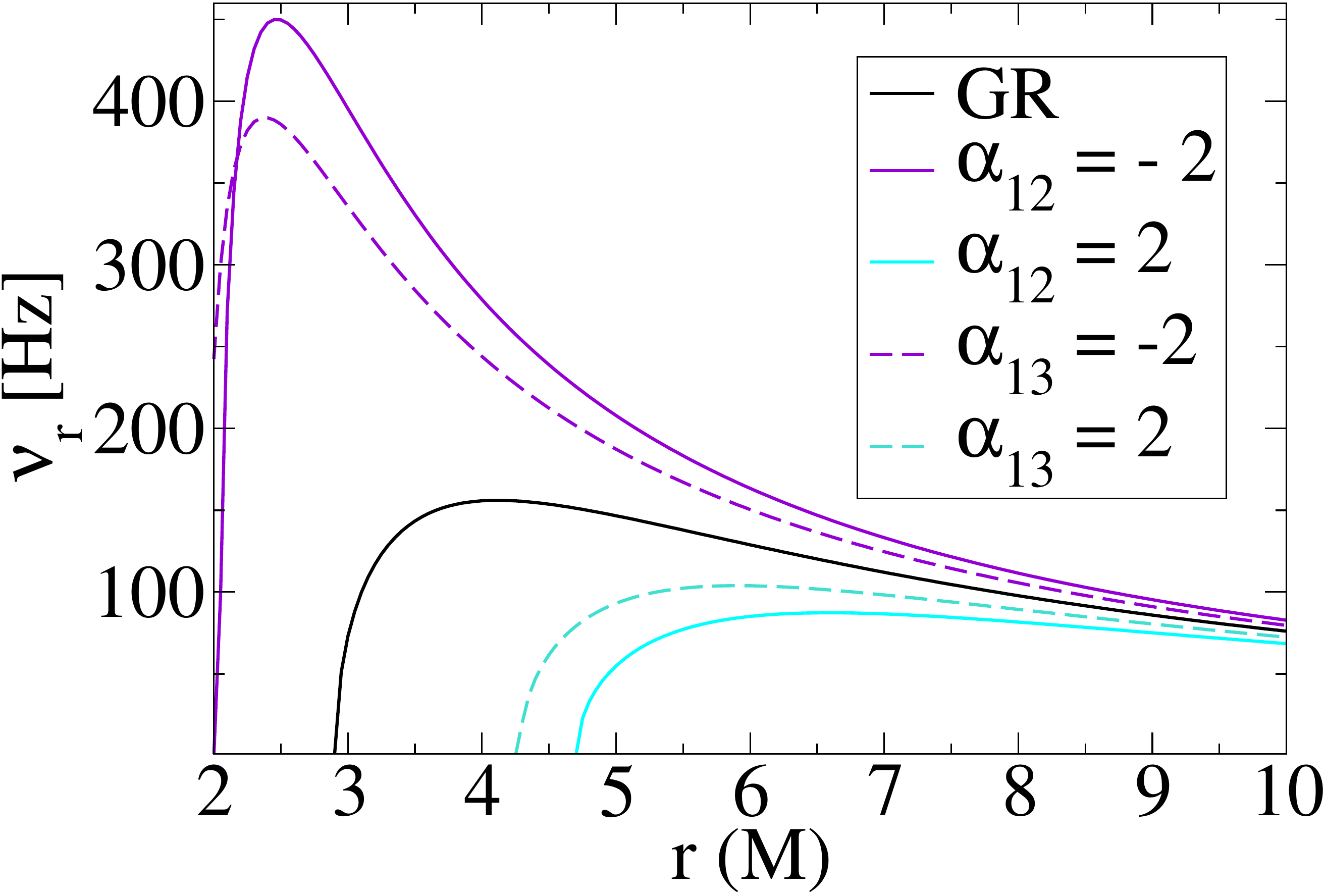}
\includegraphics[width=.3\textwidth]{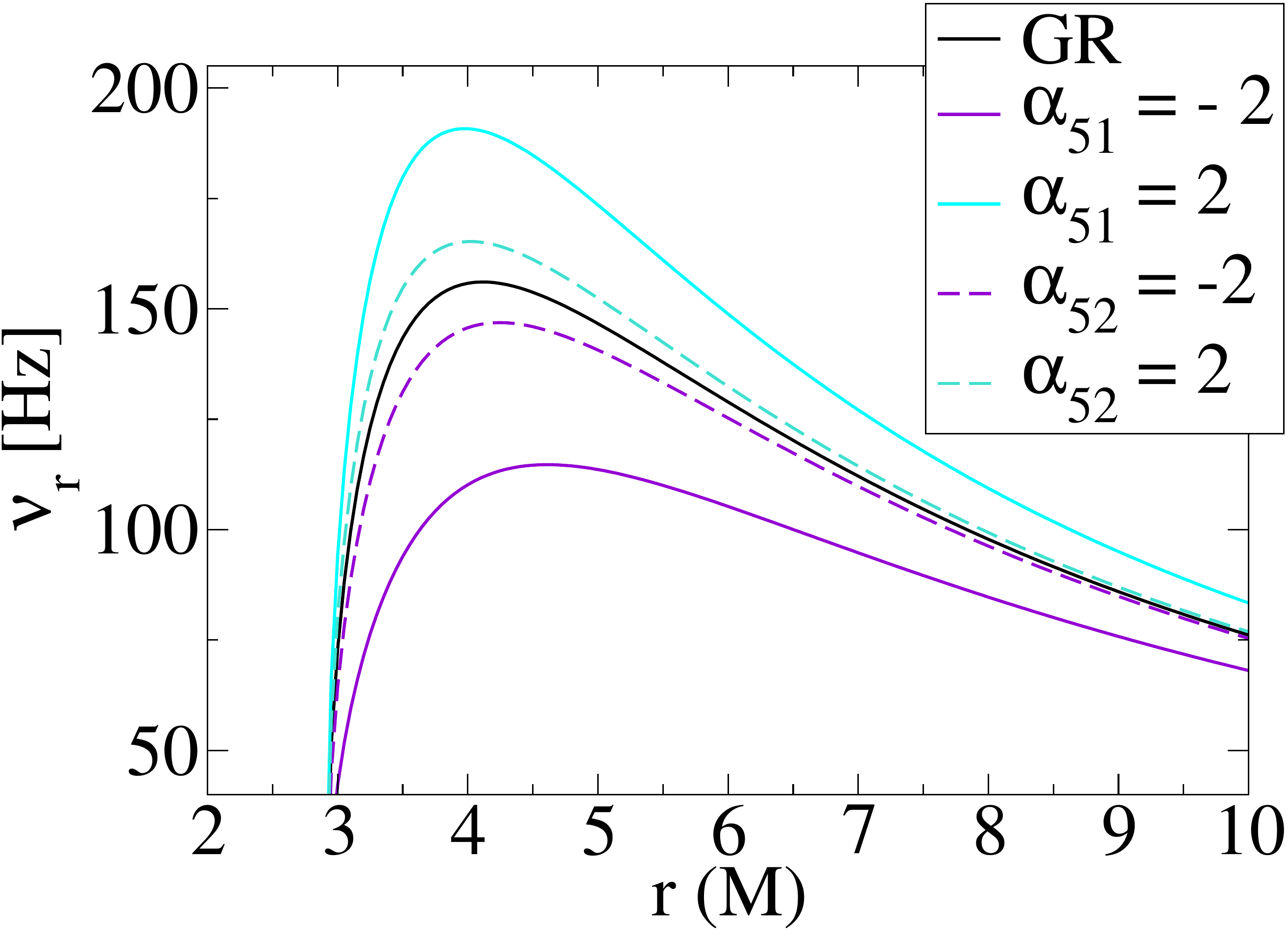}\\
\includegraphics[width=.3\textwidth]{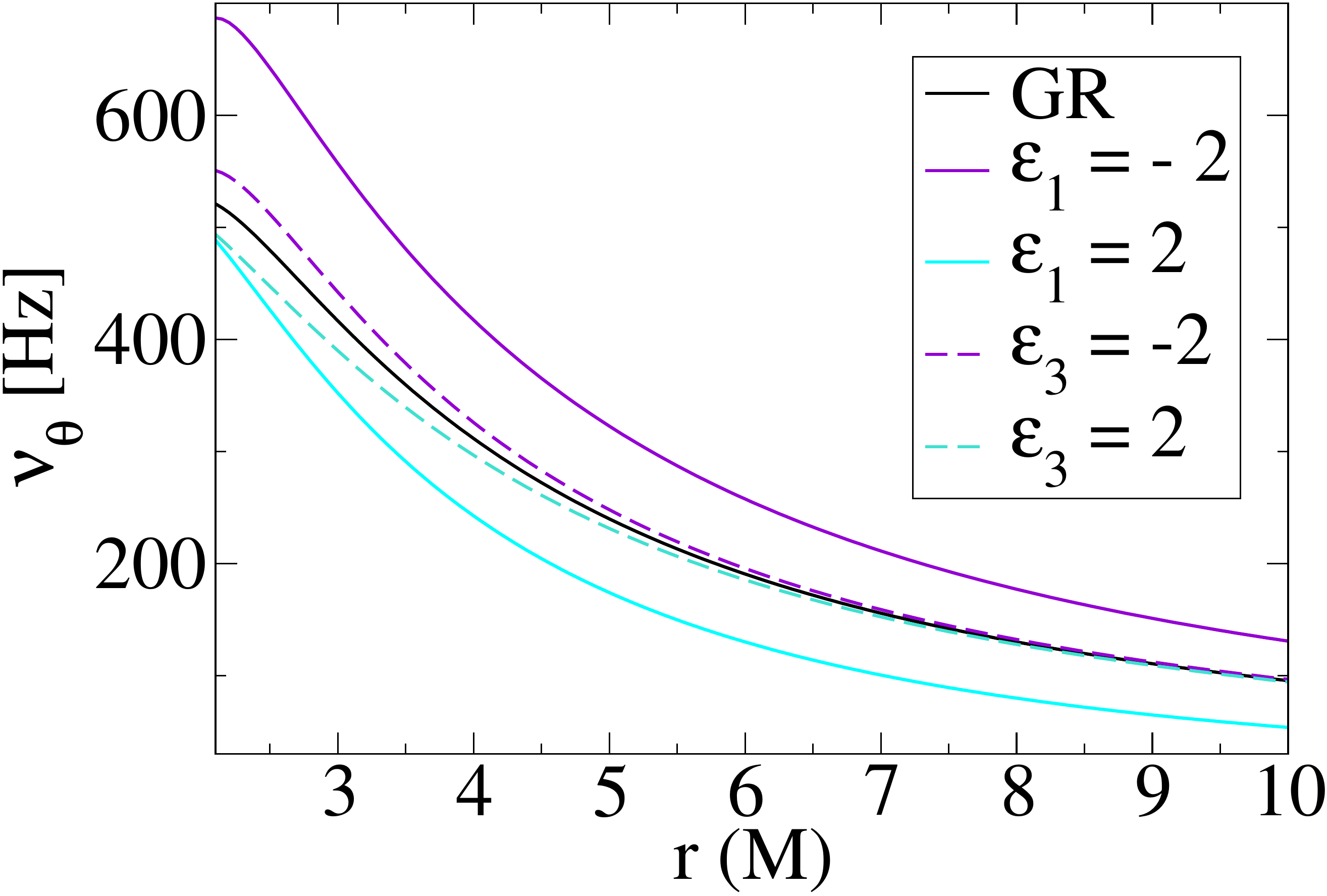}
\includegraphics[width=.3\textwidth]{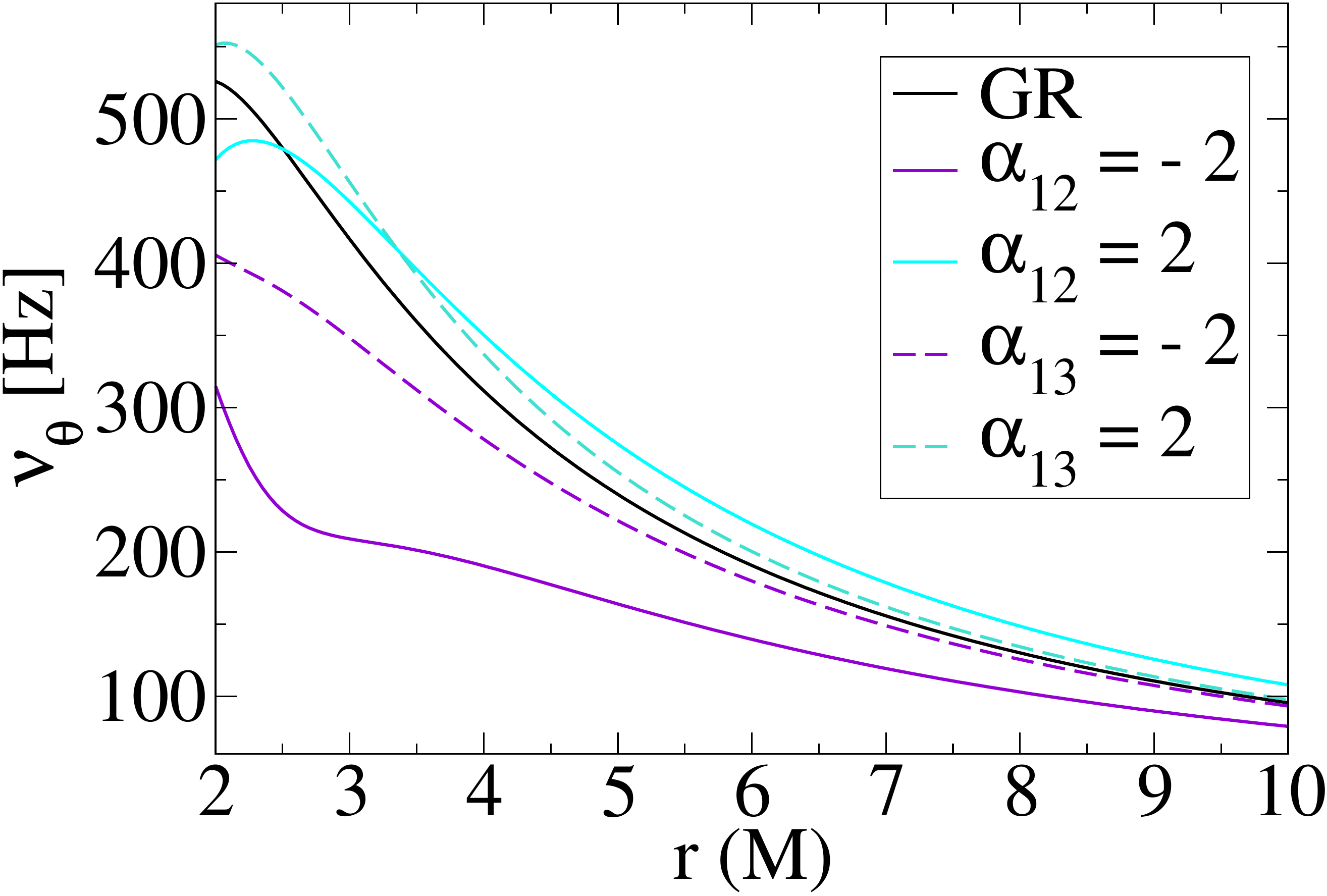}
\includegraphics[width=.3\textwidth]{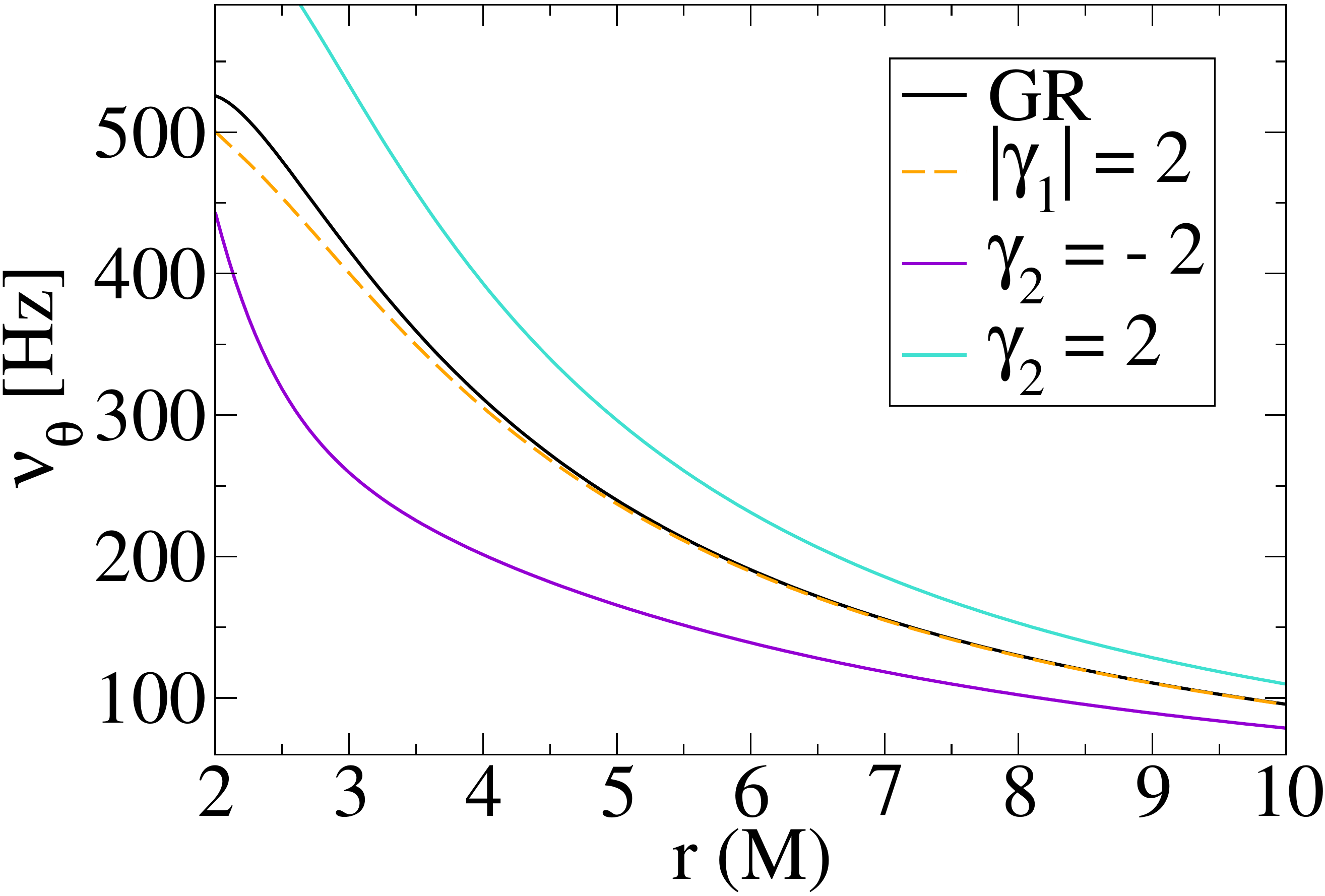}\\
\caption{Keplerian (top row), radial epicyclic (middle row), and vertical epicyclic frequencies about a BH with the lower order parameters $\epsilon_1$, $\alpha_{12}$, $\alpha_{51}$, $\gamma_1$, and $\gamma_2$ that were presumed to vanish in Ref.~\cite{Johannsen:2015pca} due to strong constraints on the ppN parameters~\cite{Williams:2004qba}.
Also shown for comparison in each case are the high-order parameters $\epsilon_3$, $\alpha_{13}$, and $\alpha_{52}$ that were used in the main analysis and~\cite{Johannsen:2015pca}.
We see that, especially for radial epicyclic frequencies, the lower order parameters impact the results strongly if they are indeed non-vanishing.
We note that when $\epsilon_1\ne0$, we have set $\alpha_{11}=\epsilon/2$ to prevent a rescaling of the BH mass.
}\label{fig:lowerOrder_nu}
\end{center}
\end{figure*}

We begin our investigation on the Keplerian and epicyclic frequencies' dependence on such lower-order parameters.
In the case of the former, we note that of the parameters we focus on in this section, only $\epsilon_1$ and $\alpha_{12}$ enter the expression (which is dependent on $g_{tt}$, $g_{\phi\phi}$, and $g_{t\phi}$) for equatorial orbits (where $\gamma_1 P_1(\cos\theta)$ vanishes entirely).
Similarly, the radial epicyclic frequencies depend only on $\epsilon_1$, $\alpha_{12}$, and now $\alpha_{51}$, due to the $g_{rr}$ dependence.
Finally, the vertical epicyclic frequencies depend on $\epsilon_1$, $\alpha_{12}$, and now $\gamma_1$ and $\gamma_2$, as a result of the $\partial_\theta$ derivatives.
In Fig.~\ref{fig:lowerOrder_nu}, we compare each of these orbital frequencies for two different cases: (i) only including the lower order parameters ($\epsilon_1$, $\alpha_{12}$, $\alpha_{51}$, $\gamma_1$ and $\gamma_2$) that were assumed to vanish in the ppN framework, and (ii) only including the next-higher-order parameters already used in the main analysis ($\epsilon_3$, $\alpha_{13}$, $\alpha_{52}$).
We observe that for every case the inclusion of the neglected lower-order parameters makes quite a large difference on the observables $\nu_\phi$, $\nu_r$, and $\nu_\theta$ as compared to the higher-order ones.
In particular, the radial epicyclic frequencies are significantly impacted upon the inclusion of $\epsilon_1$, $\alpha_{12}$, and $\alpha_{51}$.
This indicates that, if the parameters $\epsilon_1$, $\alpha_{12}$, $\alpha_{51}$, $\gamma_1$, and $\gamma_2$ are indeed non-vanishing and the ppN constraints applied to BHs are invalid, such lower-order parameters must be included for accuracy.

\begin{figure}[htb]
\begin{center}
\includegraphics[width=.4\textwidth]{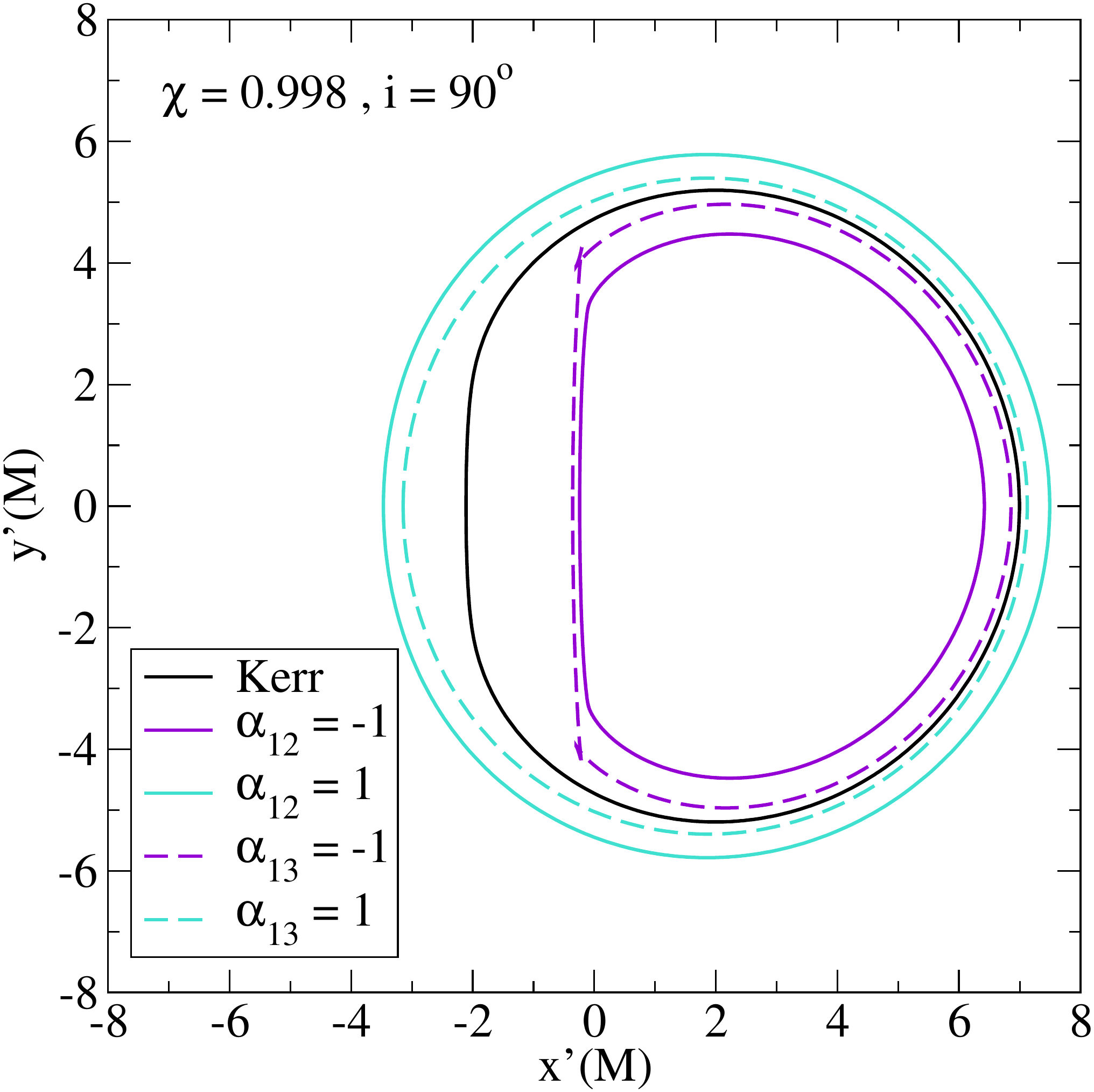}
\caption{
Comparison between the image of the photon rings about a BH when including lower order parameter $\alpha_{12}$ (presumed to vanish in Ref.~\cite{Johannsen:2015pca} due to strong constraints on the ppN parameters~\cite{Williams:2004qba}) and when instead including the next-order parameter $\alpha_{13}$.
Observe the difference made in the observable photon rings when including the lower-order parameter that may be non-vanishing if the solar-system ppN constraints are invalid in the vicinity of BHs.
}\label{fig:lowerOrder_shadows}
\end{center}
\end{figure}

Next we discuss another observable -- the BH photon rings.
We find that, of the lower-order parameters considered here, such expressions only depend on $\alpha_{12}$.
This is a result of $g(\theta)$ not entering the separated radial equations, $A_5(r)$ canceling out on each side of $R(r)=\frac{d}{dr}R(r)=0$, and a vanishing $\mu f(r)$ for photon orbits with $\mu=0$.
In Fig.~\ref{fig:lowerOrder_shadows} we plot the ensuing photon rings with (i) only $\alpha_{12}$ included, and (ii) only $\alpha_{13}$ included.
We observe that, if such lower-order parameters were non-vanishing in the ppN framework, their inclusion would make a sizable impact on the size (but not the shape) of the photon rings.

\begin{figure}[htb]
\begin{center}
\includegraphics[width=.4\textwidth]{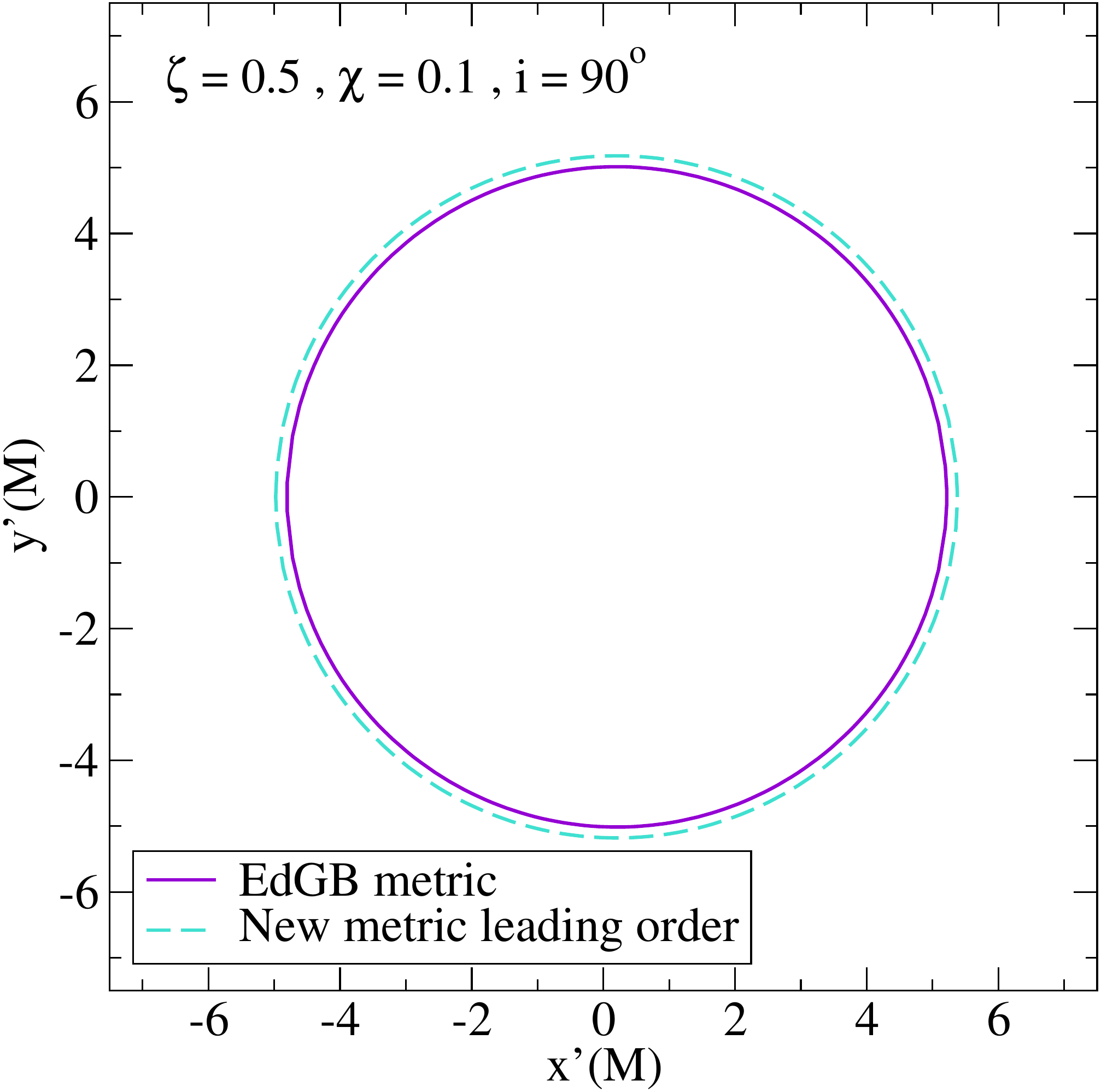}
\caption{
Comparison between photon ring images about a BH computed with two different methods: (i) using  the leading-order mapping of the new metric to EdGB found in Table~\ref{tab:maps} and (ii) using the full EdGB metric to first order in spin.
Here, the BH spin or EdGB coupling parameter $\zeta$ can not be too large or else the small-coupling or small-spin approximations begin to break down and the photon rings become nonsensical.
Observe how the two photon rings agree quite well, indicating the validity of the leading-order expansion in $1/r$.
}\label{fig:lowerOrder_EdGB_shadows}
\end{center}
\end{figure}

Finally we provide a brief analysis on the validy of including leading-order parameters in the computation of photon rings for the example of EdGB gravity.
To do this, we begin by computing the photon ring solutions in the EdGB theory of gravity, to first order in spin as described in Ref.~\cite{Ayzenberg:2014aka} (exact in the $1/r$ expansion).
Next we take the new metric and map to the EdGB theory of gravity as described in Table~\ref{tab:maps} using only the leading-order parameters in the $1/r$ expansion.
Figure~\ref{fig:lowerOrder_EdGB_shadows} compares the photon rings for these two cases for an EdGB BH with coupling parameter $\zeta=0.5$, spin $\chi=0.1$, and observer inclination $i=90^\circ$\footnote{We note that, due to the small-coupling and slow-rotation approximations used in the EdGB metric, neither $\zeta$ nor $\chi$ can be too large, else the approximations break down and the photon ring results become unreliable.}.
We observe that the two photon rings agree quite well, giving some indication to the validity of using only the leading-order terms in the $1/r$ expansion.


\section{Naked singularities in the new spacetime}\label{app:nakedSingularities}
\begin{figure}[htb]
\begin{center}
\includegraphics[width=.4\textwidth]{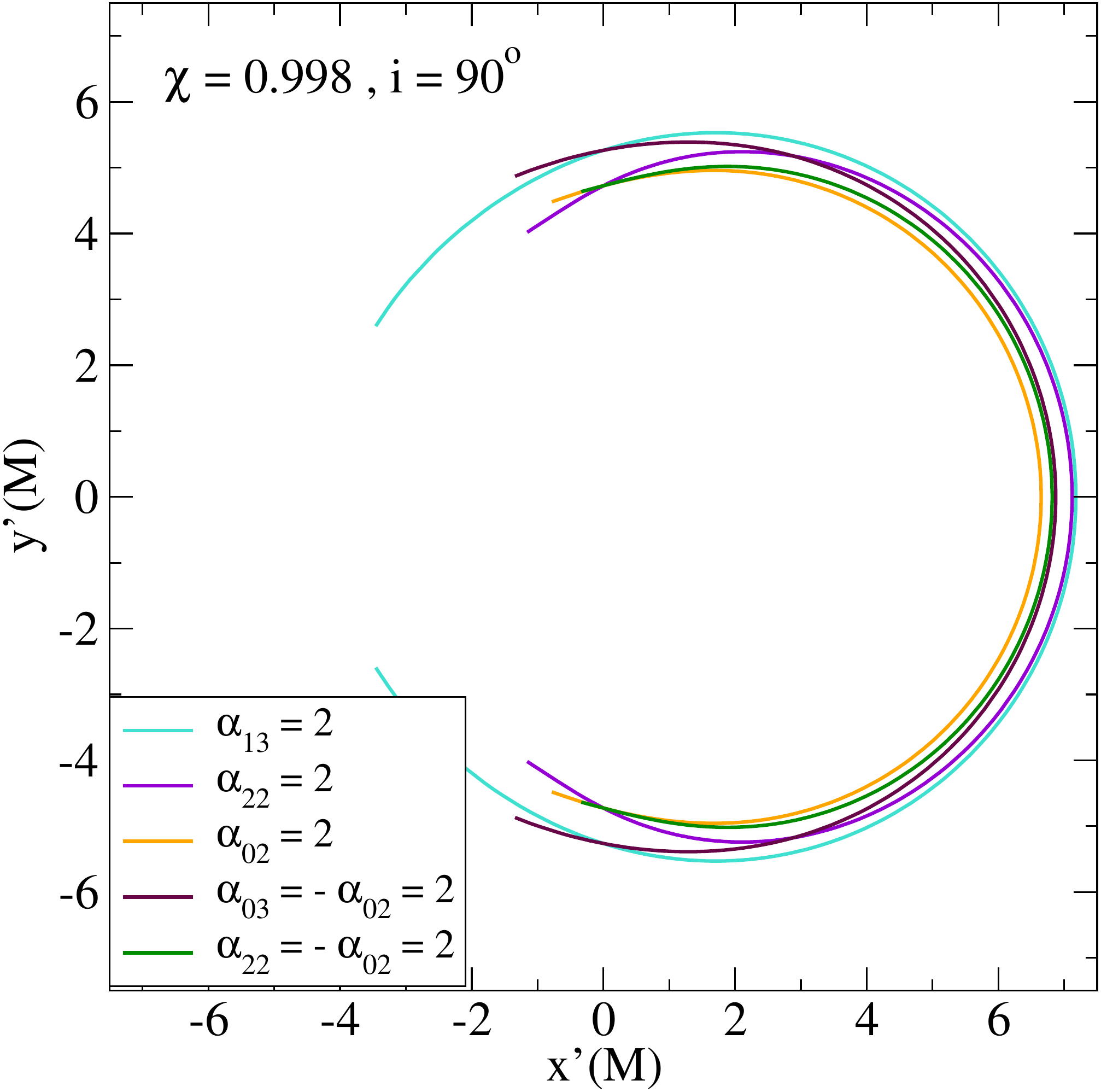}
\caption{(color online) Open photon orbits (``arcs'') about a BH with photons escaping to radial infinity.
Such orbits appear for certain parameterizations of the new metric similar to the ones presented here: $\alpha_{13}=2$, $\alpha_{22}=2$, $\alpha_{02}=2$, $\alpha_{13}=-\alpha_{02}=2$, and $\alpha_{22}=-\alpha_{02}=2$, with all other deviation parameters set to 0.
}\label{fig:nakedShadows}
\end{center}
\end{figure}

In this section we briefly discuss the emergent naked singularities for certain sections of the new metric's parameter space.
As also shown in e.g. Refs.~\cite{Hioki:2009na,Papnoi:2014aaa}, when a naked singularity is present outside of the BH event horizon $r_\EH$, closed photon rings no longer exist and the photons escape to radial infinity.
Figure~\ref{fig:nakedShadows} demonstrates this phenomena for several cases of BHs with naked singularities, such as $\alpha_{13}=2$, $\alpha_{22}=2$, $\alpha_{02}=2$, $\alpha_{13}=-\alpha_{02}=2$, or $\alpha_{22}=-\alpha_{02}=2$, with all other deviation parameters vanishing in each case.
Observe how for each case, a photon ``arc'' appears, and the orbit is not closed.
We similarly find in each such scenario that the other spacetime properties such as the orbital energy and angular momentum become discontinuous, negative, and complex.
We find that these naked singularities appear when any of $\alpha_{13}$, $\alpha_{22}$, or $\alpha_{02}$ appear as the sole non-vanishing parameter of the three, while the former two must appear alongside $\alpha_{02}$ to avoid such singularities.
We also find that when non-vanishing, the parameters $(\alpha_{13},\alpha_{02})$ or $(\alpha_{22},\alpha_{02})$ must share the same sign, else naked singularities appear.

Next we demonstrate that such open photon orbits are indeed indicative of emergent naked singularities in the new spacetime.
This is done by examining the nature of the Kretschmann invariant $K$ in both of the spacetimes that do and do not exhibit naked-singularity symptoms.
The Kretschmann invariant is given by
\begin{equation}
K=R_{\alpha\beta\gamma\delta}R^{\alpha\beta\gamma\delta}
\end{equation}
for Riemann curvature tensor given by
\begin{equation}
R^\alpha_{\beta\gamma\delta}=\Gamma^\alpha_{\beta\delta,\gamma}-\Gamma^\alpha_{\beta\gamma,\delta}+\Gamma^\mu_{\beta\delta}\Gamma^\alpha{\mu\gamma}+\Gamma^\mu_{\beta\gamma}\Gamma^\alpha{\mu\delta}
\end{equation}
with $X_{,k}$ representing a partial derivative $\frac{\partial X}{\partial x^k}$ and Christoffel symbols given by
\begin{equation}
\Gamma_{\alpha\beta\gamma}=\frac{1}{2}(g_{\alpha\beta,\gamma}+g_{\alpha\gamma,\beta}-g_{\beta\gamma,\alpha}).
\end{equation}
The scalar quantity $K$ is gauge invariant, and thus a divergence of $K$ is a sign for the presence of a true singularity.
For demonstration purposes, we pick a highly-rotating BH with $\chi=0.998$, for the two cases of $\alpha_{13}=2$, and $\alpha_{13}=\alpha_{02}=2$, where the former exhibits naked singularity behaviors, and the latter does not.
The event horizon for the latter case is located at $r_\EH=M+\sqrt{M^2-a^2}$, which reduces to $1.06$M for our given BH rotation.

\begin{figure}[htb]
\begin{center}
\includegraphics[width=.5\textwidth]{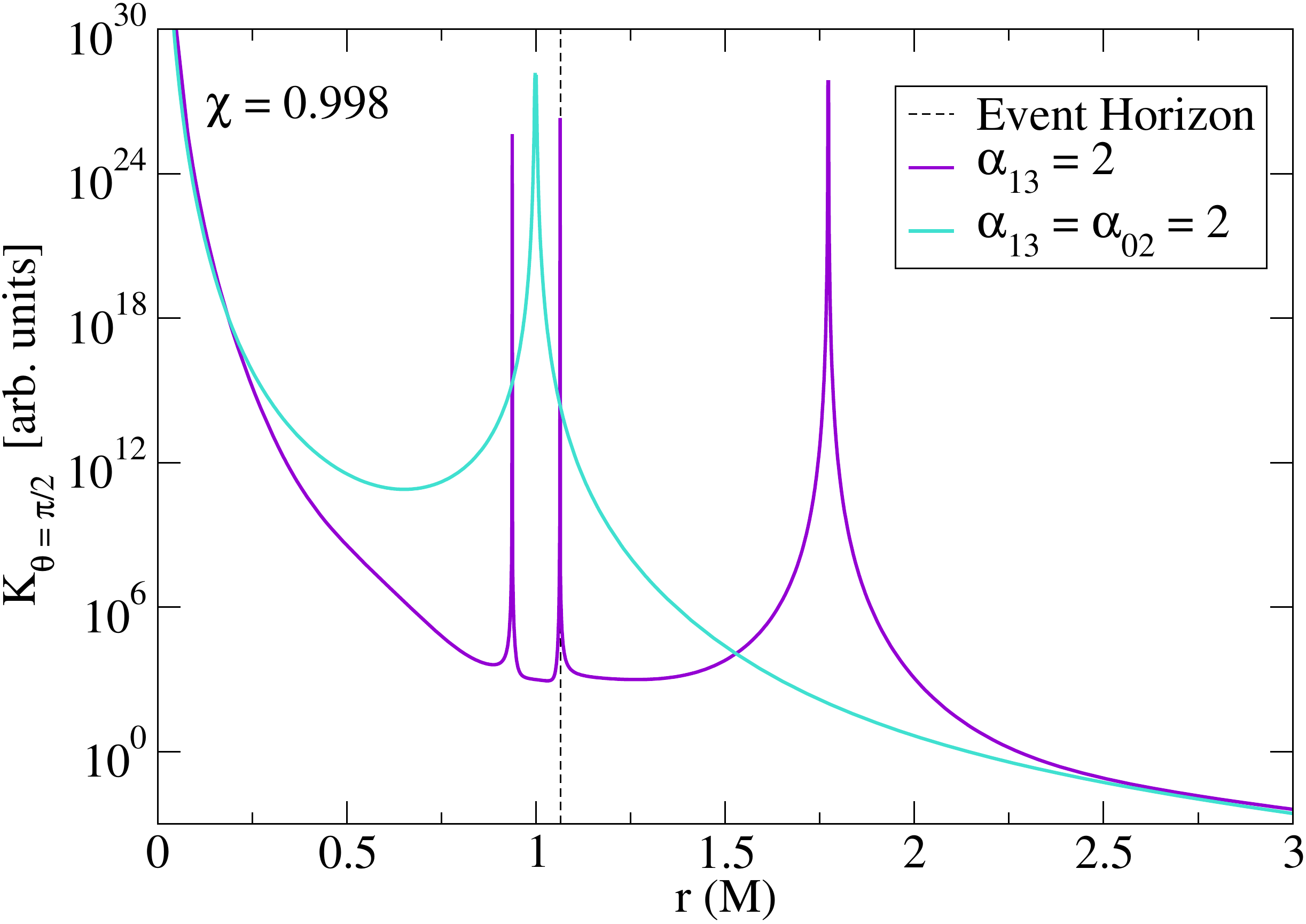}
\caption{Scalar Kretschmann invariant $K$ on the equatorial plane plotted in arbitrary units as a function of radius from the central BH described by the new metric.
This is plotted for two cases: (i) BH with non-vanishing parameter $\alpha_{13}=2$, and a (ii) BH with non-vanishing parameters $\alpha_{13}=\alpha_{02}=2$.
Also shown by the dashed vertical line is the event horizon of $r_\EH=M+\sqrt{M^2-a^2}=1.07$M for case (ii).
In case (i) we see, as predicted, that there exists a naked singularity at $r=1.77$M, well outside of the ``would-be'' event horizon.
Additionally, in this case there also exists a singularity at $r=0.93$M, and interestingly, on the ``would-be'' event horizon at $r=1.07$M.
In case (ii), we observe a singularity at $r=a=0.998$M, behind the event horizon as usual, confirming our predictions.
}\label{fig:invariant}
\end{center}
\end{figure}

Finally, we compute the Kretschmann invariant along the equatorial plane.
The results are shown in Fig.~\ref{fig:invariant} for both cases.
We observe that for the closed-photon orbit case of $\alpha_{13}=\alpha_{02}=2$, there exists a singularity at $r=a=0.998$M as usual, well behind the event horizon of $r_\EH=1.07$M.
Alternatively, for the open-orbit case of sole-non-vanishing parameter $\alpha_{13}=2$, we observe several interesting features.
First, we see a singularity behind the ``would-be'' event horizon at $r=0.93$M as one would expect.
Next, there exists a singularity directly on the ``would-be'' event horizon at $r=1.07$M, and finally we see a singularity well beyond the ``would-be'' event horizon at $r=1.77$M.
This confirms our suspicion of the existence of naked singularities, thus for the remainder of this analysis we avoid parameterizations that create such anomalies.


\clearpage
\bibliography{Zack}
\end{document}